\documentclass[preprint,showpacs,showkeys,aps]{revtex4}
\usepackage{amssymb}
\usepackage[pdftex]{graphicx}   
\usepackage{dcolumn}
\usepackage{bm}
\usepackage{latexsym}
\usepackage{ifthen}
\usepackage{lscape}
\usepackage{booktabs}
\usepackage{tabularx}
\usepackage{amssymb}
\usepackage{color}
\usepackage{hyperref}
\begin{document}
\newcommand{\ti}[1]{\mbox{\tiny{#1}}}
\newcommand{\im}{\mathop{\mathrm{Im}}}
\def\be{\begin{equation}}
\def\ee{\end{equation}}
\def\bea{\begin{eqnarray}}
\def\eea{\end{eqnarray}}
\newcommand{\tb}[1]{\textbf{\texttt{#1}}}
\newcommand{\rtb}[1]{\textcolor[rgb]{1.00,0.00,0.00}{\tb{#1}}}
\newcommand{\il}{~}
\title{Equatorial circular motion in Kerr spacetime}

\author{Daniela Pugliese$^{1,2}$, Hernando Quevedo$^{1,3}$, and Remo Ruffini$^1$}
\email{d.pugliese.physics@gmail.com, quevedo@nucleares.unam.mx, ruffini@icra.it}
\affiliation{$^1$Dipartimento di Fisica, Universit\`a di Roma "La Sapienza", Piazzale Aldo Moro 5, I-00185 Roma, Italy\\
ICRANet, Piazzale della Repubblica 10, I-65122 Pescara, Italy \\
$^2$ School of Mathematical Sciences, Queen Mary, University of London,
Mile End Road, London E1 4NS, United Kingdom\\
$^3$Instituto de Ciencias Nucleares, Universidad Nacional Aut\'onoma de M\'exico,
AP 70543, M\'exico, DF 04510, Mexico}

\date{\today}

\begin{abstract}
We analyze the properties of circular orbits of test particles on the equatorial plane of a rotating central mass whose gravitational
field is described by the Kerr spacetime. For rotating black holes and
naked singularities we explore all the spatial regions where circular orbits can exist and analyze the behavior of
the energy  and the angular momentum of the corresponding test particles. In particular, we find all the radii at
which a test particle can have zero angular momentum due to the repulsive gravity effects generated by naked singularities.
We classify all the stability zones of circular orbits. It is shown that the geometric structure of the stability zones
of black holes is completely different from that of naked singularities.
\end{abstract}

\maketitle

\section{Introduction}
\label{saxblues}
The Kerr spacetime describes the exterior gravitational field of a rotating mass $M$ with specific angular momentum $a=J/M$, where $J$ is the
total angular momentum of the gravitational source.
In Boyer--Lindquist coordinates, the Kerr metric has the form
\begin{equation} \label{alai} ds^2=dt^2-\frac{\rho^2}{\Delta}dr^2-\rho^2
d\theta^2-(r^2+a^2)\sin^2\theta
d\phi^2-\frac{2M}{\rho^2}r(dt-a\sin^2\theta d\phi)^2\ ,
\end{equation}
where
\be
\Delta\equiv r^2-2Mr+a^2,\quad\mbox{and}\quad\rho^2\equiv r^2+a^2\cos^2\theta \ .
\ee
This metric is an axisymmetric,   stationary (nonstatic) asymptotically flat solution of Einstein equations in vacuum.
The redshift infinity surface and event horizons are described
respectively by the equations
\be
g_{tt}=0, \quad g^{rr}=0 \ .
\ee
Then, the solutions of these equations are respectively
\be\label{consulta}
r^{0}_{\pm}=M\pm
\sqrt{M^2-a^2cos^2\theta}\ ,\quad\mbox{and}\quad r_{\pm}=M \pm\sqrt{M^2-a^2}\ .
\ee

Considering that $\theta\in [0,\pi]$, the radii $r^{0}_{\pm}$ and $r_{\pm}$  exist when
$|a| \leq M$ (Kerr black hole);  in particular,  for $|a|=M$ (extreme Kerr black hole)
the two horizons coincide, $r_+=r_-=M$.
The outer static limit is $r^0_+$, it corresponds to the outer boundary of the ergosphere.



A naked singularity case occurs when $|a|>M$ (for more details about the Kerr metric see, for instance,
 \cite{Chandra,Compr,GRLI} and \cite{Bardeen:1970zz,Bardeen:1972fi, Bredberg:2011hp,ZNK1980}).

The most important limiting cases are the Schwarzschild metric which is recovered for $a=0$, and
the Minkowski metric of special relativity for $a=M=0$.
The Kerr metric in Boyer--Lindquist coordinates is singular when $\rho = 0$ and when
$\Delta = 0$. However, a calculation of the Kretschmann  curvature scalar reveals that a true
curvature singularity occurs only for $\rho = 0$. Therefore, the surface represented by
 $r=0$ and $\theta=\pi/2$ corresponds to  an intrinsic curvature singularity \cite{MTW,Chandra, Compr,RuRR}.

In previous works \cite{Pugliese:2010he,Pugliese:2010ps,Mio1Que}, the motion of test particles along circular
orbits around static, spherically symmetric spacetimes was investigated in detail. We are now interested in
studying the more general case of a stationary, axisymmetric spacetimes. The study of the circular motion around
compact objects is of particular interest in the context of astrophysics. Indeed, an infinitesimal thin disk
of test particles traveling in circular orbits can be considered as an idealized model for an accretion disk
of matter surrounding the central body. Such an idealized model could be used, for instance, to estimate the amount of
energy released by matter being accreted by the central mass \cite{chandra42}. In addition,
 one can ask the question whether this hypothetical accretion
disk carries information about the nature of the central compact object. In a recent work
\cite{Pugliese:2010he,Pugliese:2010ps,Mio1Que}, this question was answered positively. Indeed, we found
that the geometric structure of the infinitesimal thin disk
around a Reissner-Nordstr\"om compact object strongly depends  on the mass-to-charge ratio.

In the present work, we generalize our previous analysis and study the motion of test particles along circular
orbits on the equatorial plane of the Kerr spacetime. We are especially interested in
studying the differences between the gravitational field of black holes and naked singularities.
Test particles moving along circular orbits are particularly appropriate to measure the effects generated
by naked singularities.
For the sake of simplicity,
we limit ourselves to the case of equatorial trajectories
because they are confined in the equatorial geodesic plane. This is a consequence of the fact
that the Kerr solution is invariant under reflections with respect to the equatorial plane.
Non-equatorial geodesics present an additional difficulty because the corresponding planes are not
geodesic. This case will not be considered in the present work.

This paper is organized as follows. In Sec. \ref{sec:orbkerr}, we use the formalism of the effective potential
to derive the conditions for the existence of circular orbits on the equatorial plane of the Kerr spacetime.
Sec. \ref{yappi} is devoted to the study of circular orbits around a rotating black hole.
In Sec. \ref{sec:orbalenskerr}, we investigate the case of naked singularities and find all the regions
where circular orbits are allowed. We analyze in detail the values of the energy and the angular momentum
as well as the stability properties of the test particles for  all the allowed regions in black holes and naked singularities.
In Sec. \ref{sec:sum}, we present a summary of behavior of the radii that determine the distribution of test particles around
the central body,  and of the radii of the last stable circular orbits. Finally, in Sec. \ref{malditesta ... :)} we
discuss our results.

\section{Circular orbits}
\label{sec:orbkerr}
We consider the circular motion of a test particle of mass $\mu$ in the
background represented by the Kerr metric (\ref{alai}). We limit ourselves
to the case of orbits situated on the equatorial plane only which are
defined by means of the conditions
\be
\theta=
\pi/2,\quad\mbox{and}\quad \frac{d\theta}{d\tau}=0\ ,
\ee
where $\tau$ is the  particle's proper time.
 We note that for $\theta=\pi/2$ the outer boundary of the ergosphere Eq.\il(\ref{consulta}) is $r^0_+=2M$ while  $r^0_-=0$.

The tangent   vector $u^{a}$ to a curve $x^\alpha(\tau)$ is
$
u^{\alpha}={dx^{\alpha}}/{d\tau}=\dot{x}^{\alpha}$.
The momentum $p^\alpha= \mu\dot x^\alpha$ of a particle with  mass $\mu$
can be normalized so that
$g_{\alpha\beta}\dot{x}^{\alpha}\dot{x}^{\beta}=-k$, where $k=0,1,-1$ for null, spacelike and timelike
curves, respectively.

Since the metric is independent of $\phi$ and $t$, the covariant
components $p_{\phi}$ and $p_{t}$ of the particle's four--momentum are
conserved along its geodesic. Thus, we  use the fact that the quantity
\be
E \equiv -g_{\alpha\beta}\xi_{t}^{\alpha} p^{\beta}
\ee
is a constant of motion, where $\xi_{t}=\partial_{t} $ is
the Killing field representing stationarity.
In general,  we may interpret $E$, for
timelike geodesics, as representing the total energy of the test particle
for a particle coming from radial infinity, as measured  by  a static observer at infinity. On the other hand, the
rotational Killing field $\xi_{\phi}=\partial_{\phi} $  yields
the following constant of  motion
\be
L \equiv
g_{\alpha\beta}\xi_{\phi}^{\alpha}p^{\beta}\ .
\ee
We interpret $L$ as the angular momentum  of the particle.
%

%
In this work, we analyze circular orbits involving a potential function $V(r)$.
It represents that value of $E/\mu$ that makes $r$ into a ``turning
point'' $(V=E/\mu)$, in other words, that value of $E/\mu$ at which
the (radial) kinetic energy of the particle vanishes \cite{Interessantissimo}.
The (positive) effective potential is
\begin{equation}\label{qaz}
V=-\frac{B}{2A}+\frac{\sqrt{B^2-4 A C}}{2A},
\end{equation}
where \cite{MTW,Chandra,Compr,GRLI, RuRR}
\bea
A&\equiv&\left(r^2+a^2\right)^2-a^2\Delta,\\
B&\equiv&-2aL\left(r^2+a^2-\Delta
\right),\\
C&\equiv& a^2L^2-\left(M^2r^2+L^2\right)\Delta\ .
\eea
The negative solution of the effective potential equation
\begin{equation}\label{qazm}
V^-\equiv-\frac{B}{2A}-\frac{\sqrt{B^2-4 A C}}{2A}
\end{equation}
can be studied by using the following symmetry
\be
V(L)=-V^{-}(-L).
\ee
%

We can note that the potential function (\ref{qaz}) is invariant under the mutual transformation
$a\rightarrow-a$ and $L\rightarrow-L$. Therefore, we will limit our analysis to the case of  positive values of $a$
for co--rotating  $(L>0)$ and counter--rotating orbits $(L<0)$.

The investigation of the motion of test particles in the spacetime (\ref{alai}) is
thus reduced to the study of motion in the effective potential $V$.
We will focus on (timelike) circular orbits for which (see also \cite{Bob})
\be\label{Eq:Kerrorbit}
\dot{r}=0,\quad V=E/\mu,\quad \partial V/\partial r=0.
\ee
Moreover, we use the following notation for the angular momentum solutions
\be\label{LPM}
\frac{L_{\pm}}{\mu M}\equiv\frac{\left|\frac{a^2}{M^2}\pm
2\frac{a}{M}
\sqrt{\frac{r}{M}}+\frac{r^2}{M^2}\right|}{\sqrt{\frac{r^2}{M^2}\left(\frac{r}{M}-3\right)\mp
2\frac{a}{M}\sqrt{\frac{r^{3}}{M^3}}}}, \ee
%
%
and the corresponding energies
\be
\frac{E^{(+)}_{\pm}}{\mu}\equiv\frac{E(L_{\pm})}{\mu}=\frac{\frac{(r^{5}M)^{1/4}\left|[a^2+(r-2M)r]\left(a \mp
\sqrt{\frac{r^{3}}{M}}\right)\right|}{\sqrt{(r-3M)\sqrt{\frac{r}{M}}\mp2a}}+
2arL_{\pm}}{r \left[r^3+a^2(r+2M)\right]},
\ee
and
%
%
\be
\frac{E^{(-)}_{\pm}}{\mu}\equiv\frac{E(-L_{\pm})}{\mu}=\frac{\frac{(r^{5}M)^{1/4}\left|[a^2+(r-2M)r]\left(a \mp
\sqrt{\frac{r^{3}}{M}}\right)\right|}{\sqrt{(r-3M)\sqrt{\frac{r}{M}}\mp2a}}-
2arL_{\pm}}{r \left[r^3+a^2(r+2M)\right]},
\ee
respectively. The investigation of the above expressions for the angular momentum and energy of the test particle
for different values of the radial coordinate allows us to extract physical information about the behavior
of the gravitational source. We mention for an  analysis of the test particle motion in Kerr spacetime  for  example  \cite{Komorowski:2009cg,Komorowski:2010we,Komorowski:2011cd,Hackmann:2009nh,Favata:2010ic,Bini:2004sy,Bini:1997eb,UJM,
DeFelice:1972ad,
DeFelice:1979mc,
Gariel:2007st,Chicone:2006rm,Lake:2010bq,Hackmann:2010zz,Bini:2004md,Bini:2004me,Bini:2005xt,Bini:2006pc,Dokuchaev:2011wm,Maeda,Gonzalez:2011fb,
Pani:2010jz}.

%
\section{Black holes}
\label{yappi}
In this section we shall consider the black hole case $0<a\leq M$.
In the non-extreme black hole case  $(0<a<M)$, it is $g_{tt}>0$ for $0<r<r^0_-$ and $r>r^0_+$. Inside the interval $r^0_-<r<r^0_+$ the metric component $g_{tt}$ changes its sign. Moreover, $g_{tt}$ vanishes for $r=r^0_{\pm}$ and $0<\cos^2\theta\leq1$, and also at $r=2M$
for $\theta=\pi/2$. The location of these hypersurfaces is such that $r^0_-<r_-<r_+<r^0_+$.

The region $r^0_-<r<r^0_+$, where $g_{tt}<0$, is called ergoregion. In this region  the Killing vector $\xi_t^{a} = (1, 0, 0, 0)$ becomes spacelike or $g_{ab}\xi_t^a\xi_t^b=g_{tt}<0$.
This fact implies in particular that a  static observer,  i.e. an observer with  four velocity  proportional
to $\xi_t^a$ so that $\dot{\theta}=\dot{r}=\dot{\phi}=0$,
cannot exist inside
the ergoregion; an observer inside this region is forced to move.

For  the extreme black hole case $(a=M)$ it holds $r_-=r_+ =M$.
Then, $g_{tt}>0$ for $0<r<r^0_-$ and $r>r^0_+$ when $0\leq\cos^2\theta<1$, and for $0<r<M$ and $r>M$ when $\cos^2\theta=1$;
moreover, $g_{tt}=0$ at $r=r^0_{\pm}$  in the interval  $0\leq\cos^2\theta<1$, and at $r=M$ for $\cos^2\theta=1$.
The location of the special radii is such that $r^0_-<r_-=r_+<r^0_+$ for $0\leq\cos^2\theta<1$ and $r^0_-=r_-=r_+=r^0_+$ for $\cos^2\theta=1$.

\begin{figure}
\centering
\begin{tabular}{c}
\includegraphics[scale=.71]{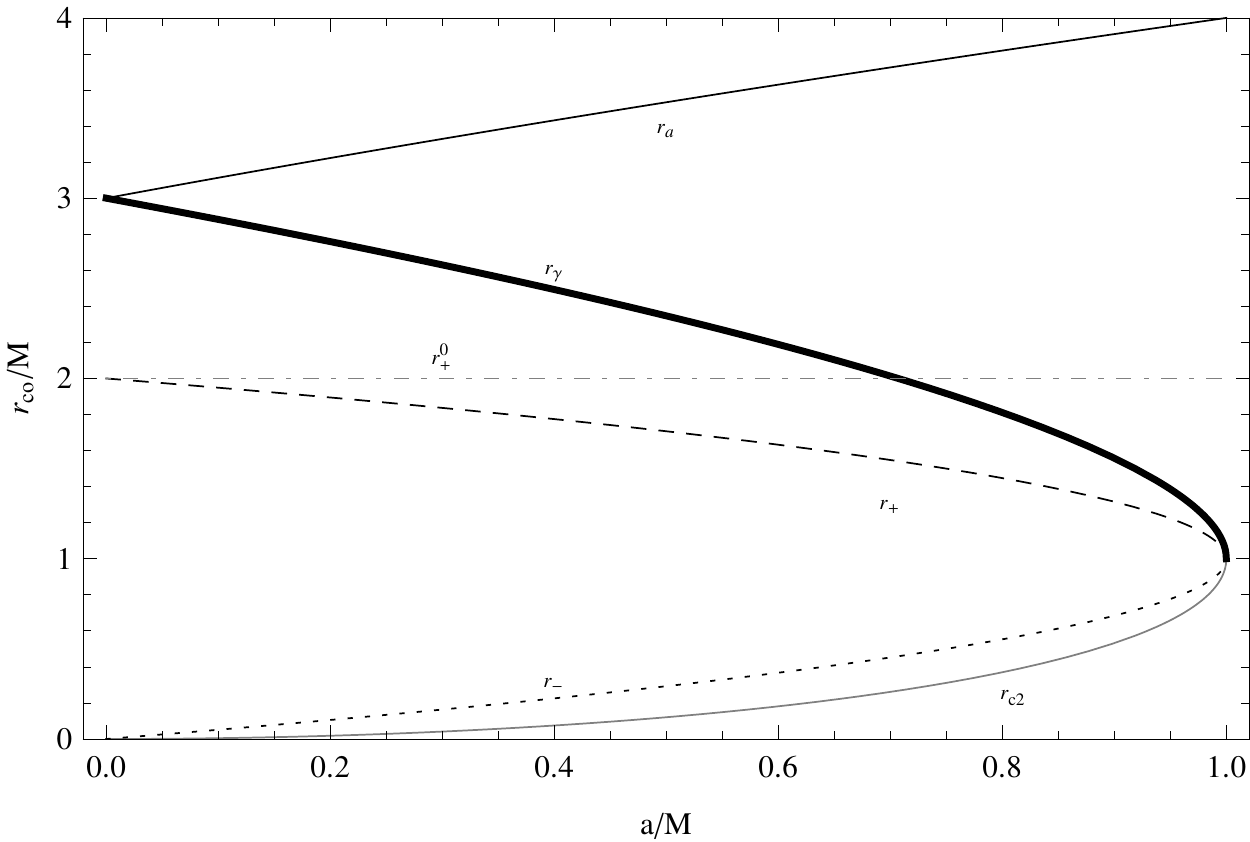}
\end{tabular}
\caption{\footnotesize{The outer horizon $r_+$ (dashed curve), the inner horizon $r_-$ (dotted curve), and $r_{a}$ (black curve), $r_{c2}$ (gray curve), and $r_{\gamma}$ (thick black curve) are plotted as function of the black hole intrinsic angular momentum $a/M$. The dotted dashed gray line represents  the outer boundary of the ergosphere $r_+^0=2M$.}}
\label{RCOKBH}
\end{figure}

To investigate the solutions of the conditions of circular motion given by Eq.\il(\ref{Eq:Kerrorbit}) it is convenient
to introduce the following radii
\bea
\label{rc1}r_{a}&\equiv& 4M \cos\left[\frac{1}{6} \arccos\left[-1+2 \frac{a^2}{M^2}\right]\right]^2,\\
r_{c2}&\equiv&4M \sin\left[\frac{1}{6} \arccos\left[1-2 \frac{a^2}{M^2}\right]\right]^2,\\
\label{rc3}r_{\gamma}&\equiv&2M \left(1+\sin\left[\frac{1}{3} \arcsin\left[1-2 \frac{a^2}{M^2}\right]\right]\right),
\eea
which have the two limiting cases
\be
r_{a}=r_{\gamma}=3M, \quad r_{c2}=0\quad\mbox{for}\quad a=0,
\ee
and
\be
r_{a}=4M, \quad r_{c2}=r_{\gamma}=M\quad\mbox{for}\quad a=M,
\ee
The dependence of these radii from the specific angular momentum is shown in Fig.\il\ref{RCOKBH}.
It is then possible to show that circular orbits can exist only for $r>r_{\gamma}$ and that there are two
regions with different values for the angular momentum, namely
\be
\label{1orbi}
r_{\gamma}<r\leq r_{a}, \quad\mbox{where }\quad  L=L_-,
\ee
and
\be\label{2orbit}
r>r_{a},\quad\mbox{where }\quad L=-L_+,\quad\mbox{and}\quad L=L_- \ .
\ee
Moreover, in the extreme black hole case, $a=M$, the circular orbits are situated at
\be
r=r_{a}=4M\ ,
\ee
with two different possible values for the angular momentum
\be
 L= \frac{13}{4 \sqrt{2}}M\mu \quad \mbox{with} \quad E =\frac{5}{4 \sqrt{2}}\mu,
\quad\mbox{and} \quad
L=- \frac{13}{4 \sqrt{2}}M\mu \quad \mbox{with} \quad E = \frac{149}{140 \sqrt{2}}\mu\ .
\ee

\begin{figure}
\centering
\begin{tabular}{c}
\includegraphics[scale=.7]{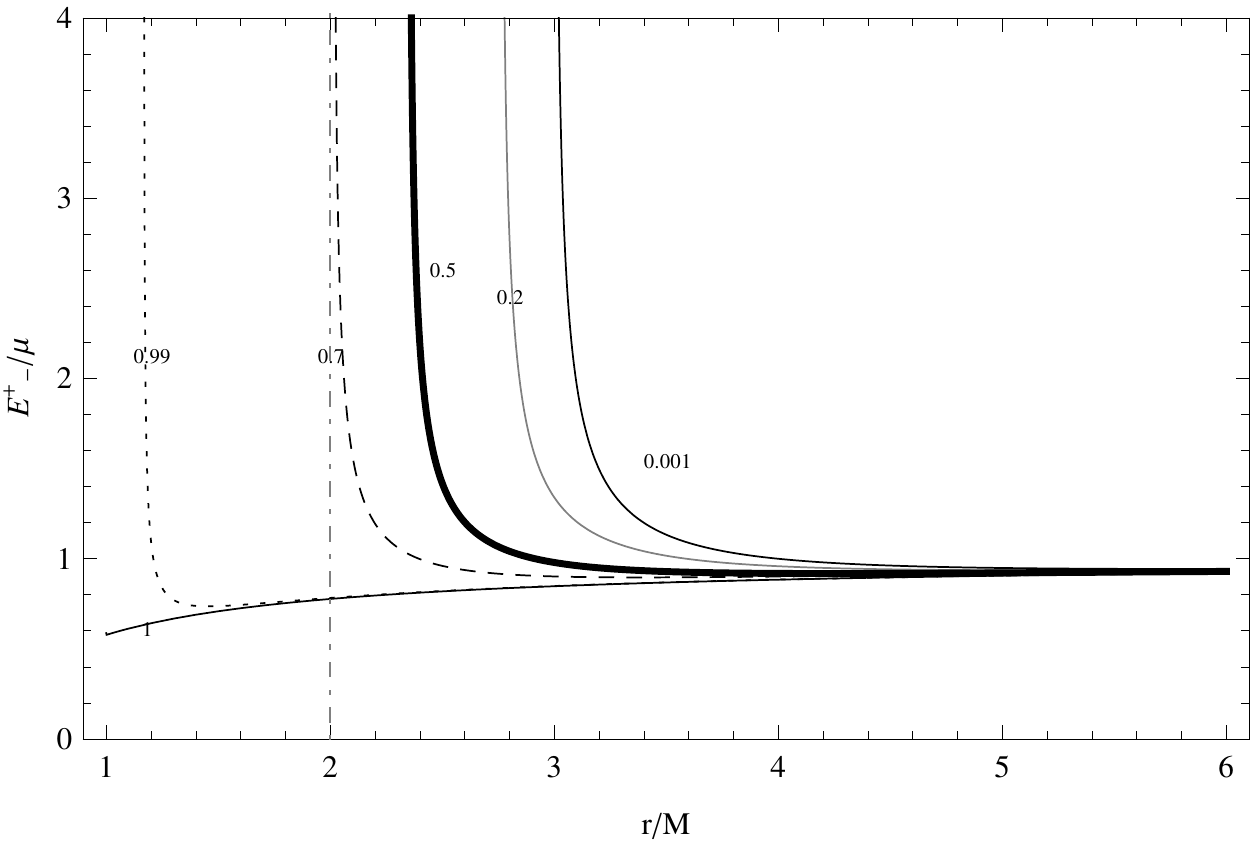}\\
\includegraphics[scale=.7]{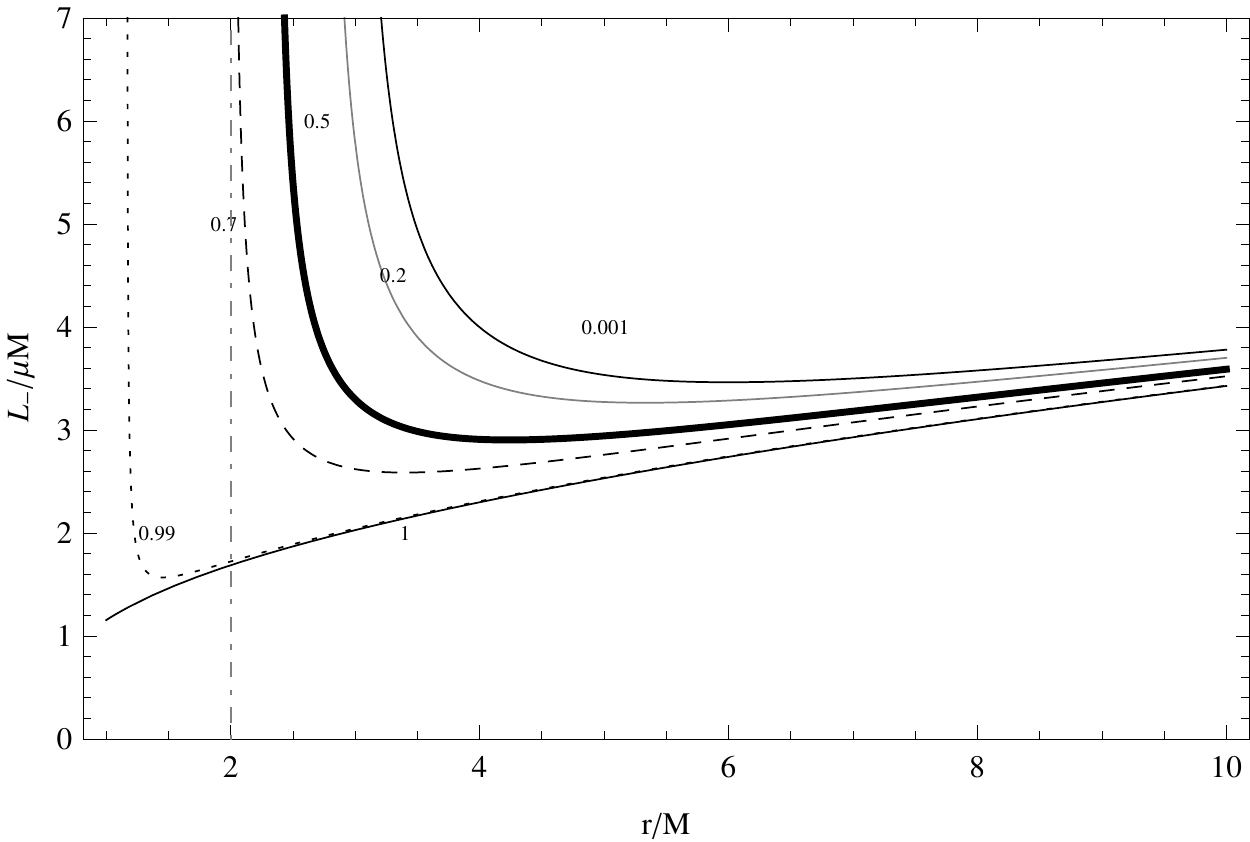}
\end{tabular}
\caption[font={footnotesize,it}]{\footnotesize{The
energy $E^{(+)}_-\equiv E(L_-)$ (upper plot) and the angular momentum $L_-$ (bottom plot) of circular orbits in a rotating Kerr black hole with  angular
momentum $0<a\leq M$, for selected values of $a/M$ in the interval $r>r_{\gamma}$.
 For $a\neq M$ the energy  $E^{(+)}_-$ is always positive and diverges as $r$ approaches $r_\gamma$. The dotted dashed gray line represents  the outer boundary of the ergosphere $r_+^0=2M$.}}
\label{Vlmplp}
\end{figure}

\begin{figure}
\centering
\begin{tabular}{c}
\includegraphics[scale=.7]{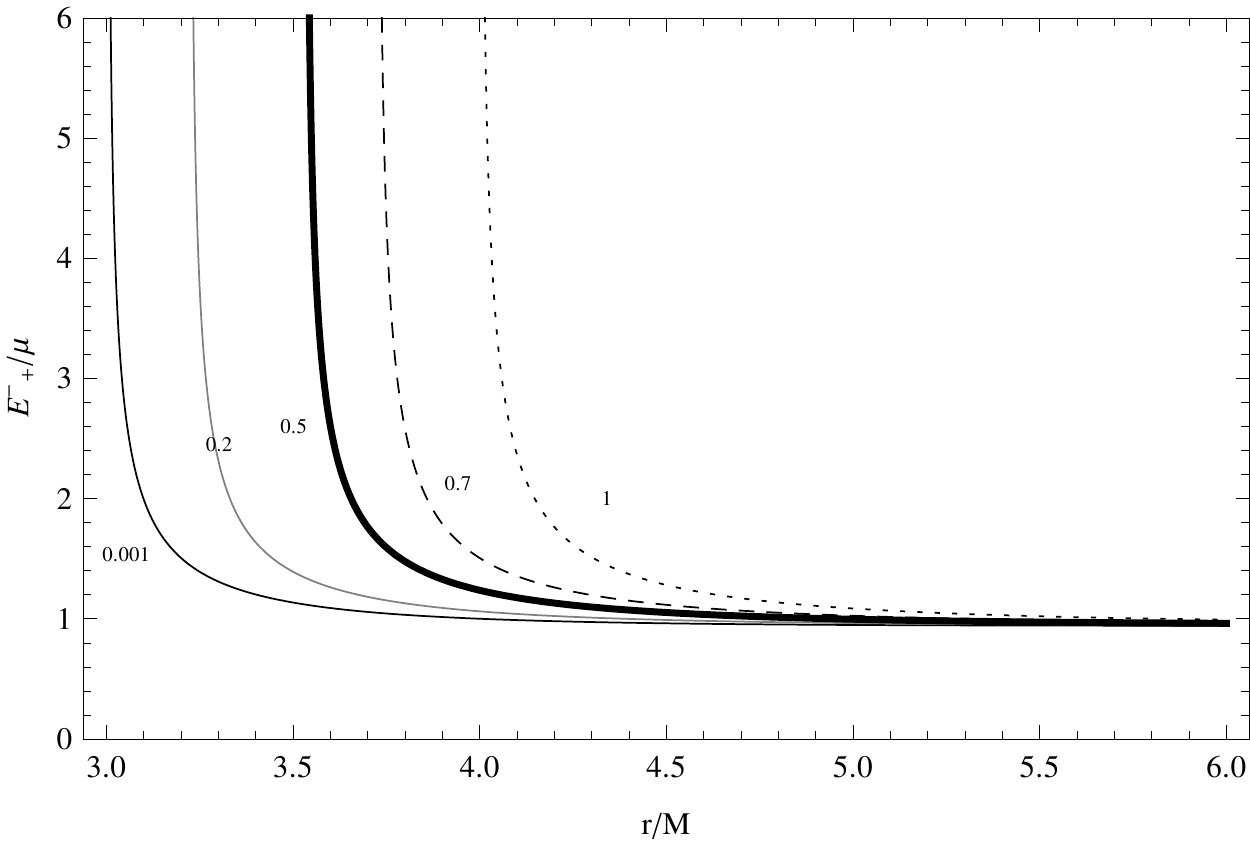}\\
\includegraphics[scale=.7]{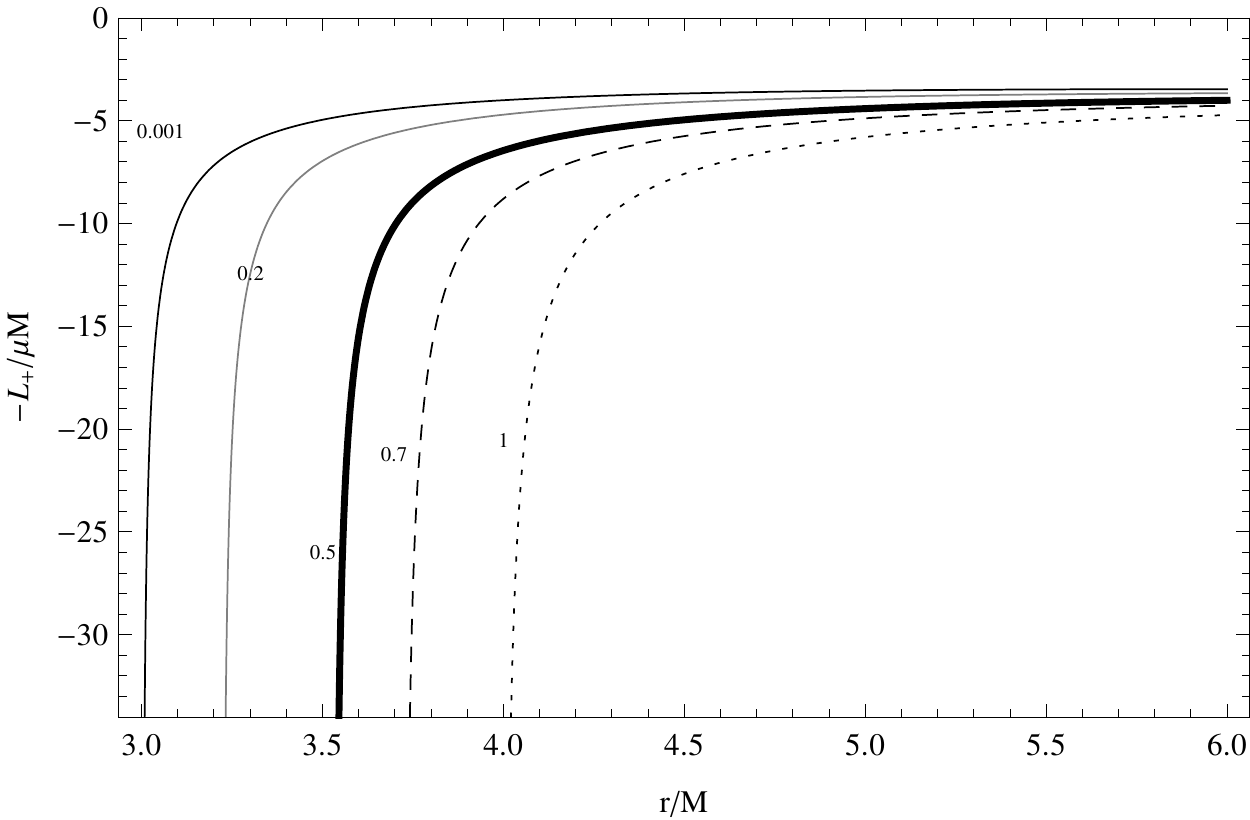}
\end{tabular}
\caption[font={footnotesize,it}]{\footnotesize{The
energy $E^{(-)}_+\equiv E(-L_+)$ and angular momentum  $L=-L_+$ of circular orbits in a Kerr black hole with angular
momentum $0<a\leq M$, for selected values of $a/M$ in the region $r>r_{a}$.
The energy $E^{(-)}_+$ is always positive and
diverges as $r$ approaches $r_{a}$.}}
\label{Vmpl}
\end{figure}

As for the first interval $r_{\gamma}<r\leq r_{a}$, the behavior of the corresponding energy and angular momentum is illustrated
in Fig. \ref{Vlmplp}. First we note that the area covered by this region increases as the specific angular momentum of the black
hole increases. Whereas $r_{a}$ and $r_{\gamma}$ coincide and equal $3M$ for nonrotating black holes $(a=0)$, their maximum separation is reached
in the case of extreme black holes $(a=M)$ for which $r_{\gamma}$ coincides with the outer horizon radius.
In the region $r>r_{a}$, circular orbits are allowed with different values of the angular momentum (the particular case with $L=-L_+$ is illustrated
in  Fig.\il\ref{Vmpl}).

We see that in the gravitational field of a black hole with $0<a<M$, particles with angular momentum $L=L_-$
can exist in the entire region $r>r_\gamma$. As the radius $r_\gamma$ is approached the angular
momentum $L_-$ and the corresponding energy $E^{(+)} _- =E(L_-)$ diverge, indicating that the hypersurface $r=r_\gamma$ is lightlike, i.e.,
it is the limiting orbit for timelike particles with $L=L_-$. Moreover, particles with angular momentum $L=-L_+$ can move along
circular orbits in the interval $r>r_a$, and the limiting lightlike counter--rotating orbit corresponds to $r=r_a$ where both $L_+$ and the energy
$E^{(-)}_+=E(-L_+)$ diverge.
%
%
\clearpage
%
\subsection{Stability}
\label{sec:bhstab}
\begin{figure}
\centering
\begin{tabular}{cc}
\includegraphics[scale=1]{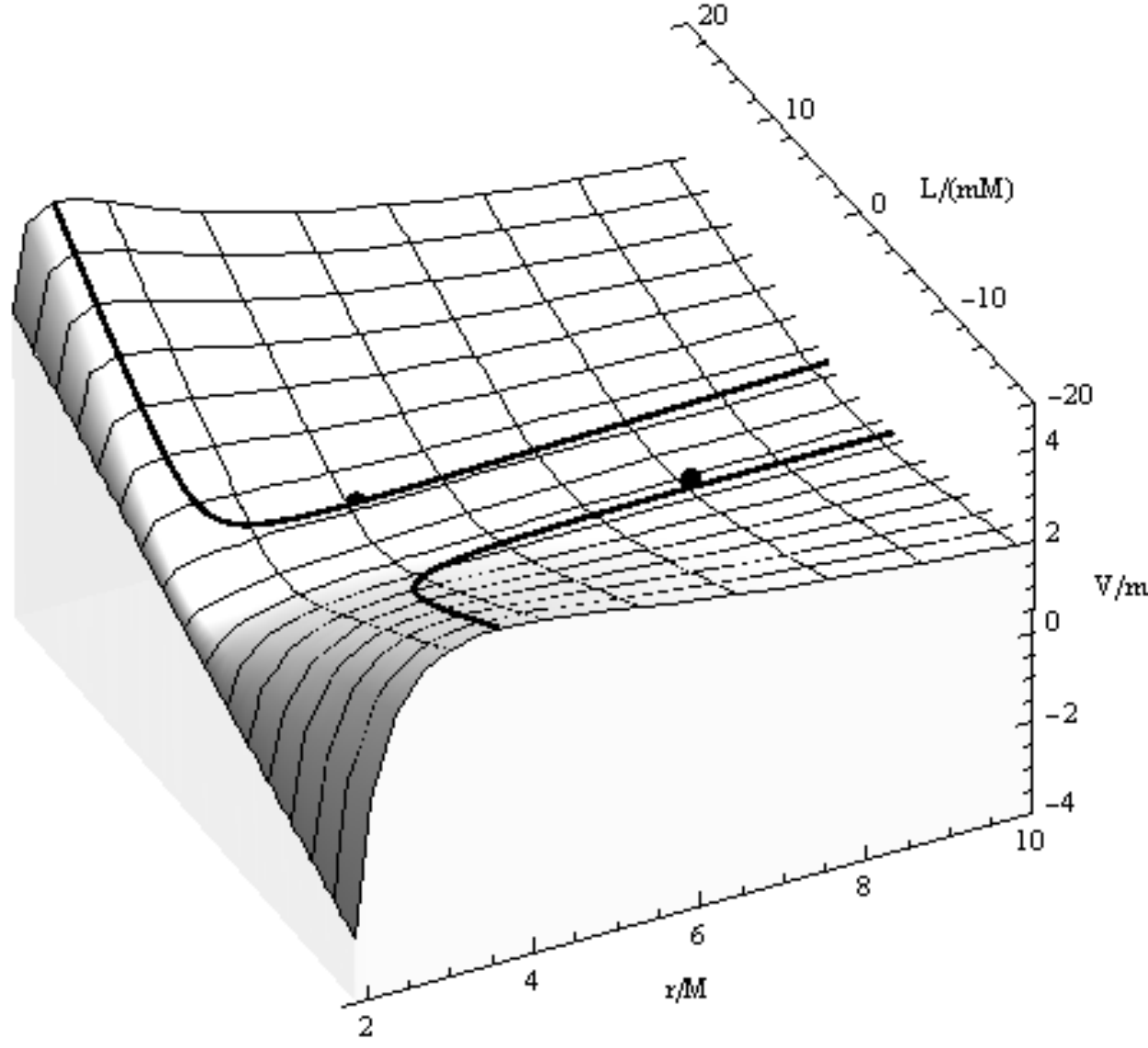}
\end{tabular}
\caption{\footnotesize{The effective
potential $V/m$  for a neutral particle of mass $\mu$ in a
Kerr black hole with   $a/M=0.5$  is
plotted as a function of $r/M$  in the range $[1.71,10]$ for the radial coordinate and
 in the range $[-10,10]$ for the angular momentum $L/(\mu M)$.
 The outer horizon is situated at
$r_{+}=(1+1/\sqrt{2})M$ (see text). Circular orbits
exist for $r>2 \left[1+\sin(\pi/18)\right]M\approx2.347M$. The solid curve represents the location of
circular orbits (stable and unstable). Stable (unstable) circular orbits are minima (maxima) of the effective potential function. The last stable circular
orbits are represented by a point. The minima are located at $r=4.233M$ with $L=2.903/(M\mu)$ and $E=0.918\mu$, and
at $r=7.554M$ with $L=-3.884/(M\mu)$ and $E=0.955\mu$}.}
\label{SV5}
\end{figure}

\begin{figure}
\centering
\begin{tabular}{cc}
\includegraphics[scale=.271]{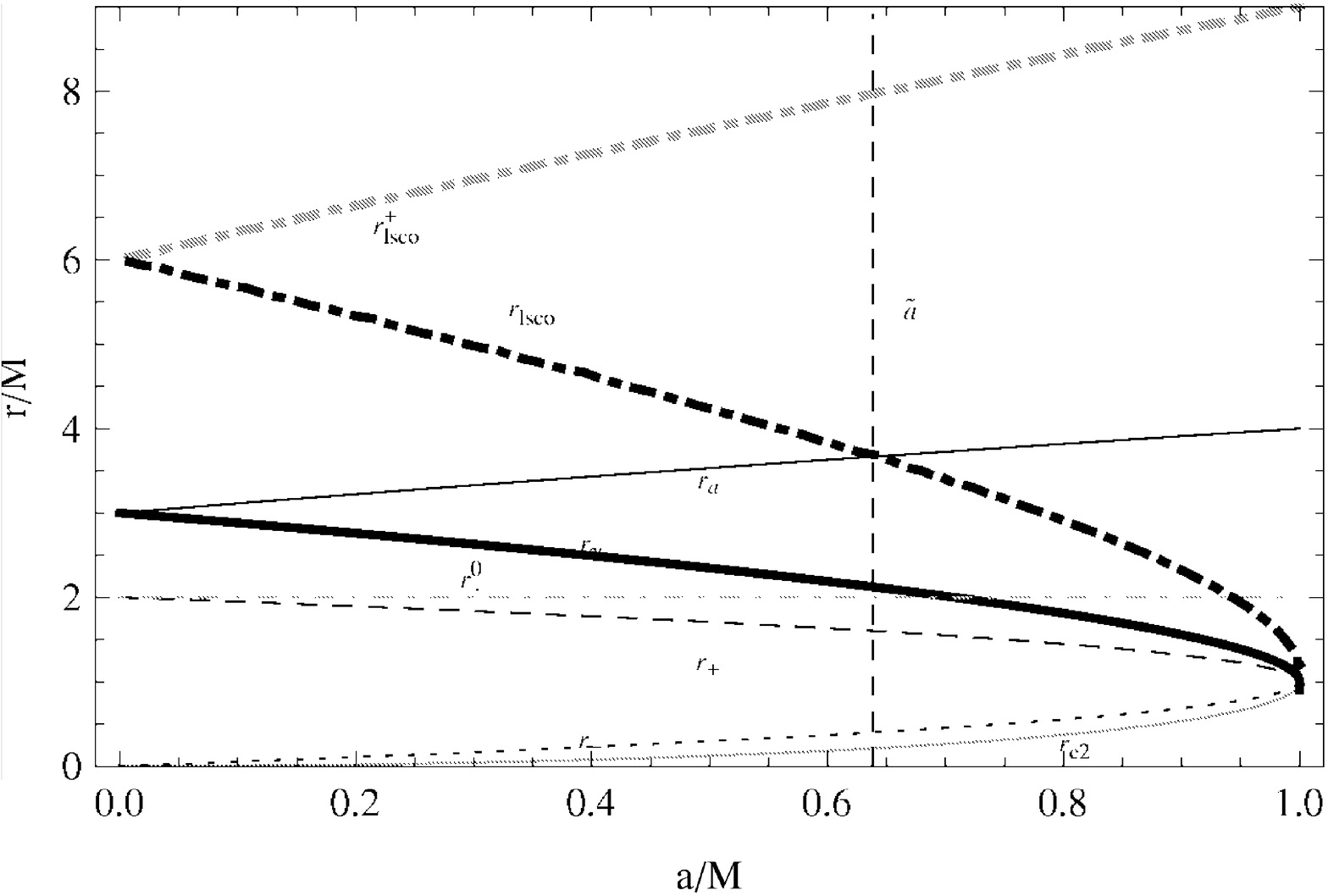}
\end{tabular}
\caption{\footnotesize{The outer horizon $r_+$ (dashed curve), the inner horizon $r_-$ (dotted curve), and $r_{a}$ (black curve), $r_{c2}$ (gray curve), $r_{\gamma}$ (thick black curve), last stable circular orbits $r_{lsco}^+$ (dotdashed black curve) and  $r_{lsco}^-$ (dotdashed gray curve)  are plotted as functions of the black hole intrinsic angular momentum $a/M$. The curves $r_{lsco}^+$  and $r_{lsco}^-$  represent the radius of the last stable circular
orbit for particles with angular momentum $-L_+$ and $L_-$, respectively.
Circular orbits with $L=L_-$ exist for $r/M>r_\gamma$ and with $L=-L_+$ for $r>r_a$. The line $\tilde{a}\approx0.638285M$ is also plotted.The dotted dashed gray line represents  the outer boundary of the ergosphere $r_+^0=2M$.}}
\label{SESMPM}
\end{figure}

From the physical viewpoint it is important to find the minimum radius for stable circular orbits which is determined by the inflection points  of the effective potential function, i.e., by  the condition
\be
{\partial^2 V}/{\partial^2 r}=0.
\ee
The behavior of the effective potential (\ref{qaz}) for a fixed value of $a/M$ and
 different values of the particle angular momentum $L/(M\mu)$ is sketched in Fig.\il\ref{SV5}.
%
The radii of the last stable circular orbits are written as \cite{RuRR,Grumiller:2009gf}
\be
\label{dicadica}
r_{\ti{lsco}}^\mp=M\left[3+Z_2\mp\sqrt{(3-Z_1)(3+Z_1+2Z_2)}\right]\ ,
\ee
where
\be\label{Eh1}
Z_1\equiv 1+\left(1-\frac{a^2}{M^2}\right)^{1/3}\left[\left(1+\frac{a}{M}\right)^{1/3}+\left(1-\frac{a}{M}\right)^{1/3}\right],
\ee
and
\be\label{Eh2}
Z_2\equiv\sqrt{3\frac{a^2}{M^2}+Z_1^2}\ .
\ee
In particular, for $a=0$ we have that $r_{\ti{lsco}}^\mp=6M$, and when $a=M$ we obtain $r_{\ti{lsco}}^-=M$
for co--rotating orbits and  $r_{\ti{lsco}}^+=9M$  for counter--rotating orbits (see also Fig.\il\ref{SESMPM}).
In general, the radii $r_{lsco}^+$ and $r_{lsco}^-$ correspond to  the last stable circular orbit of a test
particle with angular momentum $L_+$ and $L_-$, respectively.

\begin{figure}
\centering
\begin{tabular}{cc}
\includegraphics[scale=.71]{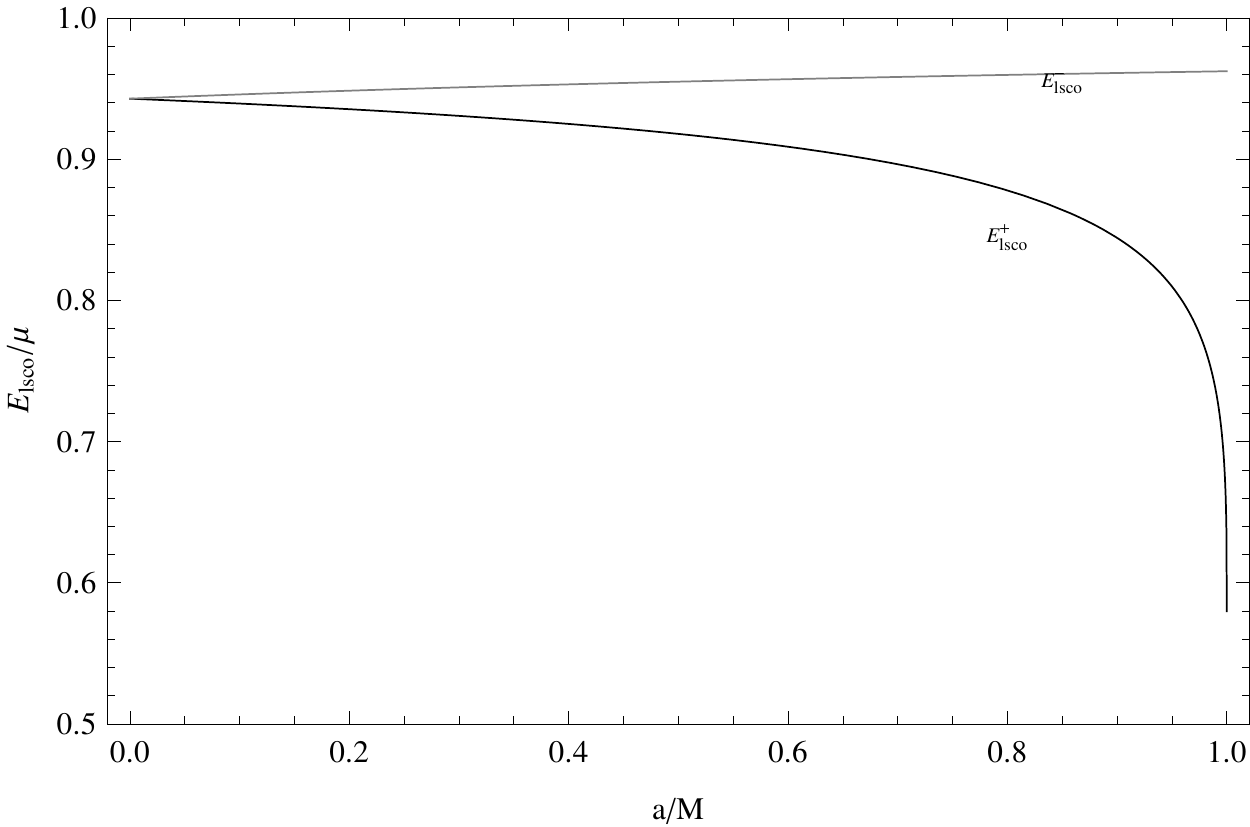}
\includegraphics[scale=.71]{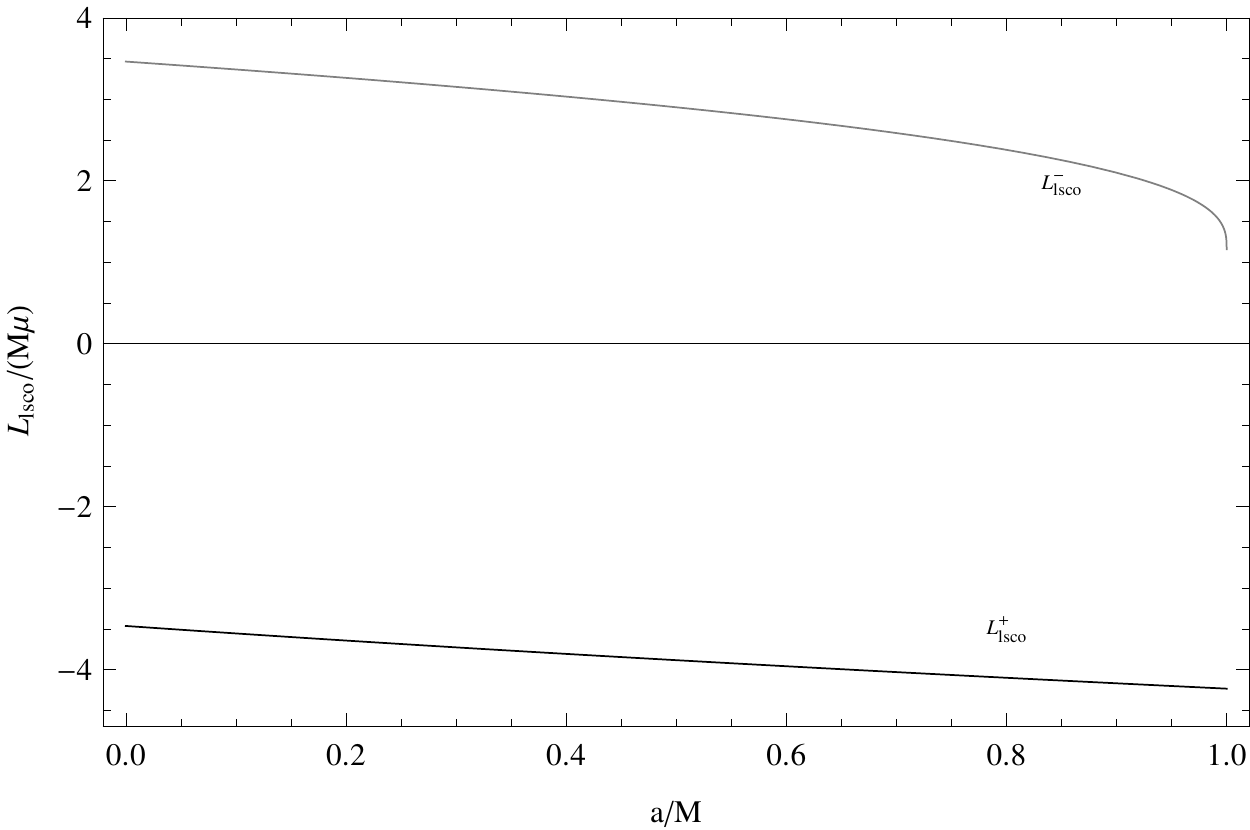}
\end{tabular}
\caption{\footnotesize{The energy $E_{lsco}/\mu$ and the angular momentum $L_{lsco}^\pm/(M\mu)$
of the last stable circular orbit as functions of the ratio $a/M\leq 1$ of a Kerr black hole.
 }}
\label{SESMPME}
\end{figure}

In Fig.\il\ref{SESMPME} the energy  $E_{lsco}^\pm/\mu=E(r_{lsco}^{\pm})/\mu$ and the angular momentum $L_{lsco}^\pm/\mu=L(r_{lsco}^{\pm})/\mu$
of the last stable circular orbits are  plotted as functions of the ratio $a/M$. One can see that
\be
E_{lsco}^+\leq E_{lsco}^-, \quad\mbox{and}\quad E_{lsco}^+=E_{lsco}^- \quad\mbox{for}\quad a=0.
\ee
Moreover, as the ratio $a/M$ increases, the energy $E_{lsco}^+  $ decreases and the energy $E_{lsco}^- $ increases.
Instead, the corresponding angular momenta of the test particles decrease as the intrinsic angular momentum increases.

To classify the circular orbits in a Kerr black hole it is convenient to distinguish two different
regions: The first  one is $a\in[0,\tilde{a}[$, where  $\tilde{a}\approx0.638285M$  is
the value at which $r_a$ and $r_{lsco}^-$ coincide, and the second one is $a\in]\tilde{a}, 1[$.

In the first region $a\in[0,\tilde{a}[$,  where $r_{a}<r_{lsco}^{-}$, we see that there exist
unstable circular orbits with $L=L_-$ in the interval $r_{\gamma}<r<r_{a}$. Moreover, in the interval $r_{a}<r<r_{lsco}^{-}$ there are
unstable circular orbits with $L=L_-$ and $L=-L_+$. In $r_{lsco}^{-}<r<r_{lsco}^{+}$ there are stable circular orbits with $L=L_-$ and unstable orbits
with $L=-L_{+}$. Finally, in the region $r>r_{lsco}^{+}$ there are stable circular orbits with $L=L_{-}$ as well as with $L=-L_{+}$.

Let us consider the second region $a\in]\tilde{a},1[$  where $r_{a}>r_{lsco}^{-}$.
We see that in the interval $r_{\gamma}<r<r_{lsco}^{-}$ there are unstable circular orbits with $L=L_-$. Moreove, in   $r_{lsco}^{-}<r<r_{a}$ there are stable orbits with $L=L_-$. In the region $r_{a}<r<r_{lsco}^{+}$ there are stable circular orbits with $L=L_-$ and unstable orbits with $L=-L_{+}$. Finally, for $r>r_{lsco}^{+}$ there are stable circular orbits with $L=L_{-}$ and $L=-L_{+}$. The classification of circular orbits in this case
is summarized in Table\il\ref{Tabdff}.
\begin{table}
\begin{center}
\resizebox{.51\textwidth}{!}{%
\begin{tabular}{llll}
&The case $0<a<M$& &
\\
\hline\hline
&Region&Angular momentum& Stability
\\
\hline
& $]r_{\gamma},r_{a}] $ &$L_-$ &$r_{lsco}^-$\\
& $]r_{a},\infty[$ &$-L_+$ ($L_-$) &$r_{lsco}^+$ ($r_{lsco}^-$)
\\
\hline
&$0<a<\tilde{a}$ ($r_{a}<r_{lsco}^-$) & &
\\
\hline\hline
& $]r_{\gamma},r_{a}[$  &$L_-$&Unstable  \\
& $]r_{a},r_{lsco}^-[$ &($L_-$, $-L_+$) &Unstable  \\
& $]r_{lsco}^-,r_{lsco}^+ [$ &$L_-$ ($-L_+$)  &Stable (Unstable) \\
& $]r_{lsco}^+,\infty[$  &($L_-$, $-L_+$) &Stable \\
\hline
&$\tilde{a}\leq a<M$ ($r_{a}\leq r_{lsco}^- $) & &
\\
\hline\hline
& $]r_{\gamma},r_{lsco}^- [$  &$L_-$&Unstable  \\
& $]r_{lsco}^-,r_{a}[$ &$L_-$ &Stable  \\
& $]r_{a},r_{lsco}^+[$ &$L_-$ ($-L_+$)  &Stable (Unstable) \\
& $]r_{lsco}^+,\infty[$  &($L_-$, $-L_+$) &Stable \\
\hline\hline
\end{tabular}
}
\caption[font={footnotesize,it}]{\footnotesize{Classification of circular orbits of test particles in a Kerr black hole.
Here $\tilde{a}\approx0.638285M$. For each region we present the value of the orbital angular momentum and the stability property. }}
\label{Tabdff}
\end{center}
\end{table}

\begin{figure}
\centering
\begin{tabular}{cc}
\includegraphics[scale=.8]{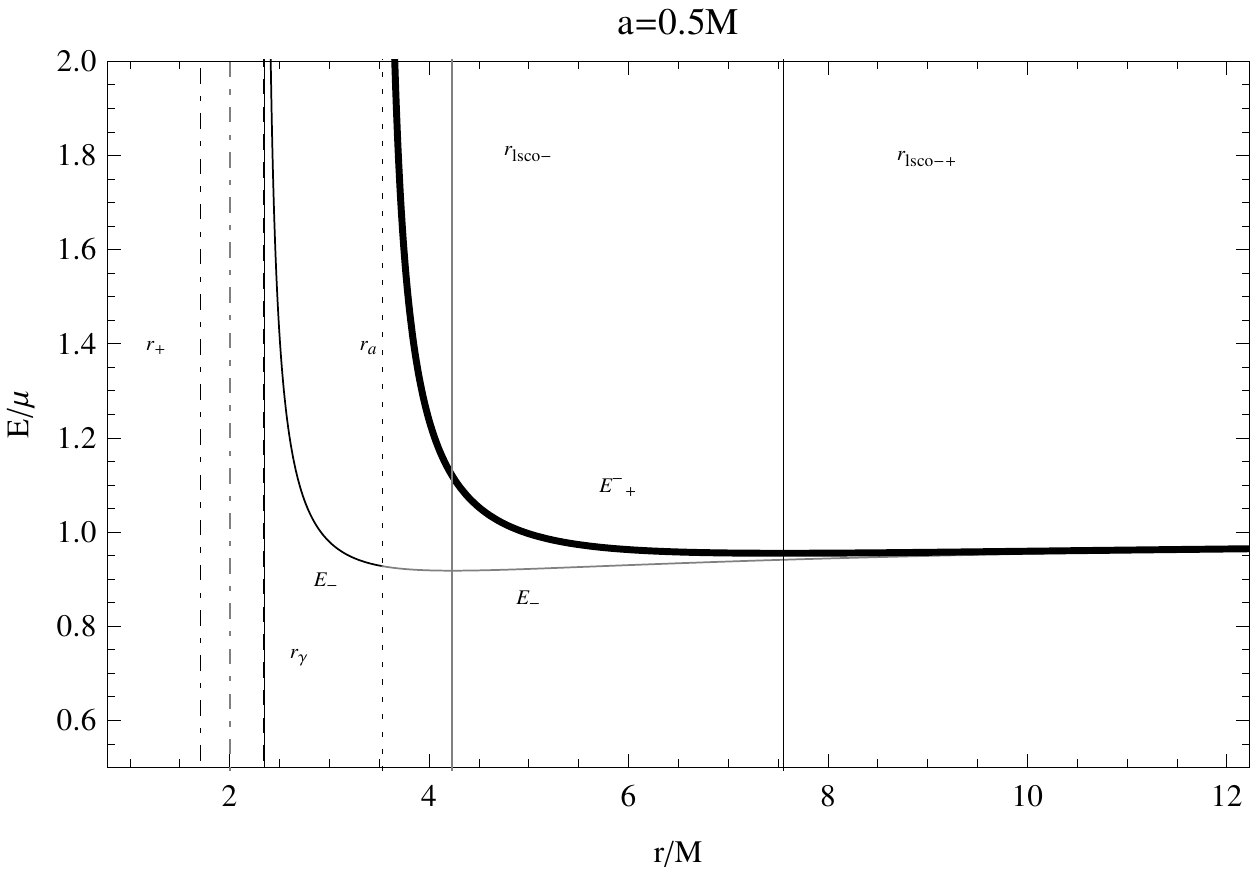}
\includegraphics[scale=.8]{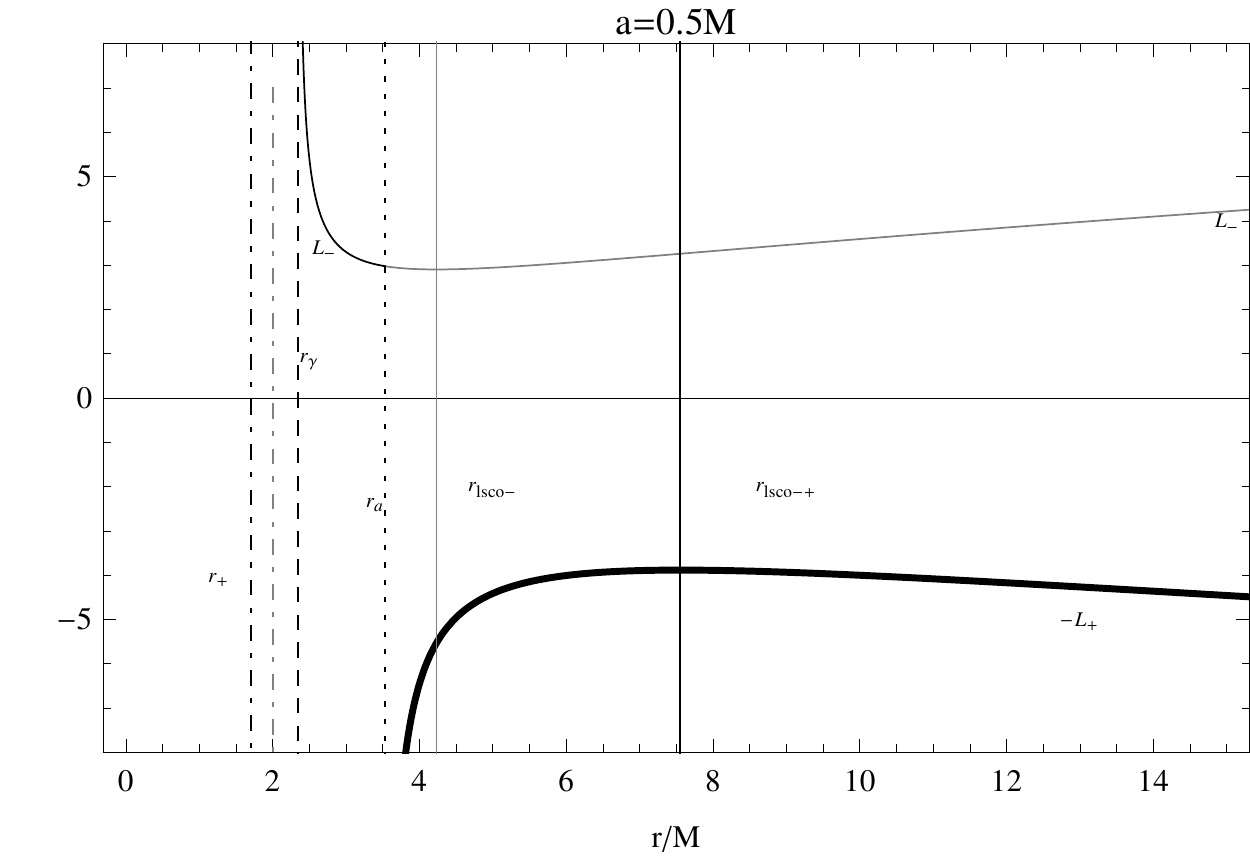}
\end{tabular}
\caption[font={footnotesize,it}]{\footnotesize{The energy  $E/\mu$ (left plot) and the angular momentum $L/(\mu M)$ (right plot)
of circular orbits in  a Kerr black hole with $a=0.5M $  as functions  of the radial coordinate $r/M$.
The energy $E^{(-)}_+\equiv E(-L+)$ and  the angular momentum $-L_+$  are represented by thick black curves,
and the  energy $E^{(+)}_-\equiv E(L-)$ and  the angular momentum $L_-$ by black curves.
In $r_{\gamma}<r<r_{a}$ there are unstable circular orbits with $L_-$. For  $r_{a}<r<r_{lsco}^{-}$ there are unstable circular orbits with
$L_-$ and $-L_+$. For $r_{lsco}^{-}<r<r_{lsco}^{+}$ there are stable circular orbits with $L_-$ and unstable with $-L_{+}$, finally
for $r>r_{lsco}^{+}$ there are stable circular orbits with $L_{-}$ and $-L_{+}$. The radii
$r_+=1.70711M$, $r_{\gamma}=2.3473M$, $r_{a}=3.53209M$ and $r_{lsco}^{-}=4.233M$,  and $r_{lsco}^{+}=7.55458M$ are also plotted.
It is evident  that $E(-L_+)>E(L-)$. The dotted dashed gray line represents  the outer boundary of the ergosphere $r_+^0=2M$. }}
\label{PLBHOLE05}
\end{figure}

\begin{figure}
\centering
\begin{tabular}{cc}
\includegraphics[scale=.7]{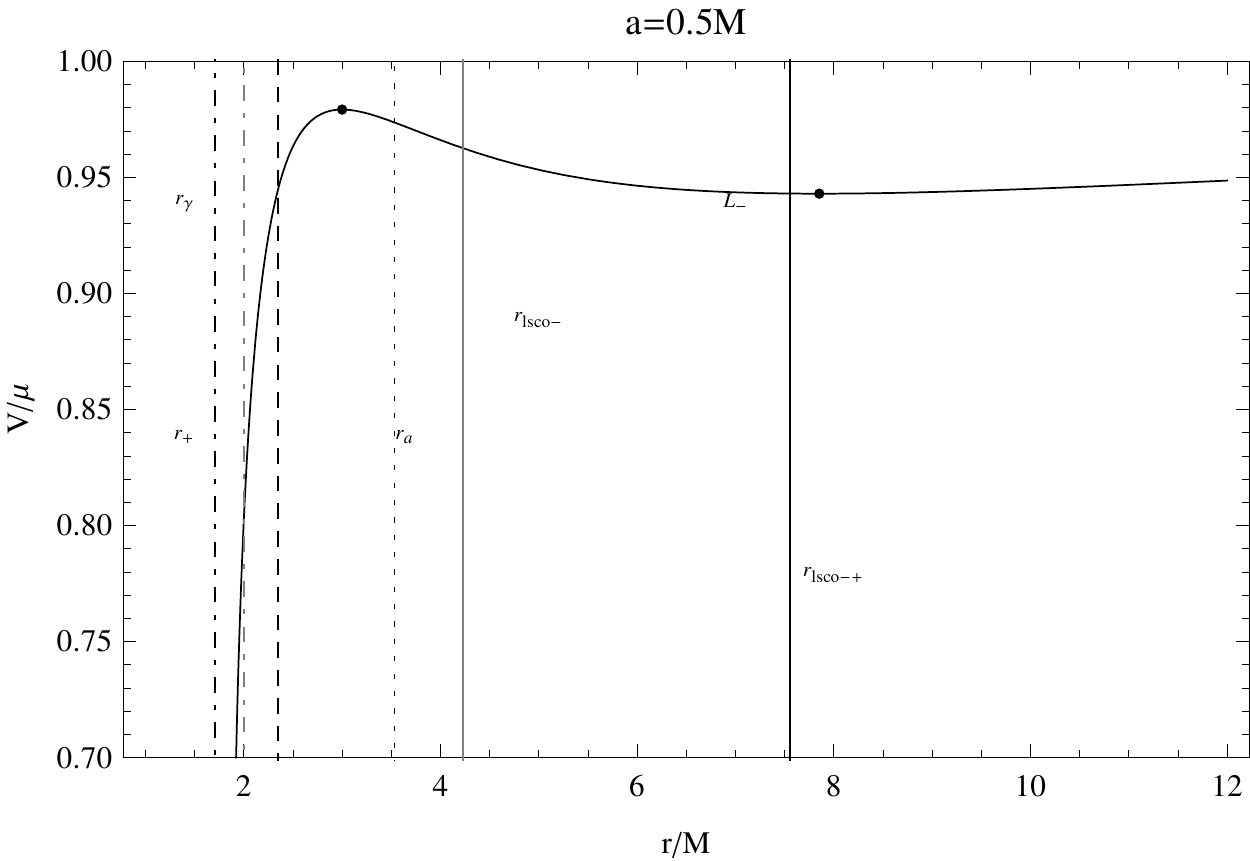}
\includegraphics[scale=.7]{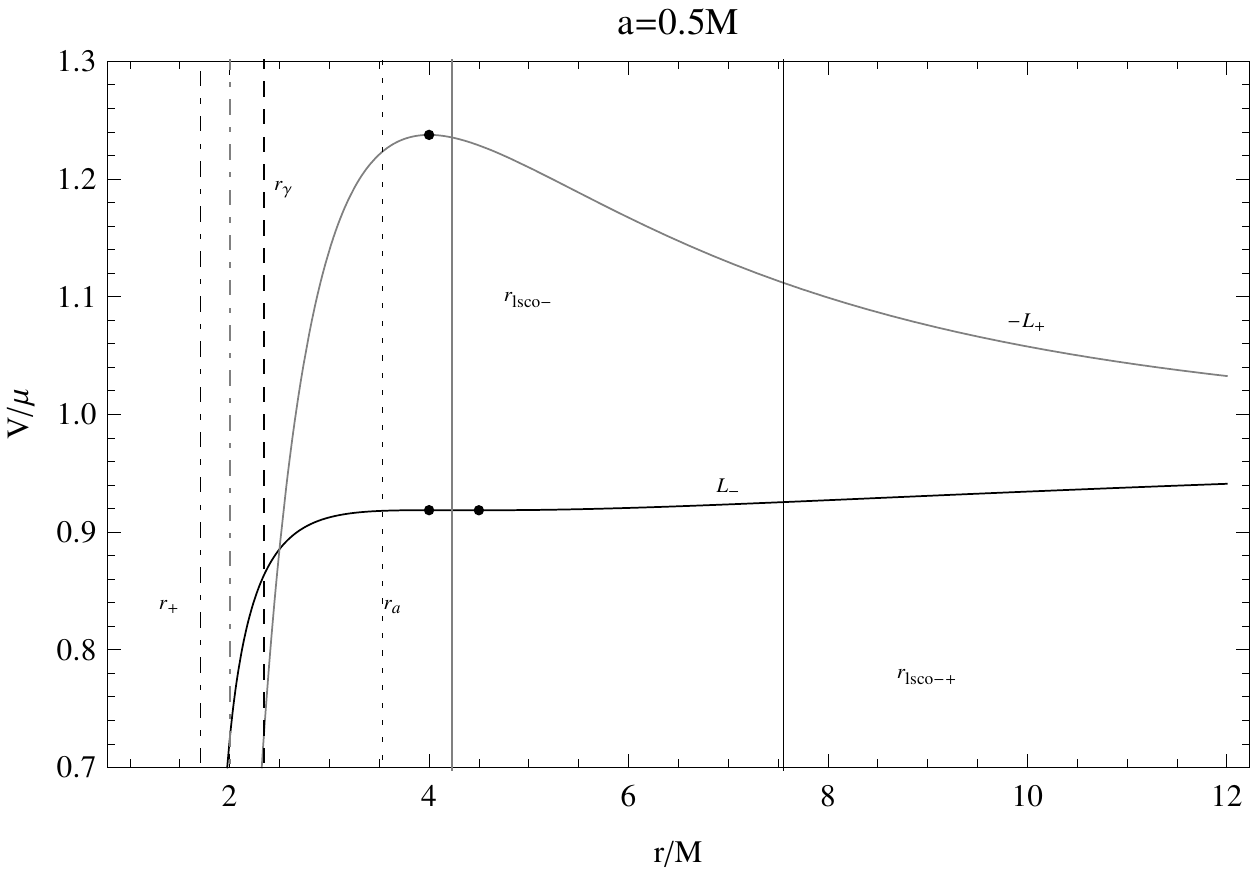}\\
\includegraphics[scale=1]{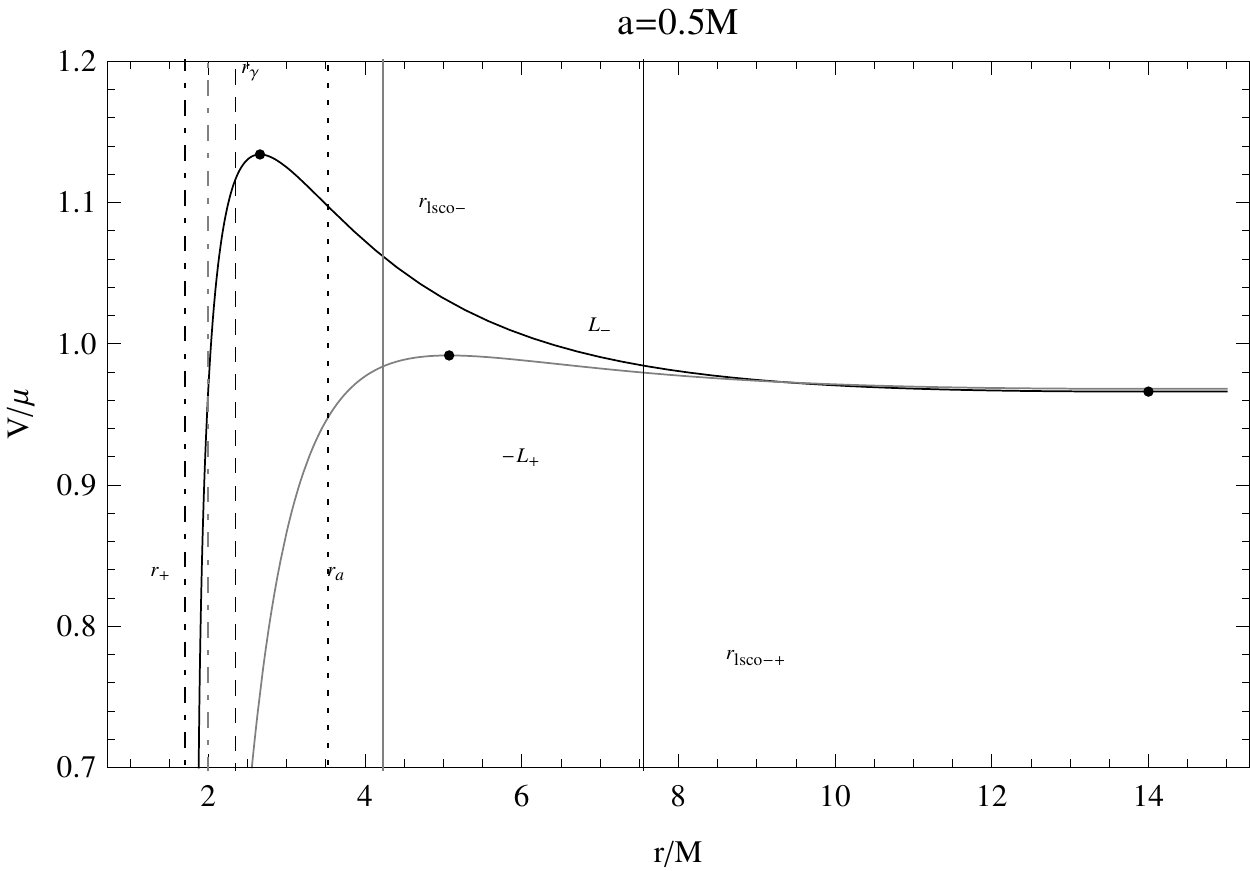}
\end{tabular}
\caption[font={footnotesize,it}]{\footnotesize{The effective potential
$V/\mu$ of a  Kerr black hole with $a=0.5M $ as a function  of $r/M$.  The radii
$r_+=1.70711M$, $r_{\gamma}=2.3473M$, $r_{a}=3.53209M$, $r_{lsco}^-   =4.233M$, and   $r_{lsco}^+   =7.55458M$ are also plotted.
The left upper plot shows the effective potential with orbital angular momentum
$L=L_-=3.29806M\mu$  for which we find a  minimum (stable orbit)  at
 $r=7.85256M$ with energy $E_-/\mu=0.942949$ and a maximum (unstable orbit) at $r=3M$ with $E_-/\mu=0.979181$.
 The right upper plot corresponds to an effective potential with orbital angular momentum
$L=L_-=2.90877\mu $ (black curve) and $L=-L_+=-6.45235M\mu$ (gray curve).
For  $L=L_-$   there is a minimum (stable orbit) at $r=4.49925M$ with $E_-=0.918487\mu$
and a maximum (unstable orbit) at $r=4M$ with $E_-=0.918559\mu$.
For  $L=-L_+$   there is  a maximum (unstable orbit) at $r=4M$ with  $E^{(-)}_+=1.23744\mu$.
The bottom plot is for an effective potential with orbital angular momentum
$L=L_-=4.09649\mu $ (black curve) and $L=-L_+=-4.36042M\mu$ (gray curve).
For  $L=L_-$   there is a minimum (stable orbit) at $r=14M$ with
$E_-=0.96609\mu$ and a maximum (unstable orbit) at $r=2.65996M$ with $E_-=1.134\mu$.
For  $L=-L_+$ there is a maximum (unstable orbit) at $r=5.07411M$ with $E^{(-)}_+=0.991686\mu$ and  a minimum (stable orbit)
at $r=14M$ with $E^{(-)}_+= 0.968052\mu$. The dotted dashed gray line represents  the outer boundary of the ergosphere $r_+^0=2M$.}}
\label{PLSIAO0145}
\end{figure}

\begin{figure}
\centering
\begin{tabular}{cc}
\includegraphics[scale=.7]{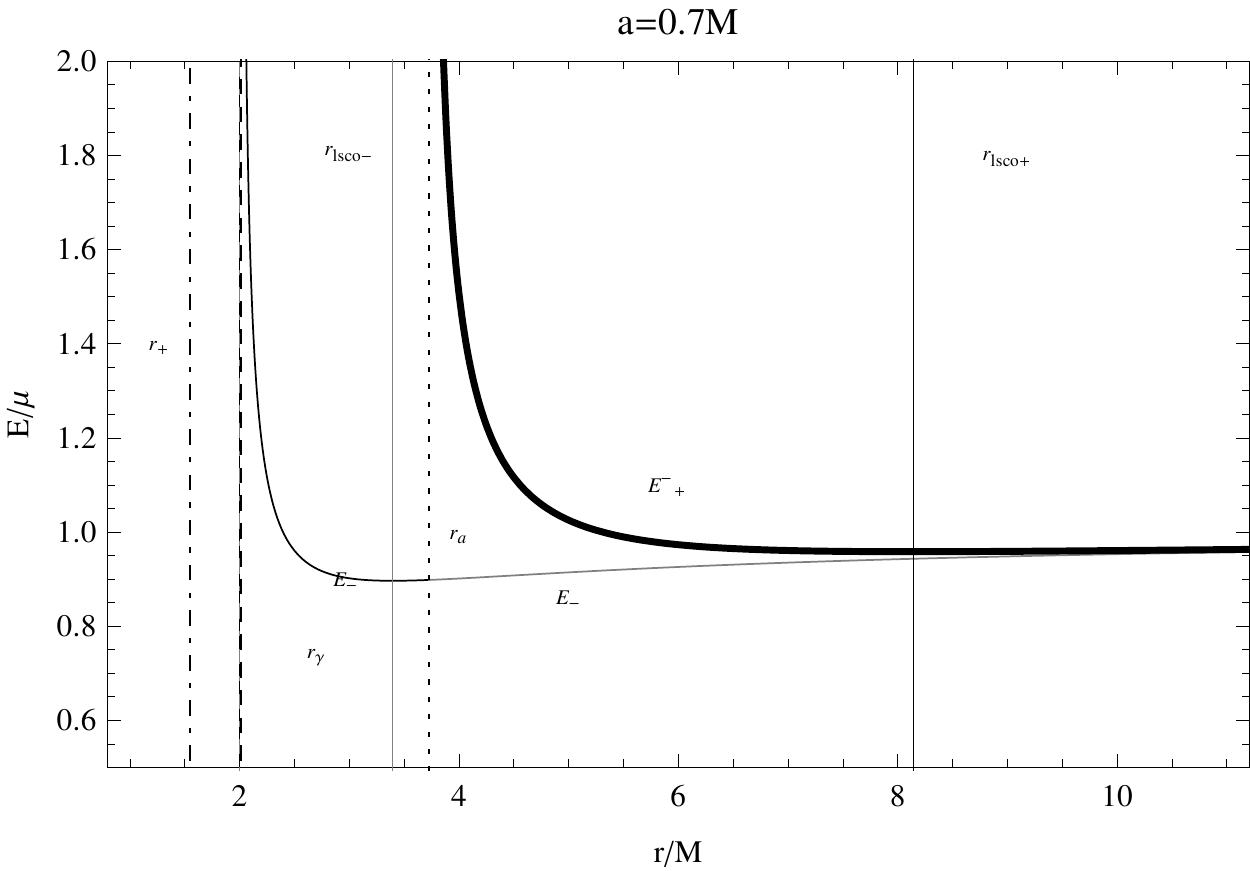}
\includegraphics[scale=.7]{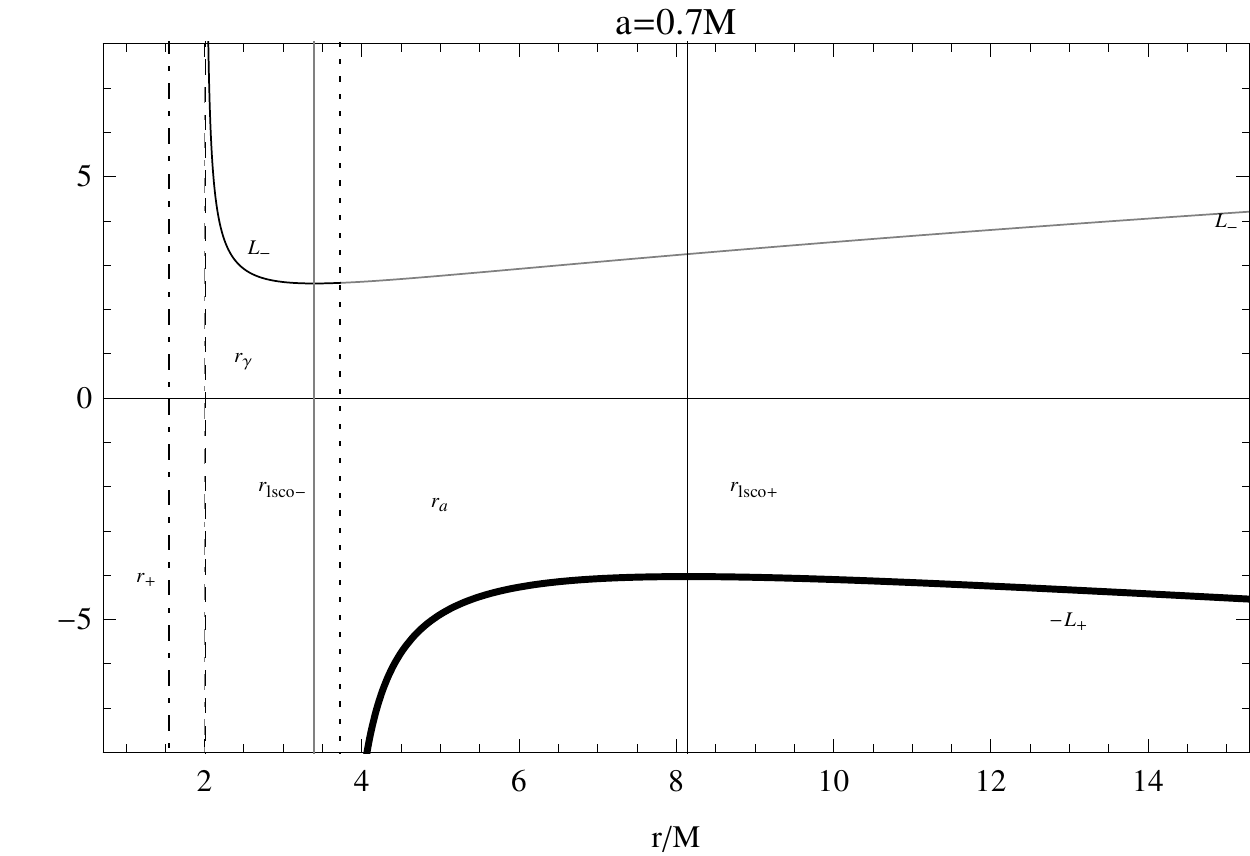}
\end{tabular}
\caption[font={footnotesize,it}]{\footnotesize{The
energy  $E/\mu$ (left plot) and the angular momentum $L/(\mu M)$ (right plot) of circular orbits in a Kerr black hole  with $a=0.7M$ as functions  of $r/M$.
The energy $E_+^-\equiv E(-L+)$ and  the angular momentum $-L_+$  are represented by thick black curves, and the  energy $E_-^+\equiv E(L-)$ and
the angular momentum $L_-$ by black curves. The radii $r_+=1.54772M$, $r_{\gamma}=2.01333M$, $r_{a}=3.72535M$,
$r_{lsco}^-   =3.39313M$,  and $r_{lsco}^+   =8.14297M$ are also plotted. The dotted dashed gray line represents  the outer boundary of the ergosphere $r_+^0=2M$.
In $r_{\gamma}<r<r_{lsco}^-   $ there are unstable circular orbits with $L_-$; in $r_{lsco}^-   <r<r_{a}$ there are stable orbits with $L_-$;
in  $r_{a}<r<r_{lsco}^+   $ there are stable circular orbits with $L_-$ and unstable with $-L_{+}$; finally, for
 $r>r_{lsco}^+   $ there are stable circular orbits with $L_{-}$ and $-L_{+}$.
It is clear that  $E(-L_+)>E(L-)$. }}
\label{PLBHOLE07}
\end{figure}

\begin{figure}
\begin{tabular}{cc}
\includegraphics[scale=.7]{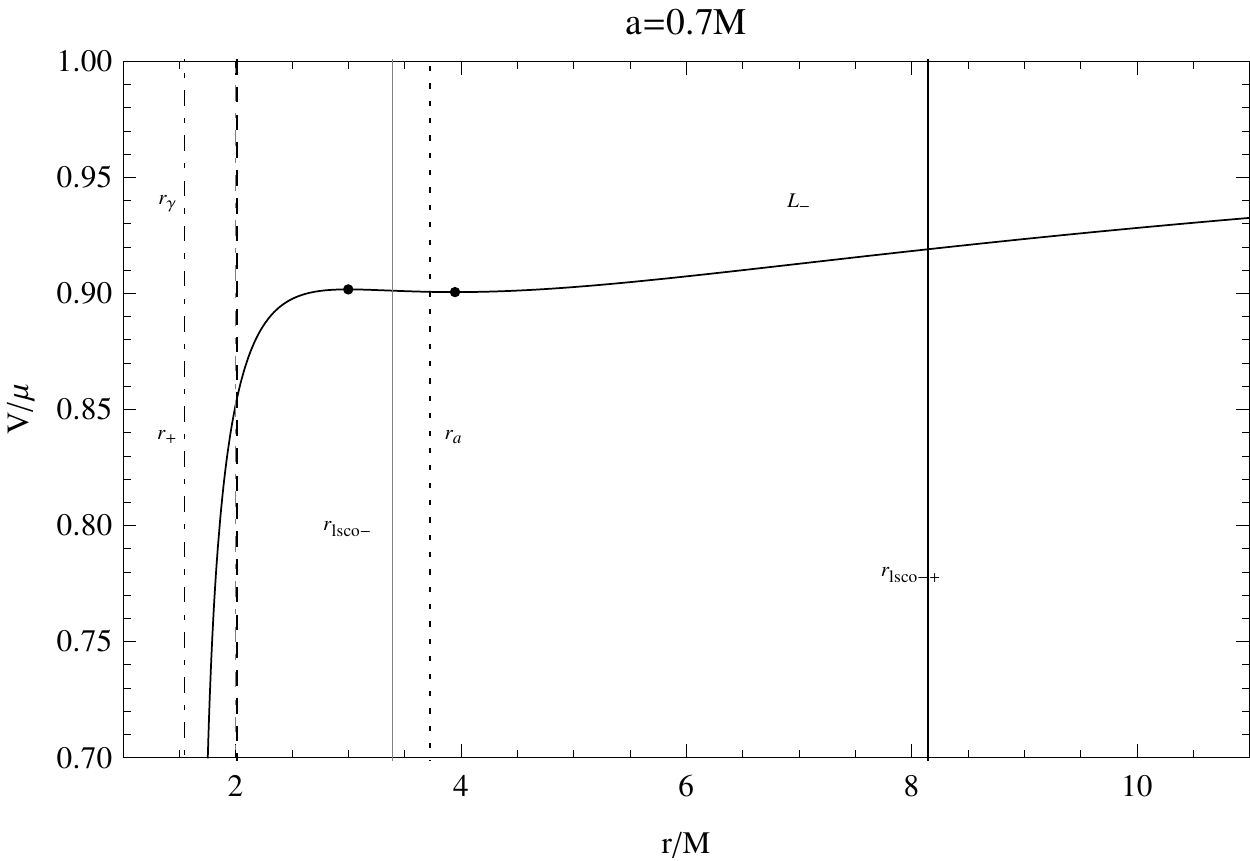}
\includegraphics[scale=.7]{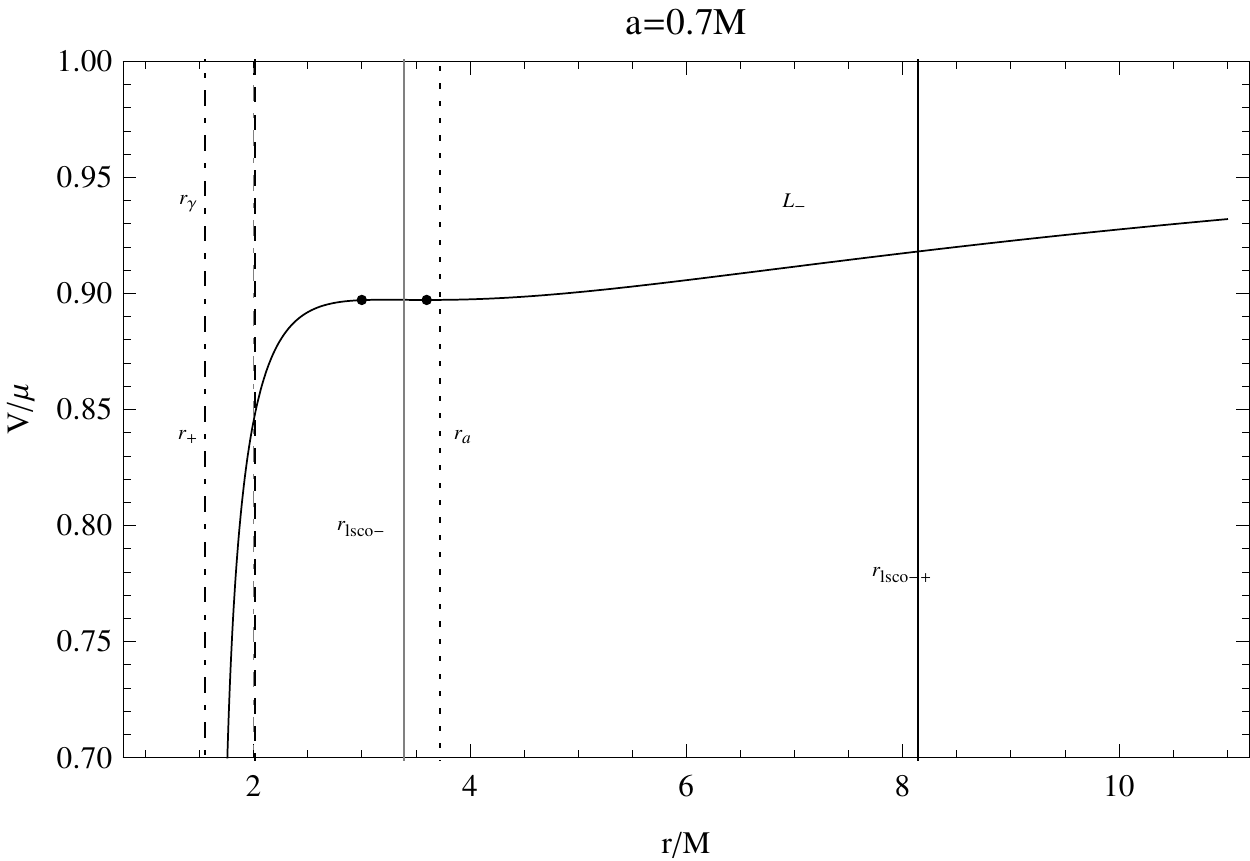}\\
\includegraphics[scale=.7]{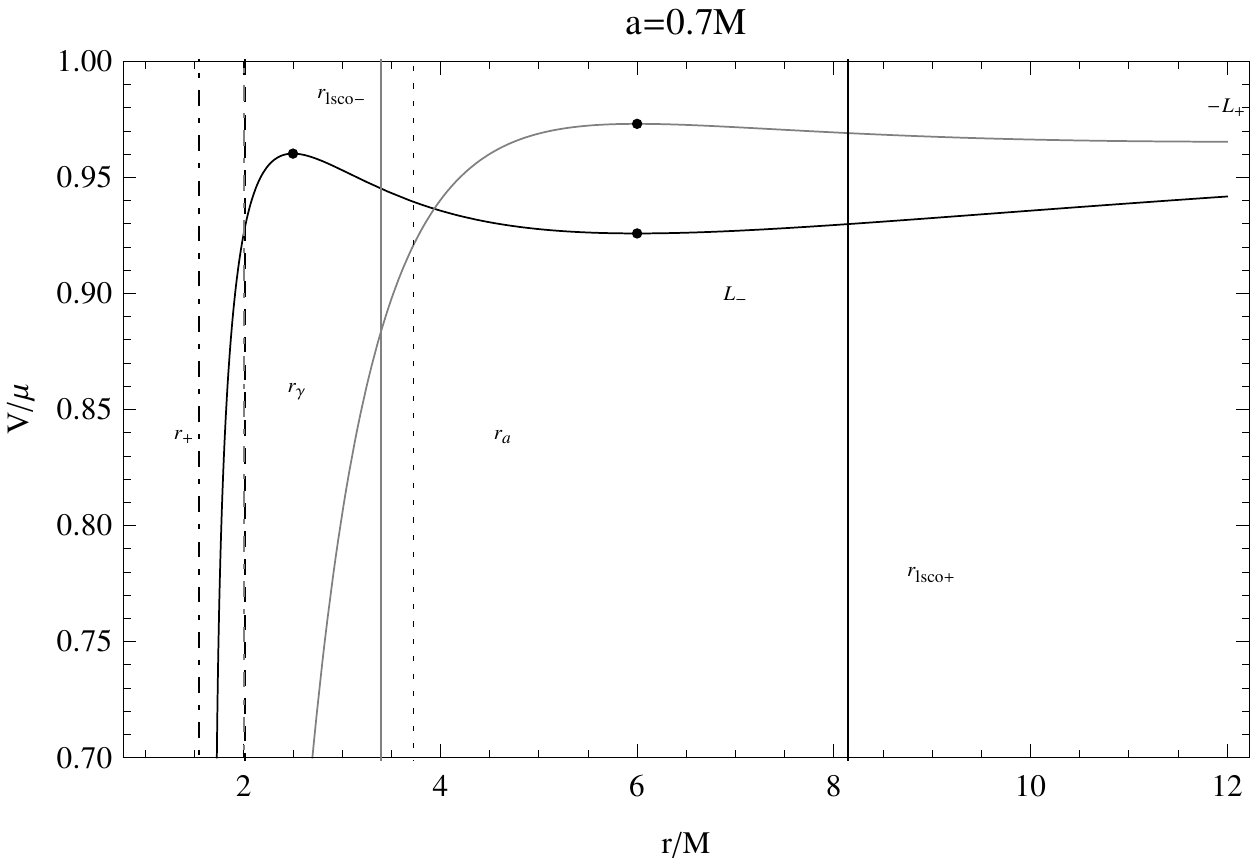}
\includegraphics[scale=.7]{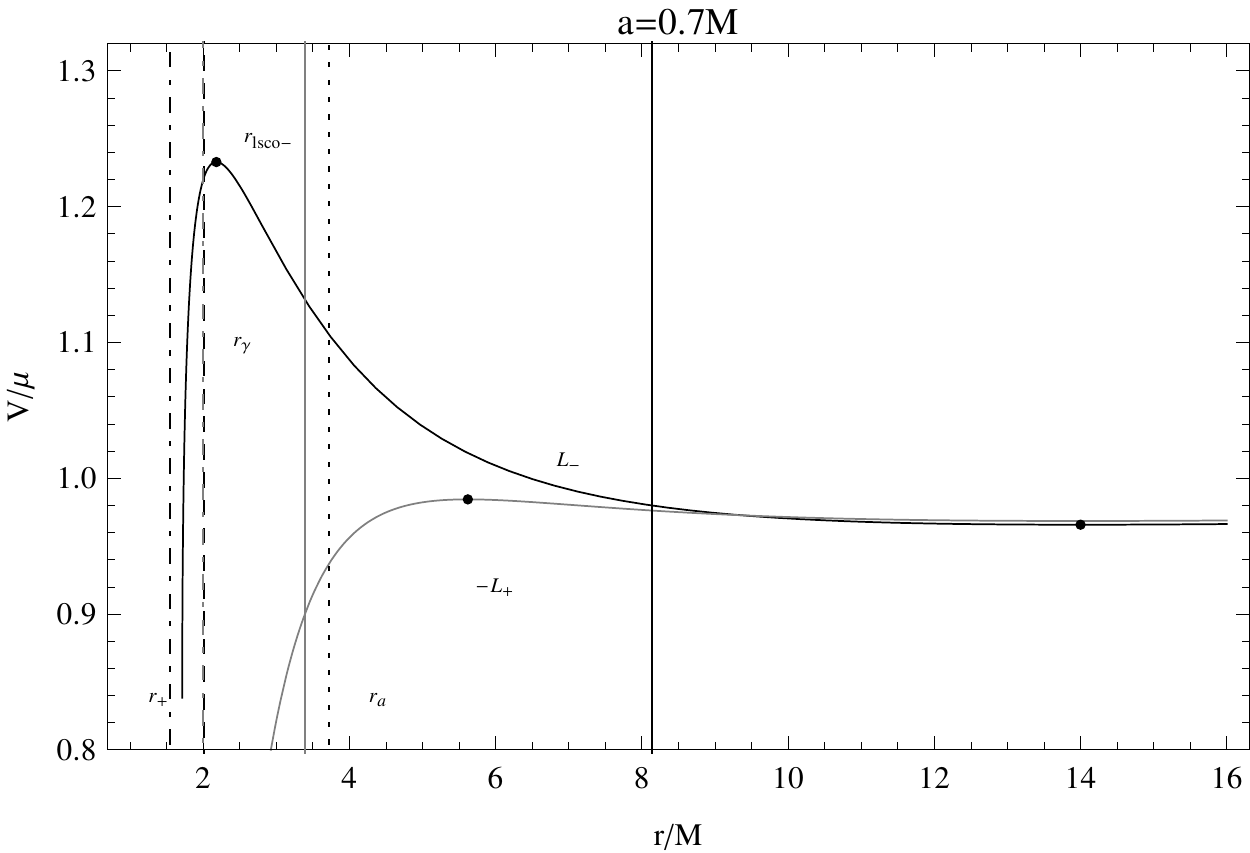}
\end{tabular}
\caption[font={footnotesize,it}]{\footnotesize{The effective potential
$V/\mu$ of a  Kerr black hole spacetime with $a=0.7M $ as function  of $r/M$.  The radii
$r_+=1.54772M$, $r_{\gamma}=2.01333M$, $r_{a}=3.72535M$, $r_{lsco}^-   =3.39313M$,  and $r_{lsco}^+   =8.14297M$ are plotted.
The left upper plot represents the effective potential with orbital angular momentum
$L=L_-=2.61948M\mu$ for which we find a minimum  (stable orbit) at $r=3.9473M$ with  $E_-=0.900551\mu$, and a maximum
(unstable orbit) at $r=3M$ with $E_-=0.901712\mu$.
The right upper plot shows an effective potential with orbital angular momentum
$L=L_-=2.59216M\mu$ for which there exists a minimum (stable orbit) at $r=3.6M$ with $E_-=0.897167\mu$,
 and a maximum  (unstable orbit) at $r=3M$ with $E_-=0.897167\mu$.
The left bottom plot corresponds to effective potentials with orbital angular momenta
$L=L_-=2.91563\mu $ (black curve) and $L=-L_+=-4.2694M\mu$ (gray curve).
For  $L=L_-$   there is a minimum (stable orbit) at $r=6.M$ with $E_-=0.925818\mu$,
 and a maximum  (unstable orbit) at $r=2.50052M$ with $E_-=0.960213\mu$.
 For  $L=-L_+$   there  is a maximum  (unstable orbit) at $r=6M$ with $E^{(-)}_+= 0.973034\mu$.
The right bottom plot is for effective potentials with orbital angular momenta
$L=L_-=4.05058\mu $ (black curve) and $L=-L_+=-4.42036M\mu$ (gray curve).
For  $L=L_-$   there is a minimum  (stable orbit) at $r=14M$ with $E_-=0.965775\mu$,
 and a maximum  (unstable orbit) at $r=2.1819M$ with $E_-=1.23283\mu$.
 For  $L=-L_+$ there is  a maximum  (unstable orbit) at $r=5.6208M$ with $E^{(-)}_+= 0.98443\mu$, and
 a minimum  (stable orbit) at $r=14M$ with $E^{(-)}_+=0.968527\mu$. The dotted dashed gray line represents  the outer boundary of the ergosphere $r_+^0=2M$.}}
\label{PLSIAO0147}
\end{figure}

\begin{figure}
\centering
\begin{tabular}{cc}
\includegraphics[scale=.7]{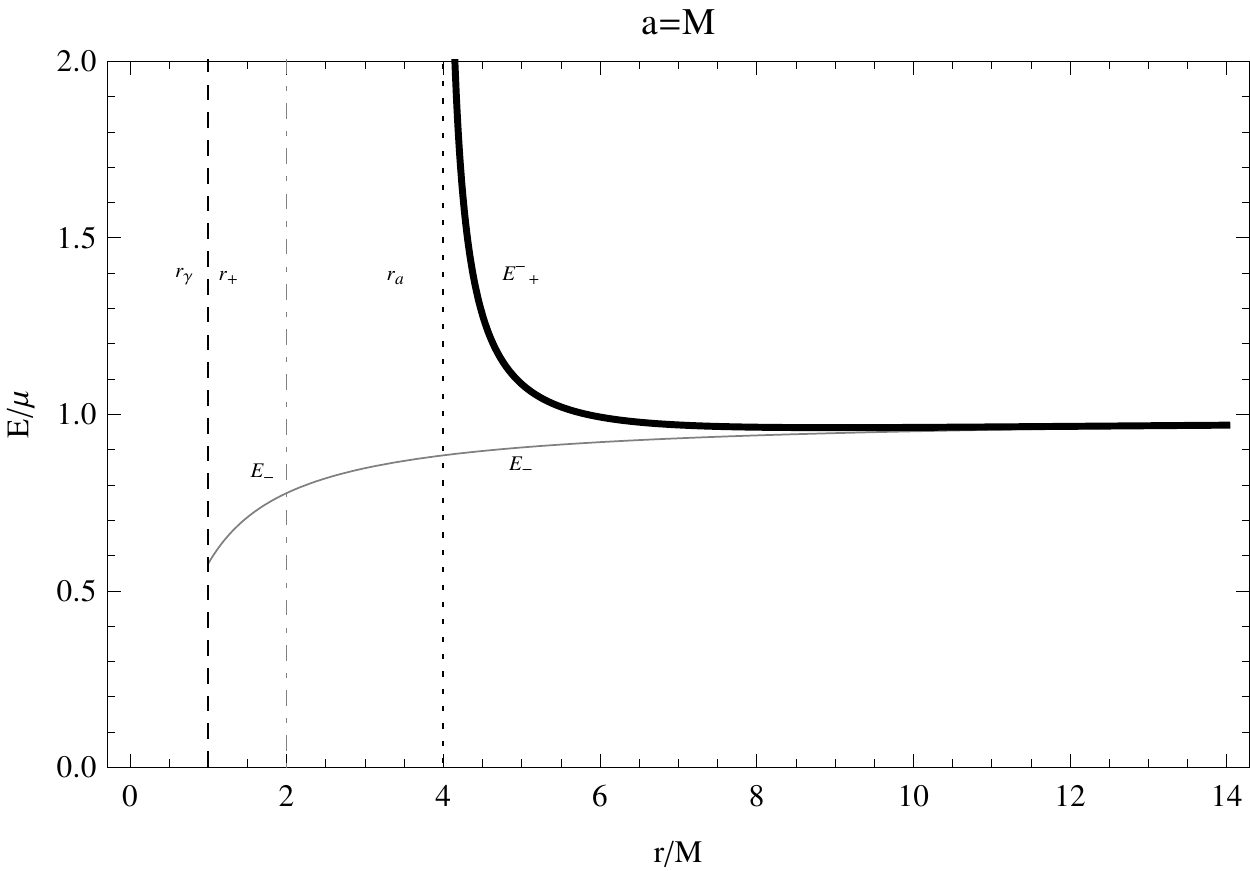}
\includegraphics[scale=.7]{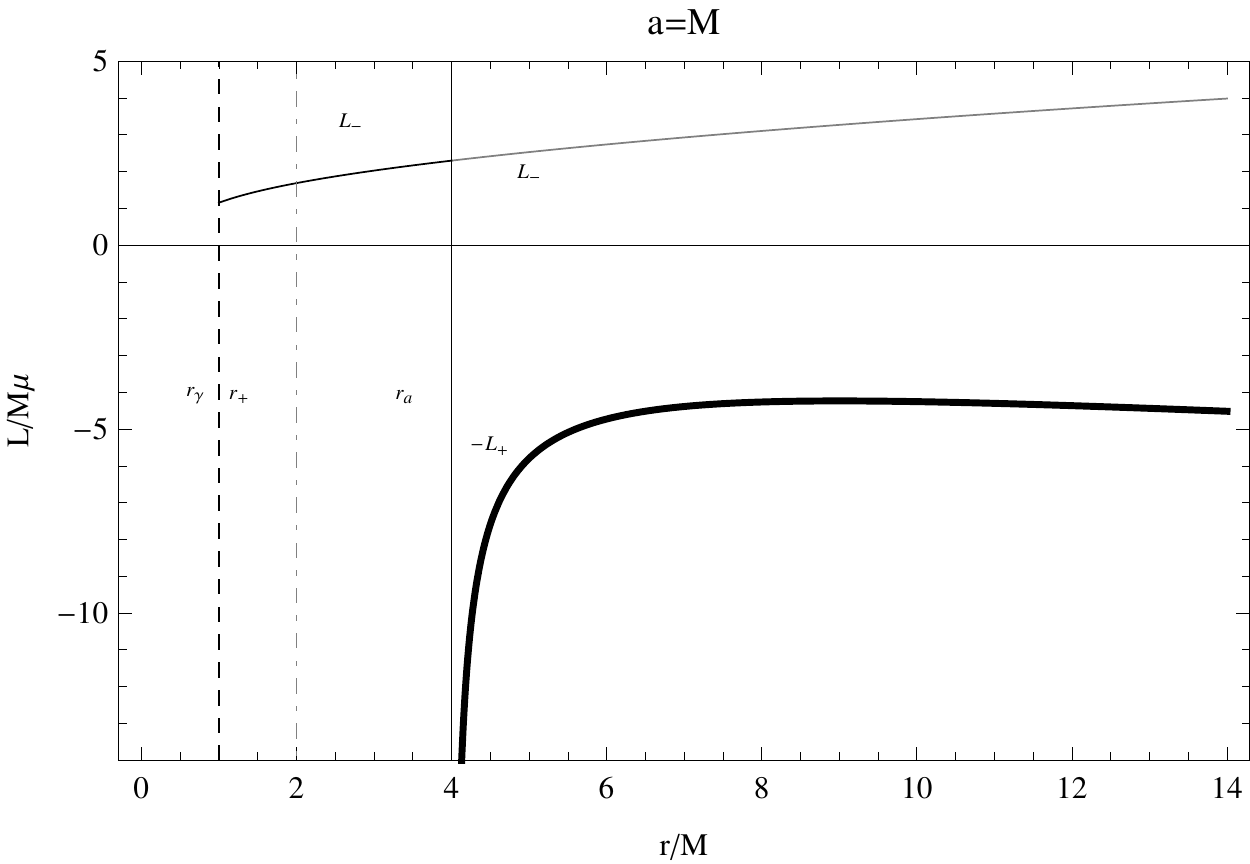}
\end{tabular}
\caption[font={footnotesize,it}]{\footnotesize{The energy $E/\mu$ (left plot) and the angular momentum $L/(\mu M)$ (right plot) of circular orbits in
an extreme Kerr black hole $(a=M)$ as functions of the radial coordinate $r/M$.
The energy $E(-L+)$ and  the angular momentum $-L_+$  are represented by thick black curves, and the  energy $E(L-)$ and  the angular momentum $L_-$ by
black curves. The radii $r_+=r_{\gamma}=M$ (dashed curve) and  $r_{a}=4M$ (black curve) are also plotted.
There exist circular orbits with $L=L_-$ in $r>r_{\gamma}$. The energy  $E(L-)$ is always positive and decreases
as $r$ approaches $r_{\gamma}$. Circular orbits with $L=-L_+$ exist also in $r>r_+$. The energy   $E(-L_+)$ is always positive and
increases as $r$ approaches $r_{a}$. It is evident that  $E(-L_+)>E(L-)$. }}
\label{PLBHOLE}
\end{figure}

\begin{figure}
\centering
\begin{tabular}{cc}
\includegraphics[scale=.7]{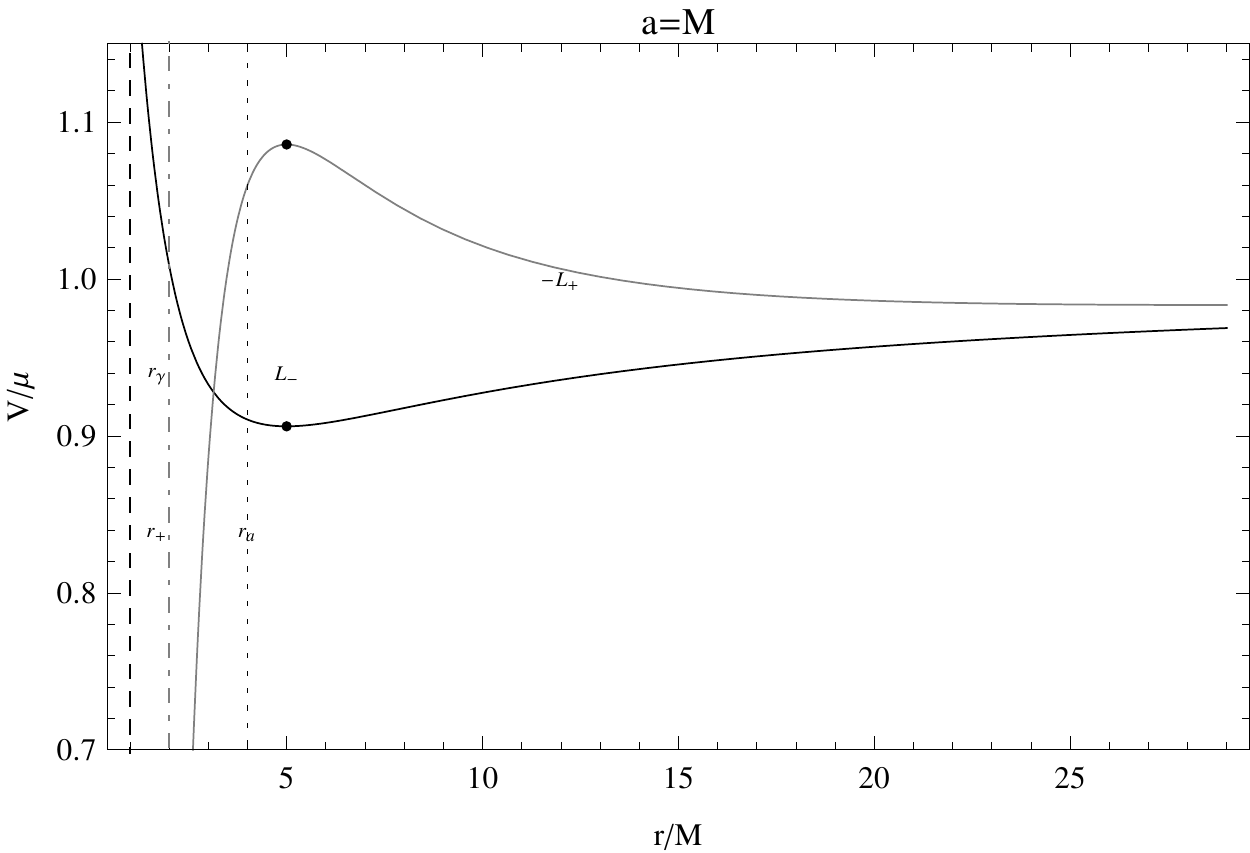}
\includegraphics[scale=.7]{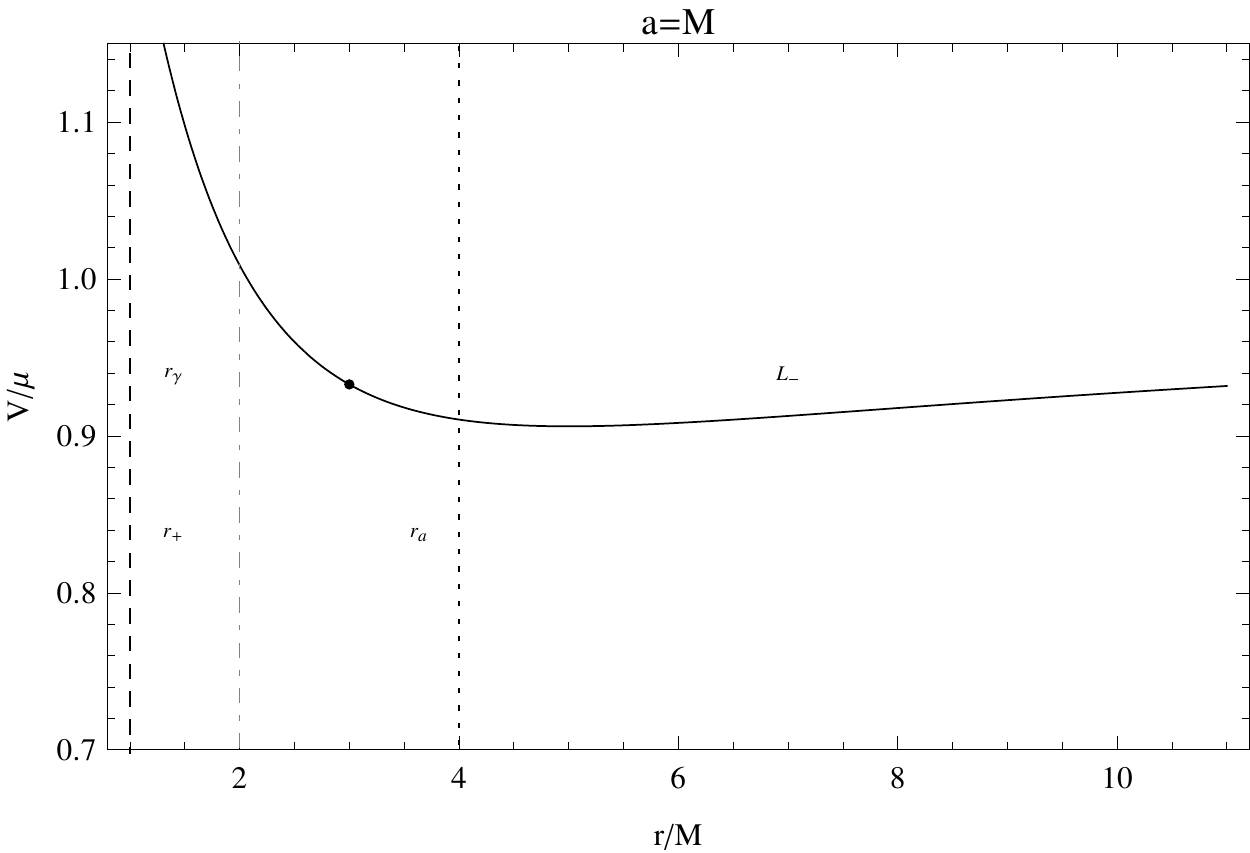}
\end{tabular}
\caption[font={footnotesize,it}]{\footnotesize{The effective potential
$V/\mu$ of an extreme Kerr black hole for a test particle with a fixed  orbital angular momentum as function of $r/M$.
The radii $r_+=r_{\gamma}=M$ (dashed curve) and  $r_{a}=4M$ (black curve) are plotted.
The dotted dashed gray line represents  the outer boundary of the ergosphere $r_+^0=2M$.
The left plot shows  the effective potentials with orbital angular momenta
$L=L_-$ (black curve) and $L=-L_+$ (gray curve).
For  $L=L_-$   there is a minimum (stable orbit) at $r=5M$ with $L_-= 2.53075 M \mu$  and $E_-=0.906154\mu$.
For  $L=-L_+$   there is a maximum (unstable orbit) at $r=5M$ with $-L_+= -5.79614M \mu$ and $E_+^-=1.08576\mu$.
There exist circular orbits with $L=L_-$ in the region $r>r_{\gamma}$,  and orbits with $L=-L_+$ in the region $r>r_+$.  The dotted dashed gray line represents  the outer boundary of the ergosphere $r_+^0=2M$. }}
\label{PSSIAO}
\end{figure}

A detailed analysis of the behavior of the energy, angular momentum and effective potential of test particles is presented
in Figs. \ref{PLBHOLE05} and \ref{PLSIAO0145} for $a/M=0.5<\tilde a$,  in Figs. \ref{PLBHOLE07} and \ref{PLSIAO0147} for $a/M=0.7>\tilde a$, and finally
in Figs. \ref{PLBHOLE} and \ref{PSSIAO} for the limiting case of an extreme black hole $a/M=1$.


\clearpage

\clearpage


To present the main result
of our analysis in a plausible manner it is convenient to introduce the idea of a hypothetical accretion disk formed by test particles on circular
orbits around the central massive object. We consider this model only in the region $r>r_+^0$.
The structure of such an accretion disk depends explicitly on the stability  properties of the
test particles. In fact, as mentioned above the radii $r_{lsco}^-$ and $r_{lsco}^+$ represent the last stable orbits for particles with
angular momentum $L=L_-$ (corotating particles) and $L=-L_+$ (counter-rotating particles), respectively. Then, in the disk contained
within the radii $[r_{lsco}^-,r_{lsco}^+]$ only the corotating particles can move along stable trajectories.
If a counter-rotating particle is located inside this disk (this is possible if the radius of the orbit is $r>r_a$),
its orbit is unstable and it must decay into an orbit with radius $r>r_{lsco}^+$. Consequently, the outer disk with $r>r_{lsco}^+$
can be build of corotating and counter-rotating particles which are both stable in this region. The size of the
inner disk $[r_{lsco}^-,r_{lsco}^+]$ depends on the value of the intrinsic angular momentum of the black hole $a$; the maximum size
is reached in the case of an extreme black hole $(a=M)$ with $r_{lsco}^+ - r_{lsco}^-= 8M$  whereas for $a=0$ the radii coincide
$r_{lsco}^+ = r_{lsco}^-$ and the disk disappears (cf. Fig. \ref{SESMPM}).

\section{Naked singularities}
\label{sec:orbalenskerr}
In  the naked singularity  case $(a>M)$, it is $g_{tt}>0$ for $0<r<r^0_-$ and $r>r^0_+$ when $0\leq\cos^2\theta <1/a^2$, for $r>0$ with $r\neq r^0_-$ when $\cos^2\theta =1/a^2$, and finally for  $r>0$ when $1/a^2<\cos^2\theta \leq 1$.
Moreover, $g_{tt}=0$ at  $r=2M$ if $\theta=\pi/2$,  at $r=r^0_{\pm}$ for  $0<\cos^2\theta <1/a^2$, and at $r=r^0_-$ for $\cos^2\theta =1/a^2$.
As in the black hole case,  in the region $(r^0_-,r^0_+)$ the Killing vector $\xi_t^{a} = (1, 0, 0, 0)$ becomes spacelike.
On the equatorial plane, $\theta=\pi/2$, it is $r^0_+=2M$ and  $r^0_-=0$. In this case, the timelike Killing vector becomes spacelike in the region $0<r<r^0_+$, for all $a>M$.

According to the results presented in Sec. \ref{sec:orbkerr},
to explore the motion of test particles along circular orbits we must solve the following equations
\be
\label{Eq:KerrorbitNS}
\dot{r}=0,\quad V=E/\mu,\quad \partial V/\partial r=0\
\ee
for the effective potential (\ref{qaz}) with $a>M$, taking into account that in this case no horizons exist.
It turns out that it is convenient to study separately the range $a\geq3 \sqrt{3}/4M$ (see Sec.\il\ref{subsec:mag})
and the range $M<a<3 \sqrt{3}/4M$ (see Sec.\il\ref{subsec:min}) for the values of the intrinsic
angular momentum of the naked singularity.

\subsection{The case $a\geq(3 \sqrt{3}/4)M$}
\label{subsec:mag}

\begin{figure}
\centering
\begin{tabular}{cc}
\includegraphics[scale=.7]{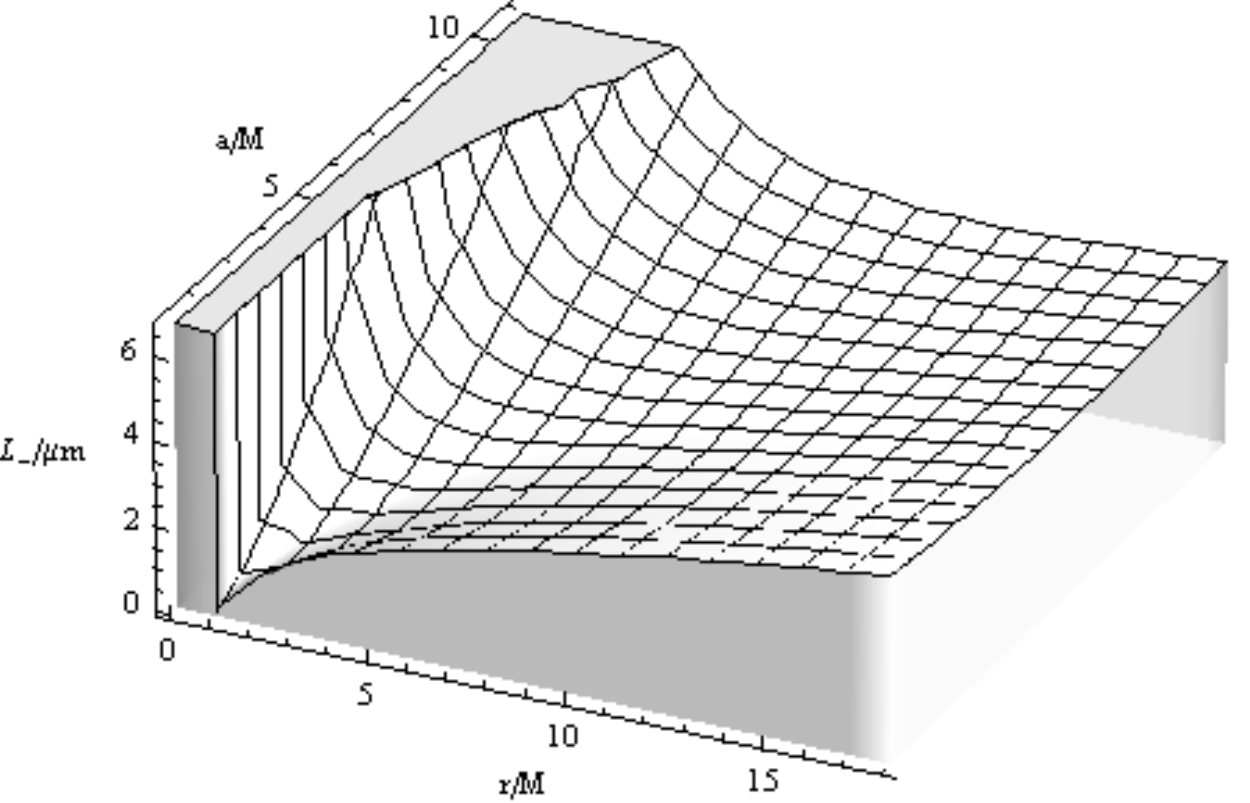}
\includegraphics[scale=.7]{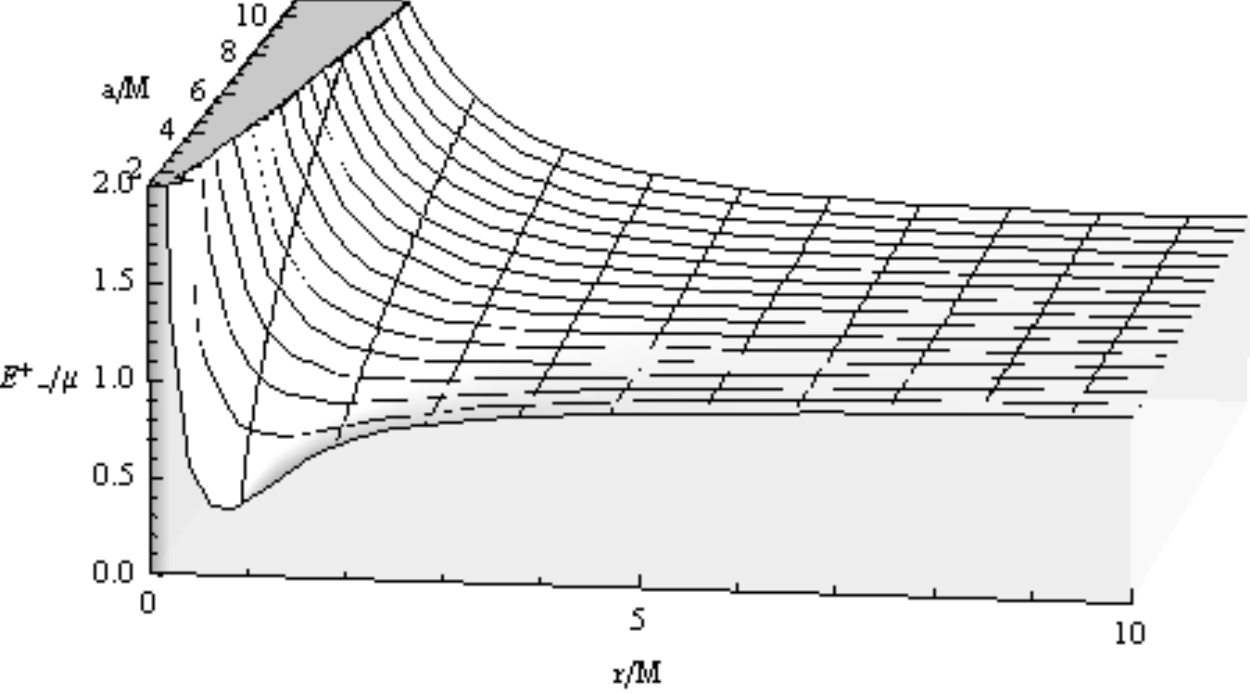}
\end{tabular}
\caption[font={footnotesize,it}]{\footnotesize{Angular momentum and energy of test particles in a Kerr naked singularity with $a\geq(3 \sqrt{3}/4)M$.
The angular momentum $L=L_{-}$ (left plot) and the energy $E_-^{(+)}\equiv E(L_-)$ (right plot) of circular orbits are plotted as functions of $r>0$  and
$a\geq(3 \sqrt{3}/4)M$. The particle's energy is always positive. It is possible to note a region of minima for the energy corresponding to the  minima
of $L_{-}$.}}
\label{Sec2Lm}
\end{figure}

In this case we find that for all  $r>0$ there exist circular orbits with angular momentum $L=L_{-}$    and energy
$E^{(+)}_-=E(L_{-})$. In Fig. \ref{Sec2Lm} we illustrate the behavior of the energy and angular momentum of
test particles for this case.

\begin{figure}
\centering
\begin{tabular}{cc}
\includegraphics[scale=.7]{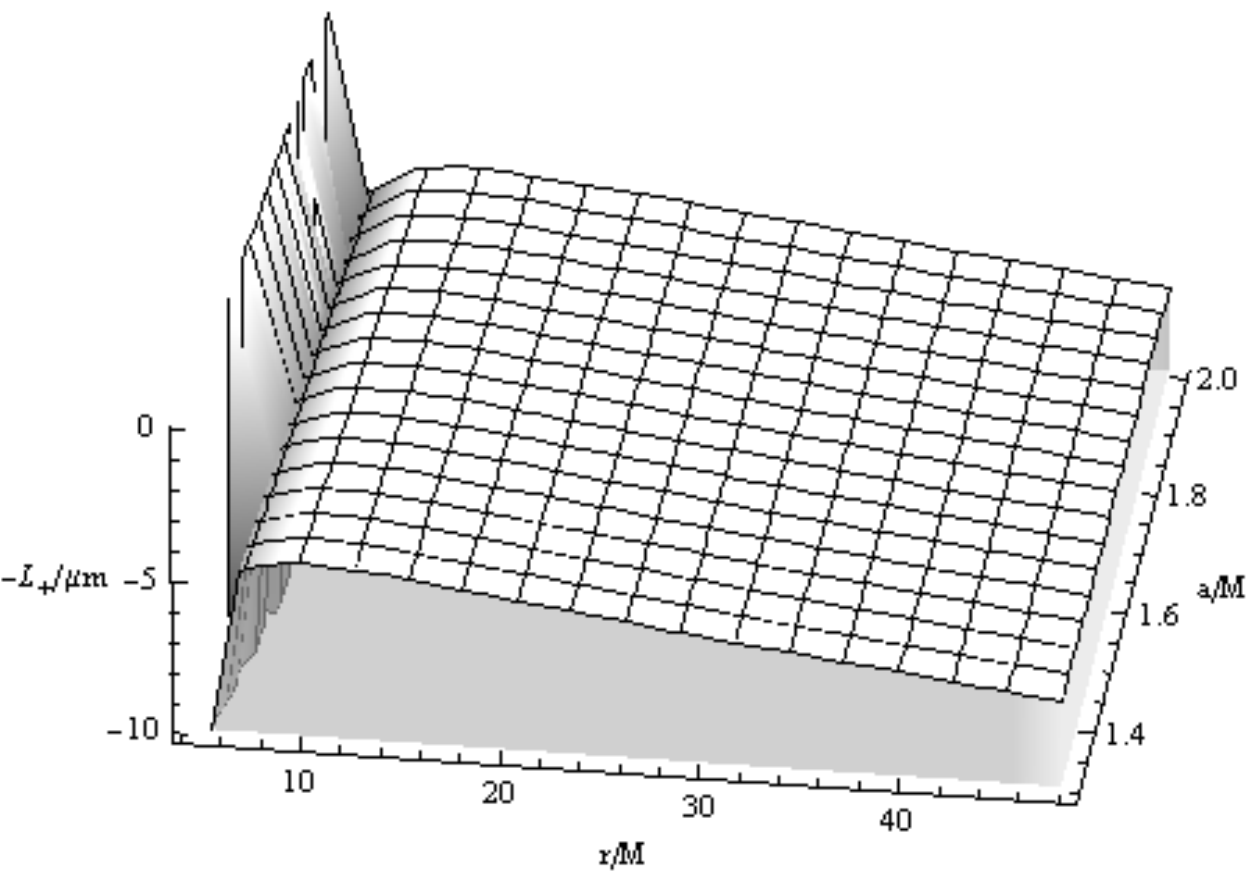}
\includegraphics[scale=.7]{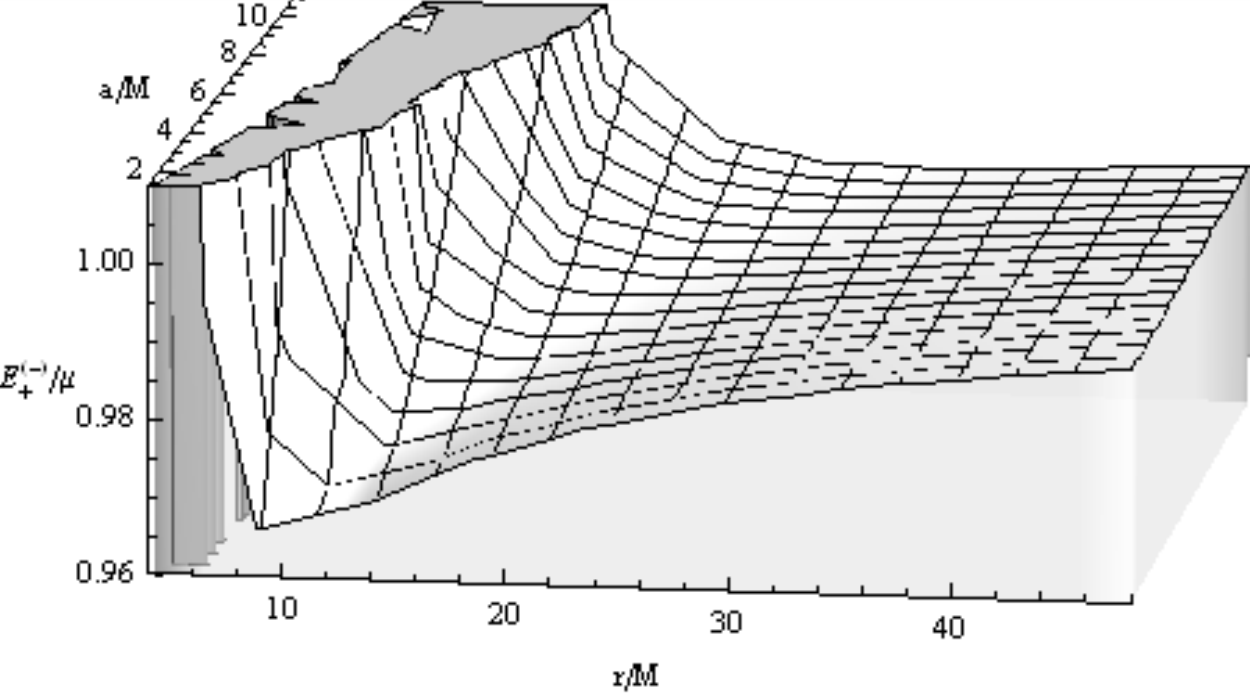}
\end{tabular}
\caption[font={footnotesize,it}]{\footnotesize{Angular momentum and energy of test particles in a Kerr naked singularity with $a\geq(3 \sqrt{3}/4)M$.
The angular momentum $L=-L_{+}$ (left plot) and the energy $E_+^{(-)}\equiv E(-L_+)$ (right plot) of circular orbits are plotted as functions of $r>r_a$  and
$a\geq(3 \sqrt{3}/4)M$. The particle's energy is always positive. It is possible to note a region of minima for the energy corresponding to the minima
of $-L_{+}$. }}
\label{Sec2mLp}
\end{figure}

An analysis of the effective potential shows that
a second class of circular orbits with $L=-L_{+}$ and energy $E^{(-)}_+=E(-L_+)$ can be found in the region $r>r_a$ where
\be\label{rag}
\frac{r_a}{M}\equiv2+\frac{1+\left(2\frac{a^2}{M^2}-1+2
\sqrt{\frac{a^4}{M^4}-\frac{a^2}{M^2}}\right)^{2/3}}{\left(2\frac{a^2}{M^2}-1+2
\sqrt{\frac{a^4}{M^4}-\frac{a^2}{M^2}}\right)^{1/3}}\ .
\ee
The expression for the energy and angular momentum of the test particles in this region is depicted in Fig. \ref{Sec2mLp}.

\begin{figure}
\centering
\begin{tabular}{cc}
\includegraphics[scale=.7]{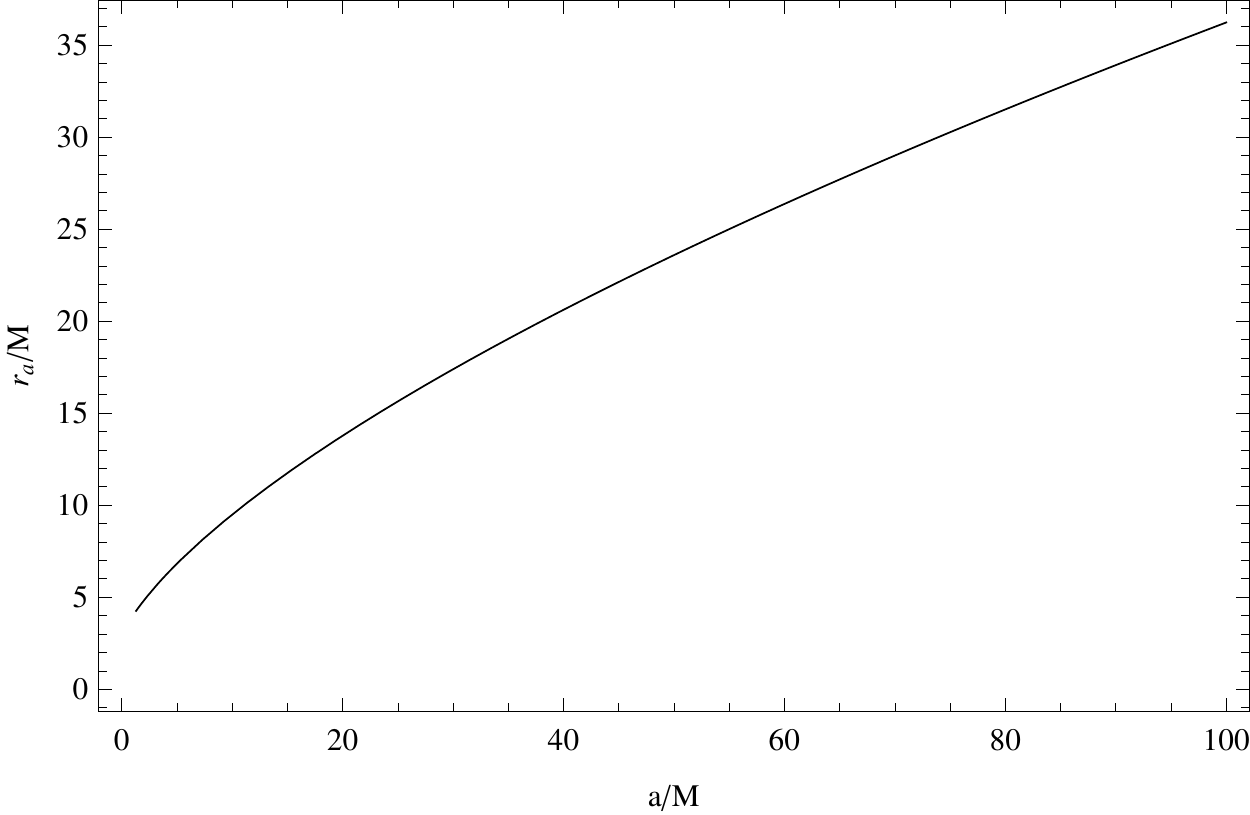}
\includegraphics[scale=.7]{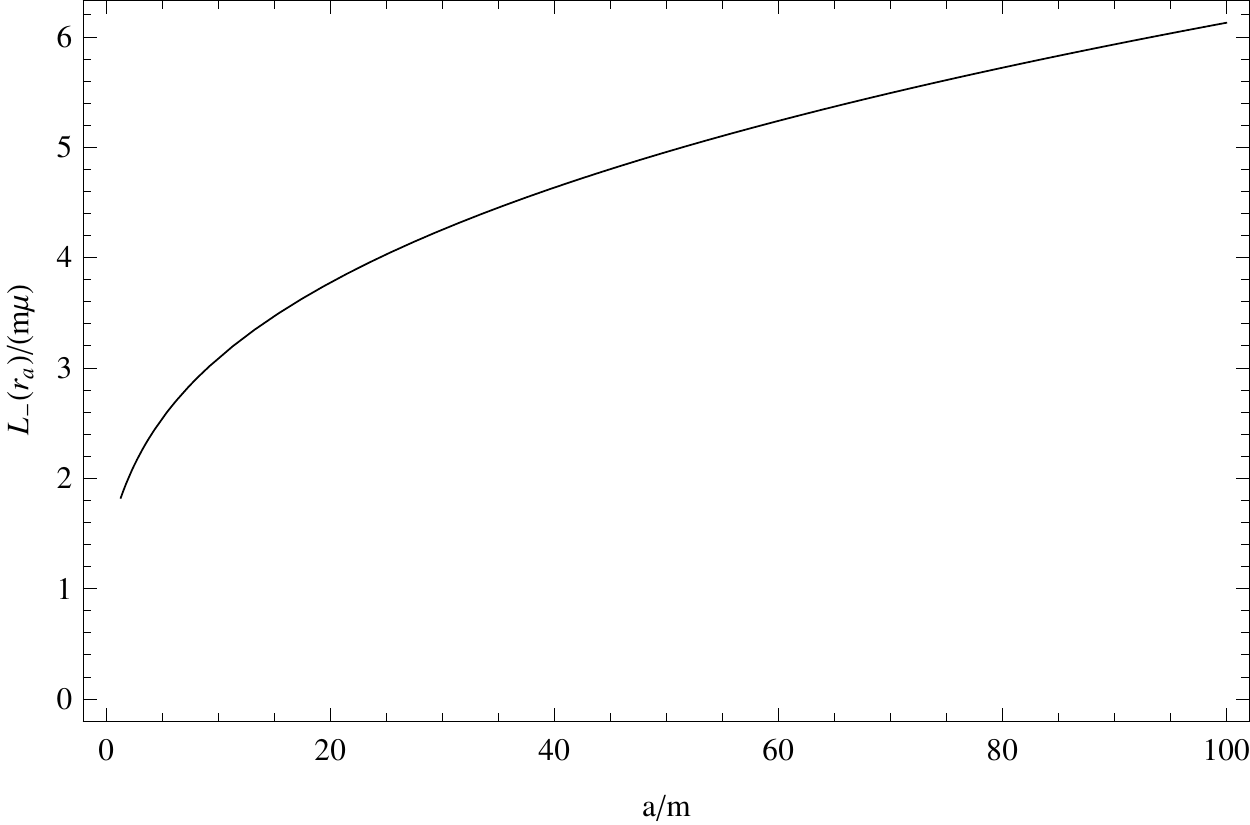}
\end{tabular}
\caption[font={footnotesize,it}]{\footnotesize{The graphic shows the radius $r_a/M$ (left plot) as a function of the
intrinsic angular momentum parameter $a/M$ in the interval $(3\sqrt{3}/4,100)$,
and the particle orbital angular momentum
$L(r_a)/(\mu M)$ (right plot) as a function of $a/M$. The dotted dashed gray line represents  the outer boundary of the ergosphere $r_+^0=2M$.}}
\label{Plotcira21lk}
\end{figure}

The special radius $r_a$ and the angular momentum for this radius $L(r_a)/(\mu M)$ increase as the intrinsic angular
momentum of the naked singularity increases, as shown in Fig.\il\ref{Plotcira21lk}. Notice that we are using
the same notation $r_a$ for the radius (\ref{rc1}) of a black hole and the radius (\ref{rag}) of a naked singularity.
Although these radii are different in their definitions, we use the same notation because
in the limiting case  $a=M$ they both have the same limiting value $r_a=4M$. This will turn out later on to be convenient
when we compare
the results of black holes with those of naked singularities.

\begin{figure}
\centering
\begin{tabular}{c}
\includegraphics[scale=.7]{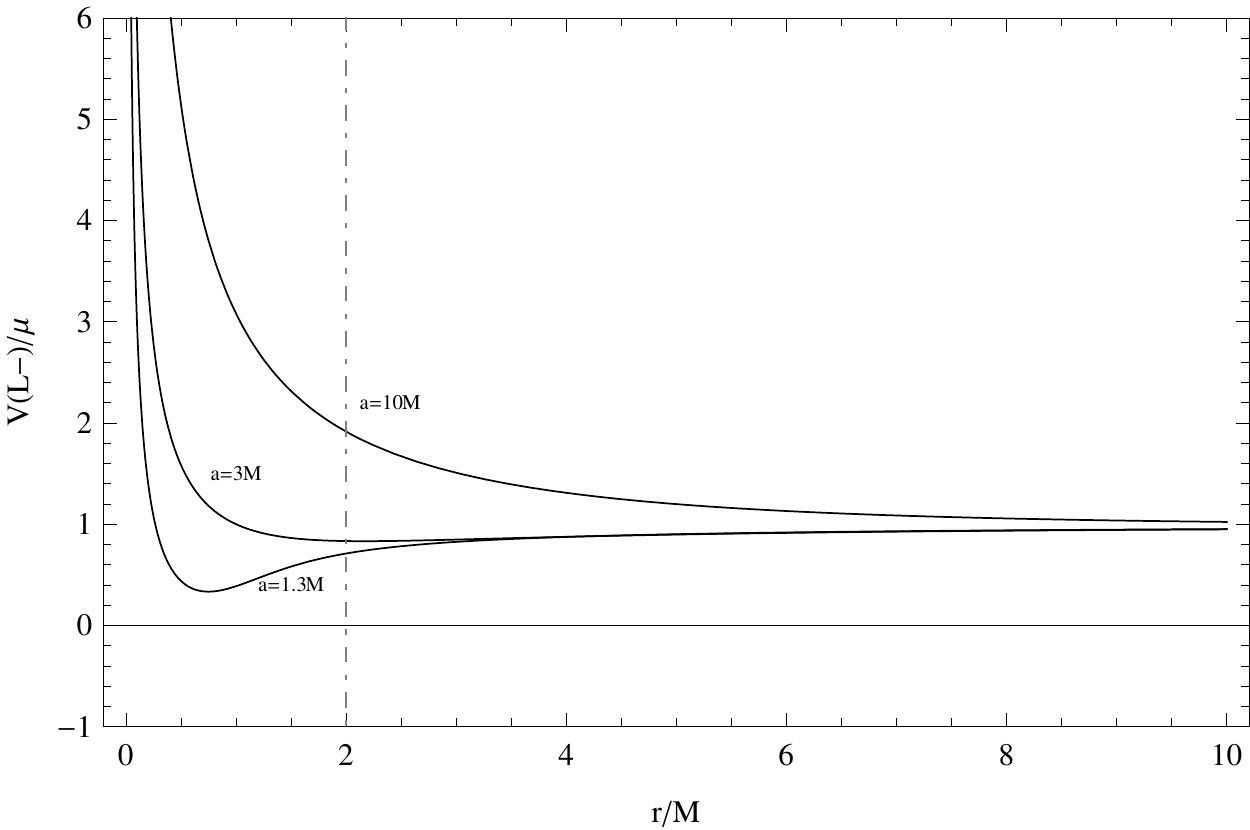}
\includegraphics[scale=.7]{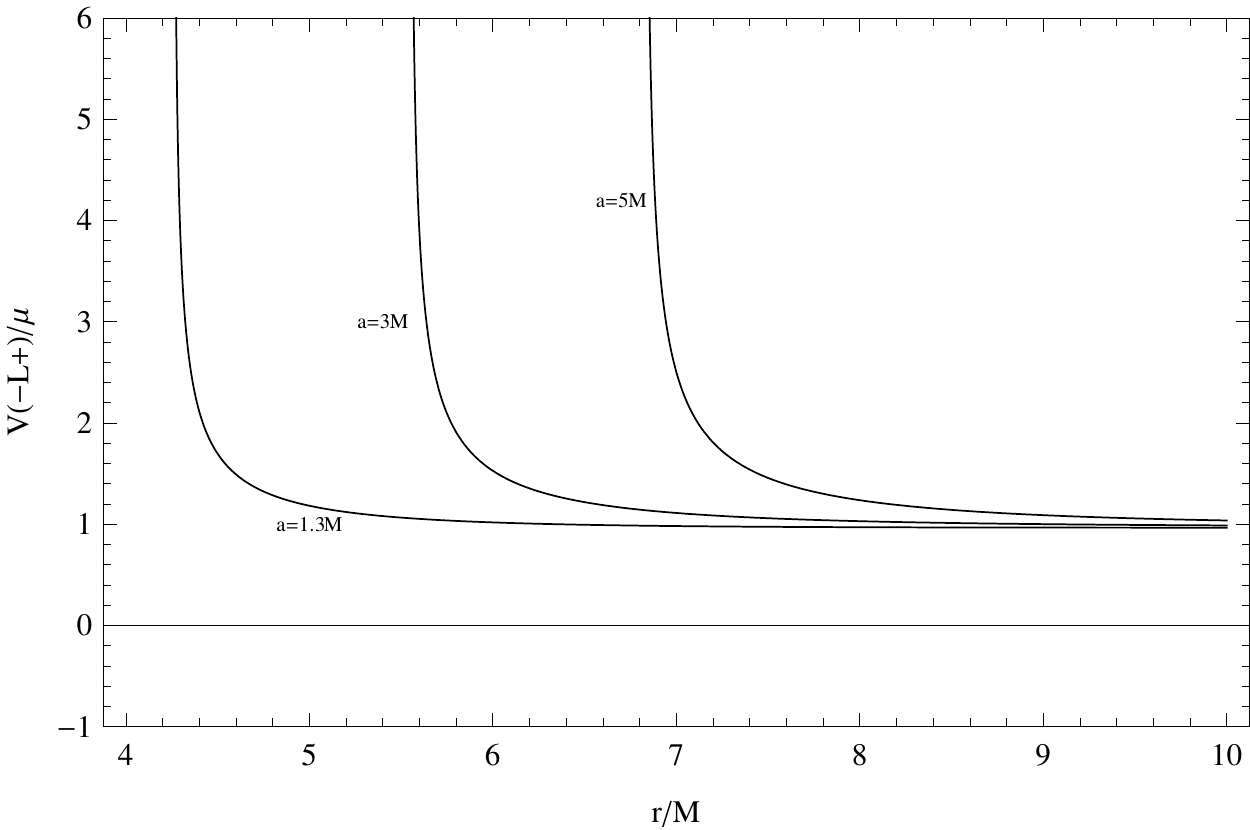}
\end{tabular}
\caption[font={footnotesize,it}]{\footnotesize{The
energy of circular orbits in a Kerr naked singularity with source angular
momentum $a\geq(3\sqrt{3}/4)M$ is plotted in terms of the radial coordinate $r/M$ for selected values of $a/M$.
The left plot corresponds to particles with $L=L_-$, and the right plot is particles with
$L=-L_+$. The energy is always positive and diverges as the limiting radius is approached. The dotted dashed gray line represents  the outer boundary of the ergosphere $r_+^0=2M$.}}
\label{Pemp}
\end{figure}

The energies $E(L_-)$ and $E(-L_+)$ for the two classes of test particles allowed in this are compared in Fig.\il\ref{Pemp}.
For particles with angular momentum $L=L_-$ we see that the energy diverges as the limiting value $r\rightarrow 0$ is approached.
Similarly, for particles with $L=-L_+$ the energy diverges as the radius approaches the limiting value $r\rightarrow r_a$, indicating
that the orbit located at $r=r_a$ is lightlike.

\begin{figure}
\centering
\begin{tabular}{c}
\includegraphics[scale=1]{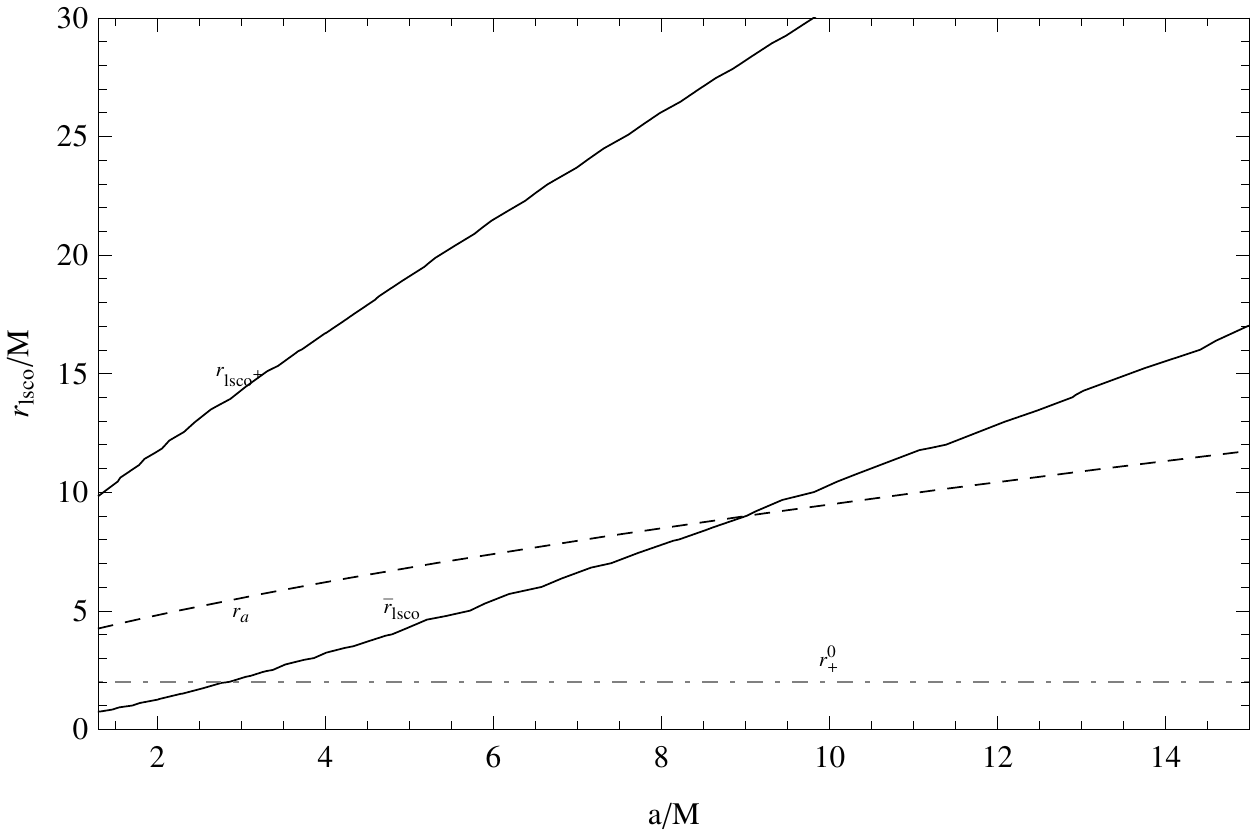}
\end{tabular}
\caption[font={footnotesize,it}]{\footnotesize{Radius of the last
stable circular orbits for test particles in a Kerr naked
singularity with $a\geq\frac{3\sqrt{3}}{4}M$. The dotted dashed gray line represents  the outer boundary of the ergosphere $r_+^0=2M$. The radius
$\bar{r}_{lsco}$ ($r_{lsco}^+$) is the limiting minimum radius of stability for
particles with $L=L_-$ ($L=-L_+$).
}} \label{PSdd}
\end{figure}

We now study the stability of the test particles in this specific case. An
analysis of the turning points of the potential (\ref{qaz}) indicates that
the radius of the last stable circular orbit for particles with $L=L_-$ (located in the region $r>0$)
is given by
\be
\bar{r}_{lsco}  \equiv M\left(3-{Z_2}+\sqrt{(3-{Z_1})(3+{Z_1}-2{Z_2})}\right)\ ,
\label{rbar}
\ee
where $Z_{1}$ and $Z_2$ were defined in Eq.\il(\ref{Eh1}) and Eq.\il(\ref{Eh2}), respectively.
Moreover, for particles with $L=-L_+$ located at $r>r_a$ there exists a minimum radius
$r= r_{lsco}^+$ for the last stable circular orbit. The expression for $r_{lsco}^+$ is given in Eq.\il(\ref{dicadica}).
The behavior of this limiting radii in terms of the intrinsic angular momentum of the naked singularity is
depicted in Fig.\il\ref{PSdd}. If follows that both radii increase as the value of $a/M$ increases.

It turns out that it is necessary to distinguish two different regions, namely ${a}/{M}\in [{3 \sqrt{3}}/{4},9]$
and ${a}/{M}\in ]9,+\infty[$.

\subsubsection{ The region ${a}/{M}\in [{3 \sqrt{3}}/{4},9]$}

\begin{figure}
\centering
\begin{tabular}{cc}
\includegraphics[scale=1]{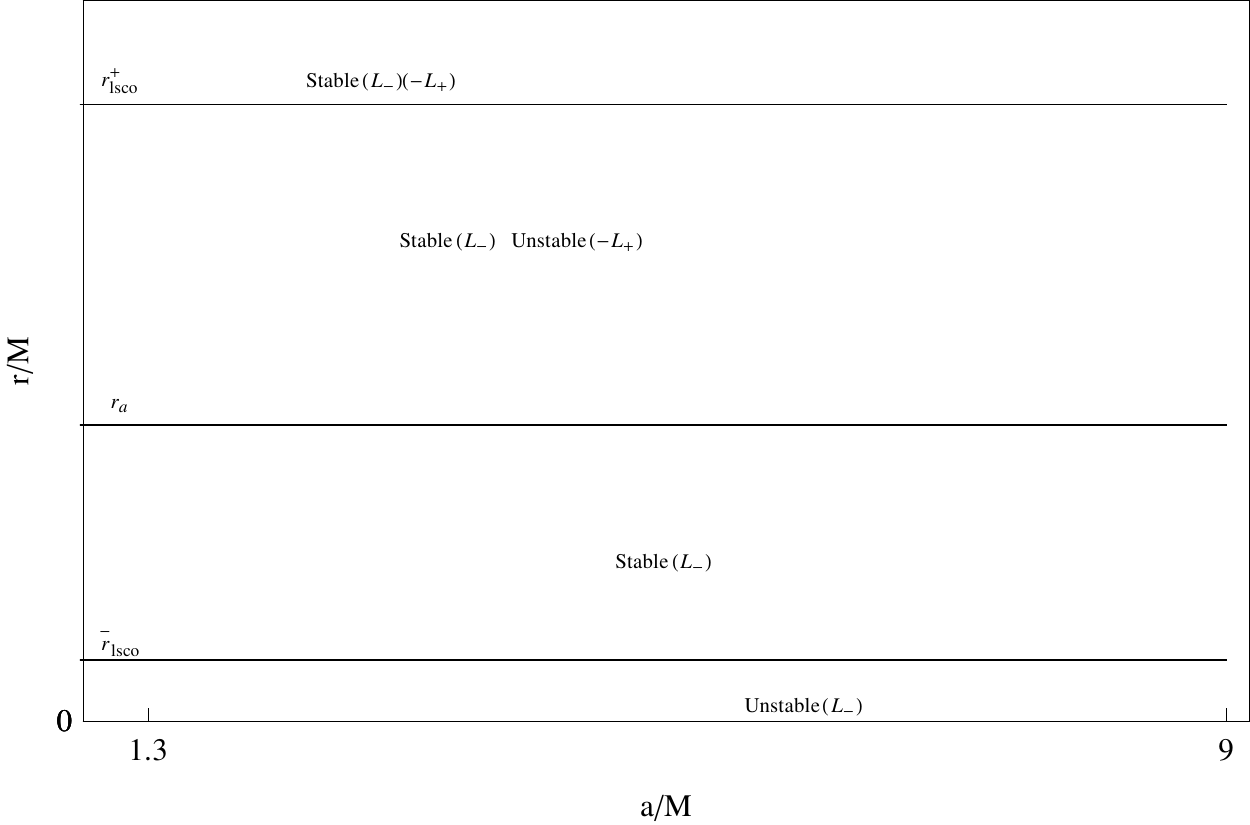}
\end{tabular}
\caption[font={footnotesize,it}]{\footnotesize{Orbits stability in
a Kerr naked singularity with $\frac{3 \sqrt{3}}{4}M\leq a\lesssim9$.
The different radii of the last stable circular orbits  $r_{lsco}$ are
plotted in terms of the intrinsic angular momentum $a/M$. }}
\label{Sta09a}
\end{figure}

In the first region  ${a}/{M}\in [{3 \sqrt{3}}/{4},9]$ which is characterized by
\be \bar{r}_{lsco} <r_a<r_{lsco}^+ ,\quad\mbox{and}\quad
\bar{r}_{lsco} =r_a\quad \mbox{for}\quad a\approx9M,
\ee
there exist unstable circular orbits with $L=L_-$  in the interval $0<r<\bar{r}_{lsco}$,
and stable orbits with $L=L_-$ in the interval $\bar{r}_{lsco} <r<r_a$. Moreover,
in the region $r_a<r<r_{lsco}^+ $ there are stable orbits with
angular momentum $L=L_-$, and unstable orbits with angular momentum $L=-L_+$.
Finally, for $r>r_{lsco}^+ $ there are stable orbits with $L=L_-$ and
$L=-L_+$. In Fig.\il\ref{Sta09a} we present a summary of this case.

\begin{figure}
\centering
\begin{tabular}{cc}
\includegraphics[scale=.7]{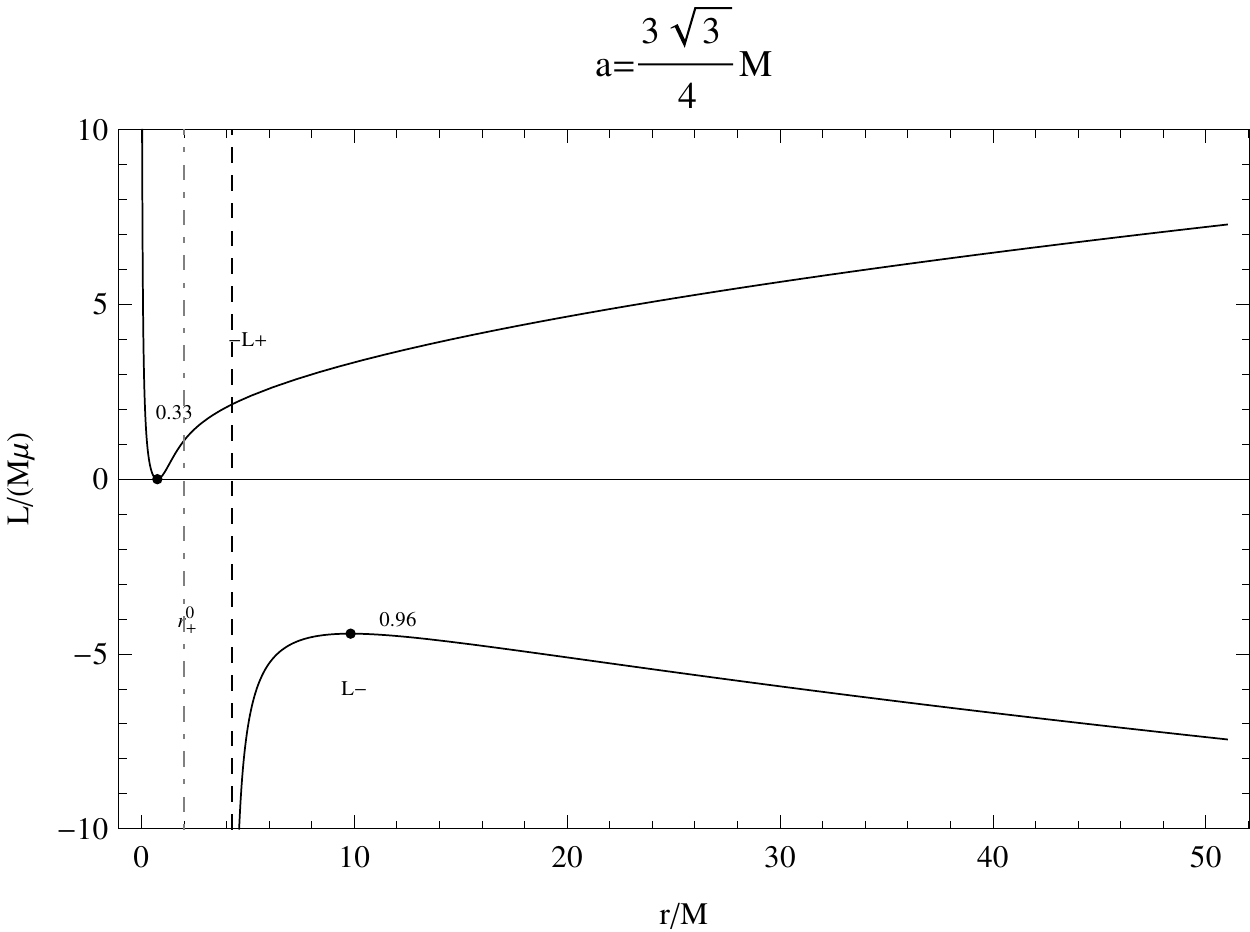}
\includegraphics[scale=.7]{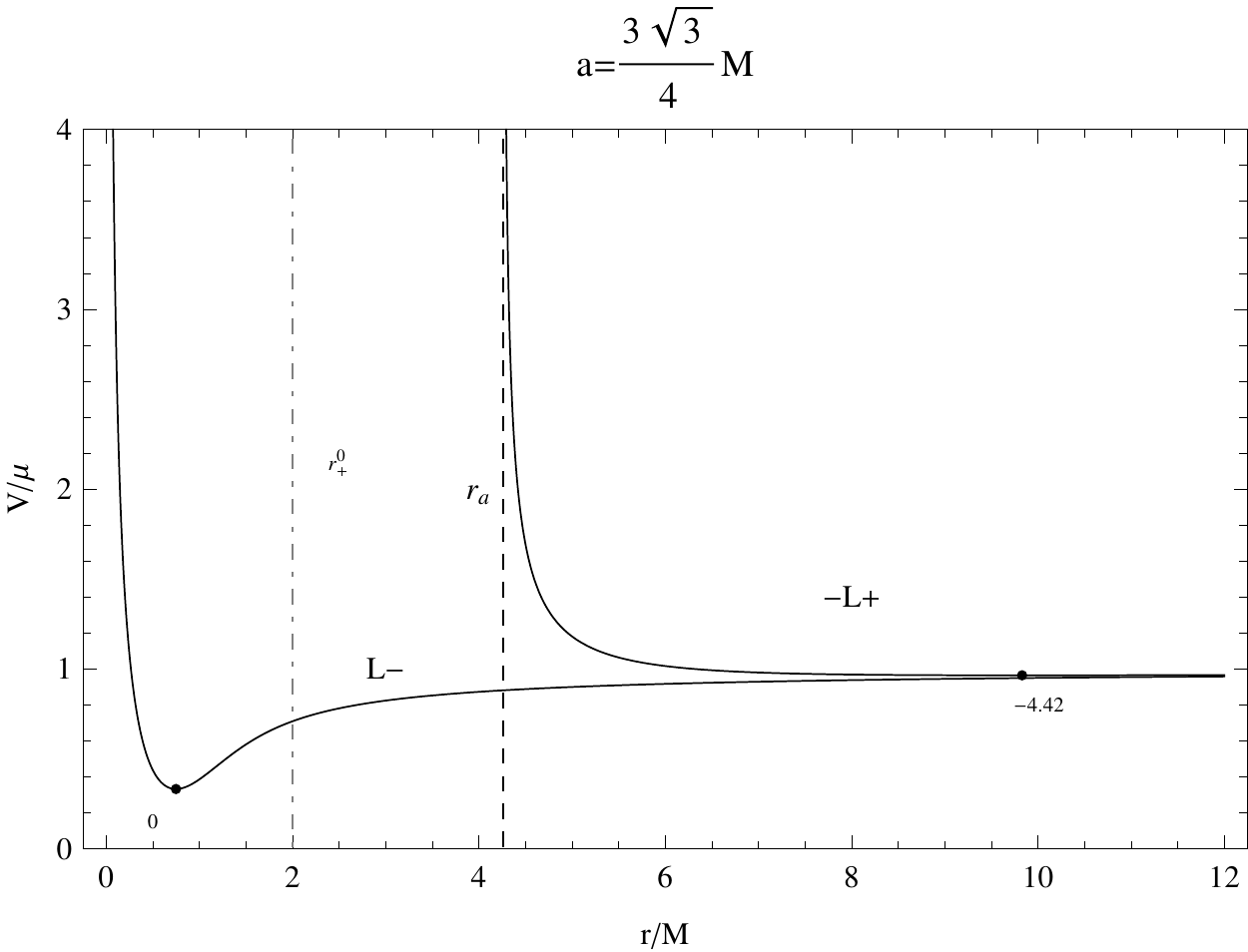}
\end{tabular}
\caption[font={footnotesize,it}]{\footnotesize{The
angular momentum and the energy of test particles in the field of a Kerr
naked singularity with $a=3\sqrt{3}/4 M \approx 1.30M$ are plotted as functions of the
radial coordinate $r/M$. The dots represent  the last stable
circular orbits; numbers close to the points denote the energy  $V/\mu$ of
the last stable circular orbit. For $r>r_{a}\approx4.2592M$ there exist
circular orbits  with angular momentum $L=-L_+$,
and for all $r>0$  with $L=L_-$ (see text).
For $r=0.75M$ the particle has $L=0$ and energy $E\approx0.333M$. The dotted dashed gray line represents  the outer boundary of the ergosphere $r_+^0=2M$.
}}
\label{Plotcira219V}
\end{figure}

As a concrete example for this case we consider now the motion of test particles around
a naked singularity with $a=\frac{3 \sqrt{3}}{4}M$. In this case, circular orbits with orbital angular momentum $L=L_-$ exist in the
range  $r>0$, and with $L=-L_+$ in the range $r>r_{a}\approx4.259M$.
The energy and angular momentum of these circular orbits are plotted in  Figs.\il\ref{Plotcira219V}.

\begin{figure}
\centering
\begin{tabular}{cc}
\includegraphics[scale=.7]{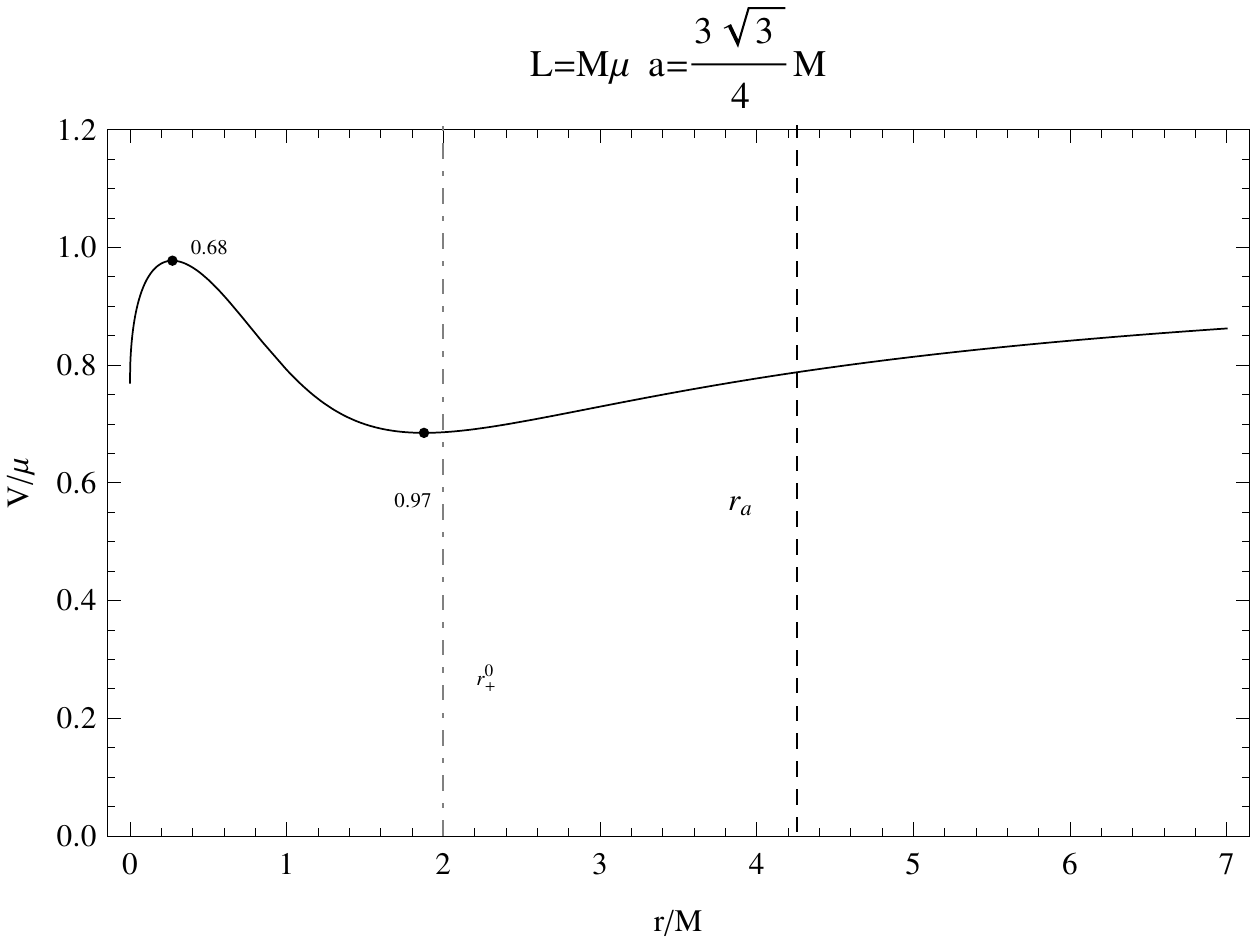}
\includegraphics[scale=.7]{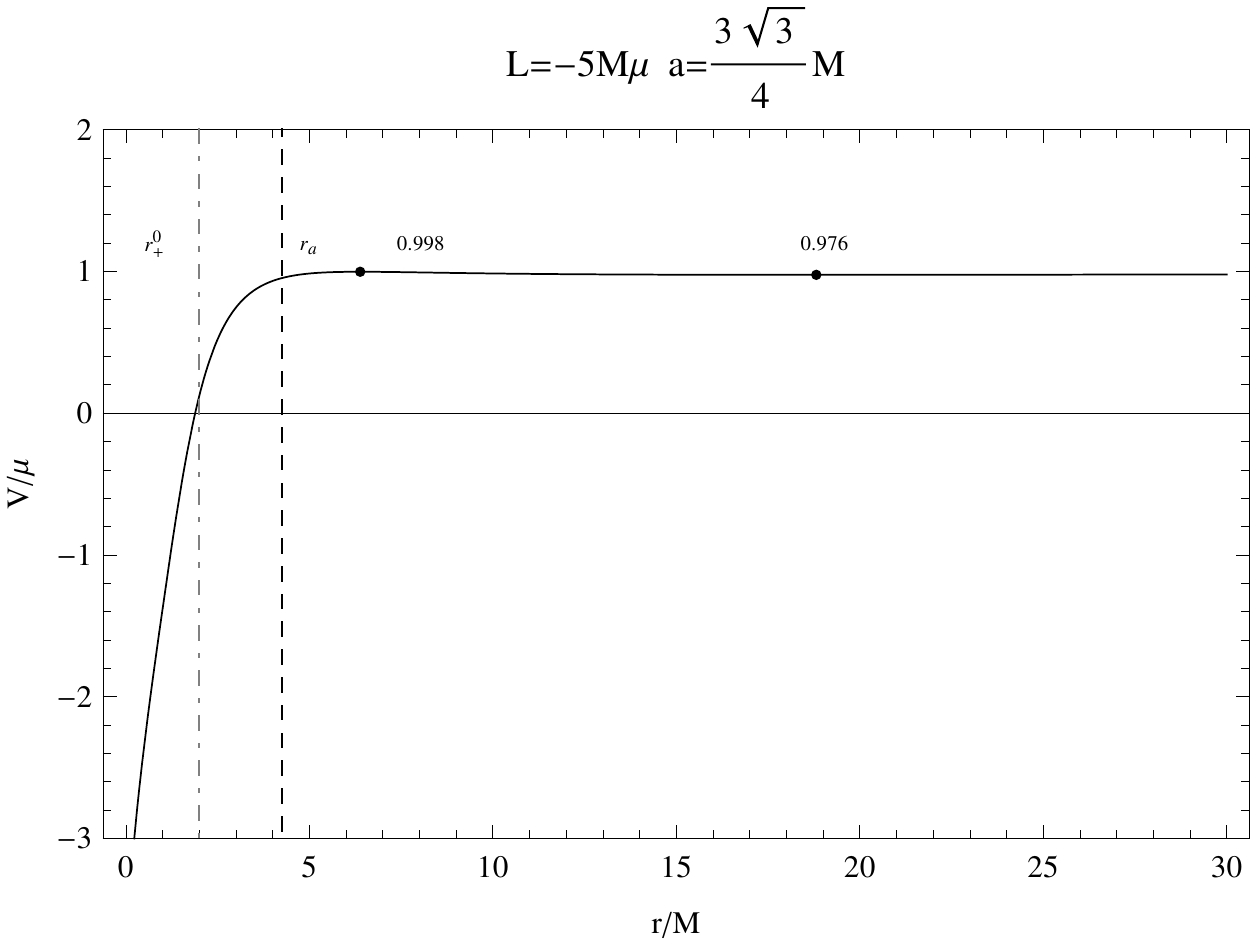}
\end{tabular}
\caption[font={footnotesize,it}]{\footnotesize{The effective
potential of a naked singularity with $a=\frac{3 \sqrt{3}}{4}M$
for fixed values of the particle angular momentum $L/(M\mu)$. The dotted dashed gray line represents  the outer boundary of the ergosphere $r_+^0=2M$.
The radius $r_a$ is also plotted (see text).
The dots denote the critical points of the potential.
Numbers close to the dots denote the energy $V/\mu$ of the maxima and minima
of the effective potential.} }
\label{PlotVa13}
\end{figure}

In  Fig.\il\ref{PlotVa13} the effective potential is plotted for different values of the orbital angular momentum.
In particular, an ``orbit" with zero angular momentum $(L=0)$ and energy $E\approx0.333M$ exists
for $r=0.75M$ (see also Sec.\il\ref{sec:L0}).

From the analysis of the effective potential it follows that the turning points are located at
$r_{lsco}^+   \approx 9.828M$ where $L_{lsco}^+   \approx-4.421\mu M$ and $V_{lsco}^+   \approx0.96\mu$.
Moreover, in the interval  $0<r<r_{lsco}^-   $ the orbits with angular momentum $L=L_-$ are unstable;
in the interval  $r_{lsco}^-   <r<r_{a}$ the orbits with $L=L_-$
are stable; and for $r_{a}<r<r_{lsco}^+   $ we see that the orbits with
$L=L_{-}$ are stable and those with $L=-L_+$ are unstable. Finally,
in the range $r>r_{lsco}^+   $, the
orbits with $L=-L_{+}$ and $L=L_{-}$ are both stable.

%

\subsubsection{The region $\frac{a}{M}\in\left]9,+\infty\right[$}
\begin{figure}
\centering
\begin{tabular}{c}
\includegraphics[scale=1]{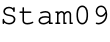}
\end{tabular}
\caption[font={footnotesize,it}]{\footnotesize{Orbits stability in
a Kerr naked singularity with $a\gtrsim9M$.
The radii $r_{lsco}$ of the last stable circular orbits are
plotted as functions of the intrinsic angular momentum $a/M$.
The radius $r=r_a$ is also plotted.
}}
\label{Sta09b}
\end{figure}

In the second region ( $a\gtrsim9M$) which  is characterized by
\be
r_a<\bar{r}_{lsco} <r_{lsco}^+ \ ,
\ee
there are unstable orbits with angular momentum $L=L_-$ in the interval
$ 0<r<r_a$ and with $L=L_-$ and $L=-L_+$ in the  interval
$ r_a<r<\bar{r}_{lsco}$.
Moreover, for $\bar{r}_{lsco} <r<r_{lsco}^+ $ there are stable orbits    with $L=L_-$ and
unstable ones  with $L=-L_+$.
Finally,  for $r>r_{lsco}^+ $ there are
stable orbits with  both $L=L_-$ and   $L=-L_+$.
In Fig.\il\ref{Sta09b} a schematic summary of this case is presented.

\begin{figure}
\centering
\begin{tabular}{cc}
\includegraphics[scale=.7]{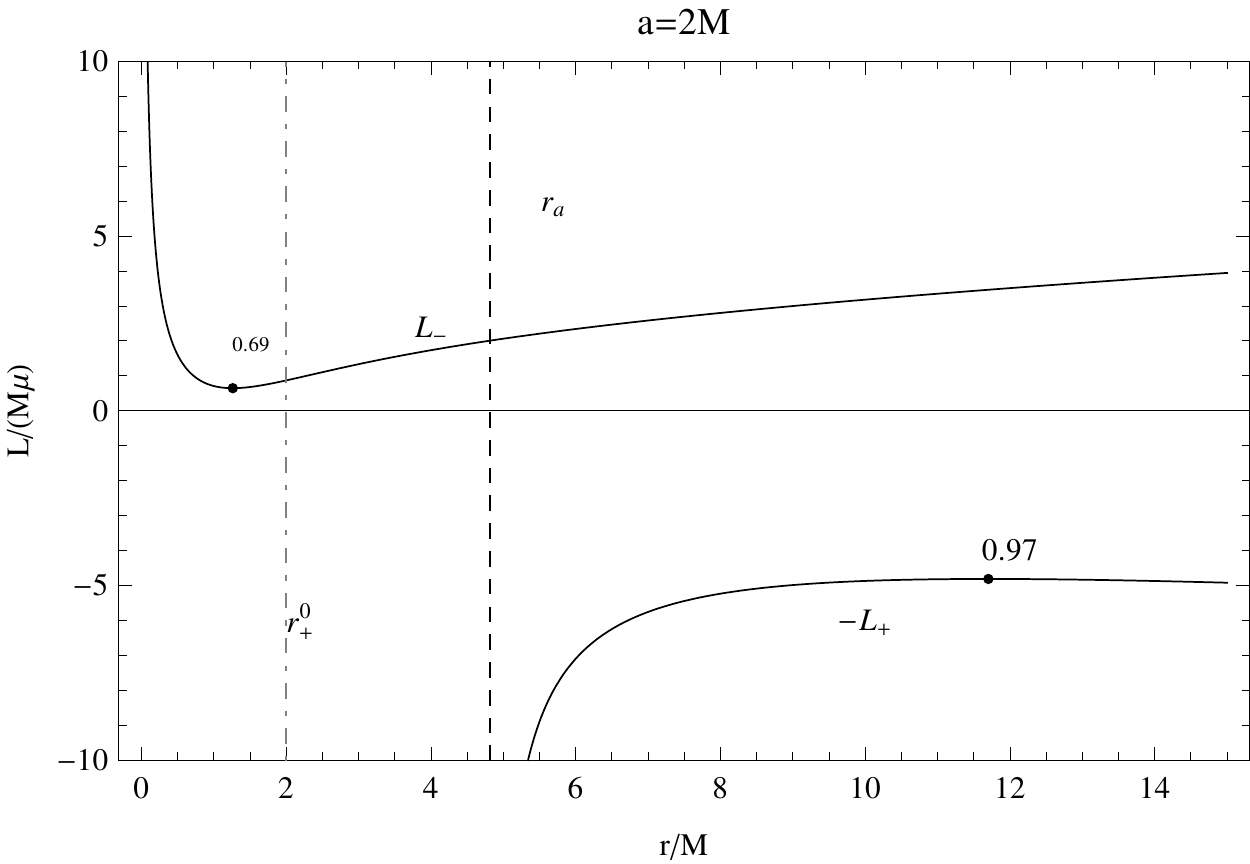}
\includegraphics[scale=.7]{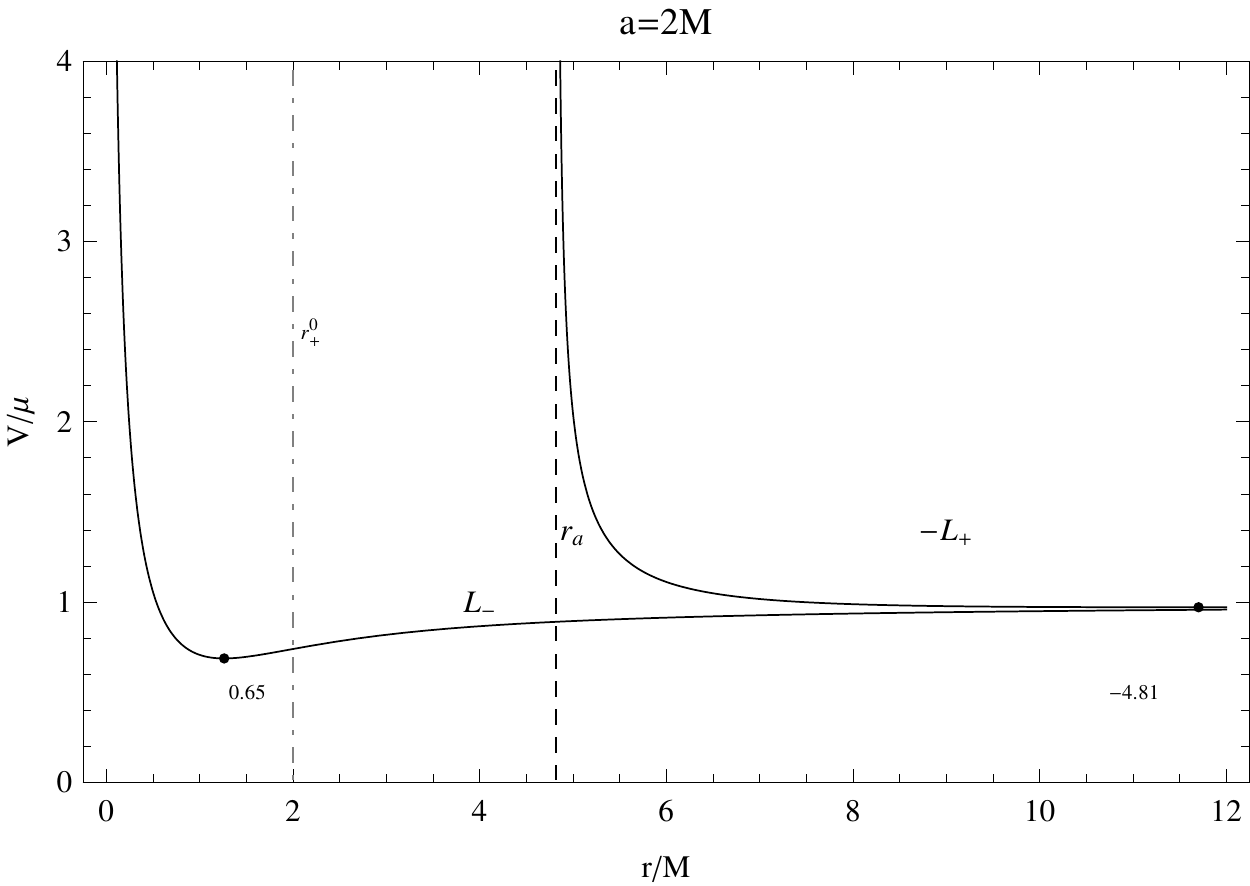}
\end{tabular}
\caption[font={footnotesize,it}]{\footnotesize{The angular momentum and
the energy of circular orbits in a Kerr naked singularity with   $a=2M$,
as functions of the radial distance $r/M$. The dotted dashed gray line represents  the outer boundary of the ergosphere $r_+^0=2M$.
The dots denote the position of the last stable
circular orbits, and the numbers close to the dots indicate the value of the corresponding
energy $V/\mu$ or angular momentum $L/(M\mu)$. In the range  $r>r_{a}\approx4.822M$ there exist circular
orbits with $L=-L_+$, and in $r>0$ with   $L=L_-$ (see text).} }
\label{Plotcira2}
\end{figure}

As a concrete example of this case we now analyze the circular motion of test particles around
a naked singularity with $a=2M$. In this case, circular orbits with angular momentum $L=L_-$ exist in
the entire range  $r>0$, and with $L= -L_+$ in the range $r>r_{a}\approx4.822M$.
The energy and the angular momentum of the circular orbits are plotted in
Figs.\il\ref{Plotcira2}.

\begin{figure}
\centering
\begin{tabular}{cc}
\includegraphics[scale=.7]{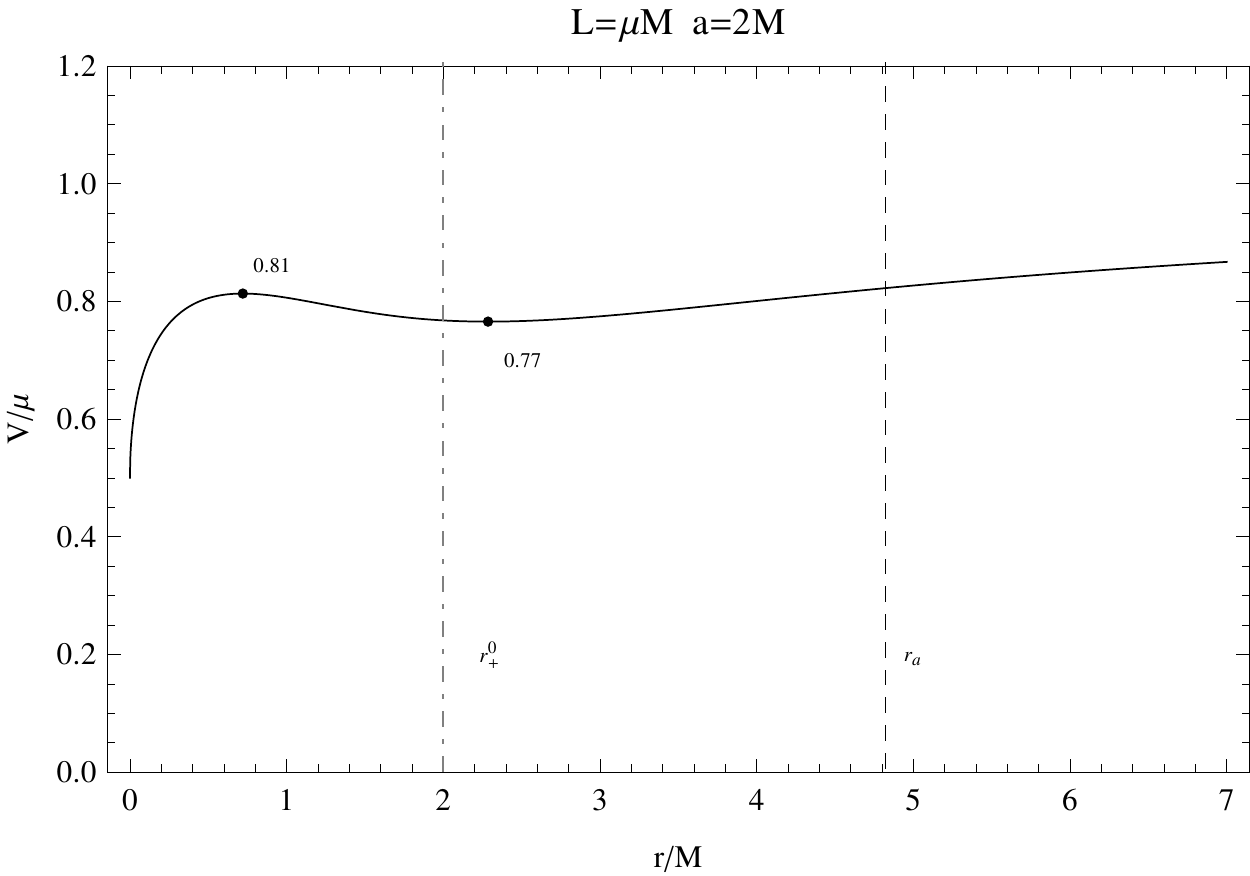}
\includegraphics[scale=.7]{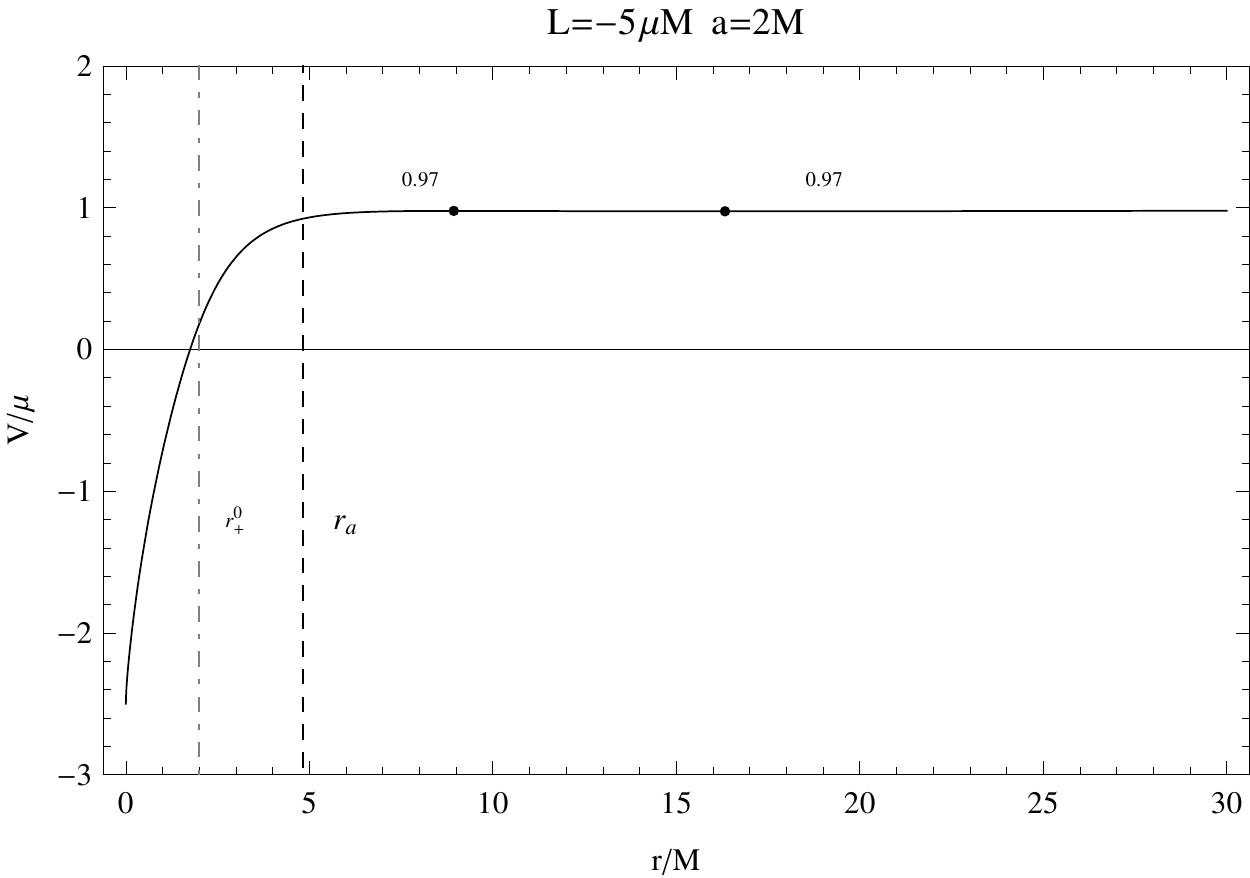}
\end{tabular}
\caption[font={footnotesize,it}]{\footnotesize{The effective
potential of a naked singularity with $a=2M$  for fixed values of
the particle angular momentum $L/(M\mu)$. The radius $r_a$ is also plotted (see text). The dotted dashed gray line represents  the outer boundary of the ergosphere $r_+^0=2M$. The dots represent
the critical points of the potential. Numbers close to the dots indicate the energy $V/\mu$ of
the maxima and minima of the effective potential.
} }
\label{PlotVa2}
\end{figure}

In Fig.\il\ref{PlotVa2}, the effective potential of circular orbits is plotted for selected values of the orbital angular momentum in terms
of the radial distance.
The turning points of the effective potential are
$r_{lsco}^+   \approx 11.702M$ for which $L_{lsco}^+   \approx-4.814\mu M$
and $V_{lsco}^+   \approx0.971\mu$,
and $r_{lsco}^-   \approx1.263M$
with $L_{lsco}^-   \approx0.645\mu M$ and
$V_{lsco}^-   \approx0.687\mu$.

%
%

The distribution of circular orbits is as follows:
In the interval $0<r<\bar{r}_{lsco}   $  there exist unstable
orbits with $L=L_{-}$ which become stable for $\bar{r}_{lsco}   <r<r_{a}$; in the
interval   $r_{a}<r<r_{lsco}^+   $ the orbits with
$L=L_{-}$ are stable while those with $L=-L_{+}$ are unstable. In the outer region
 $r>r_{lsco}^+   $ orbits with  $L=-L_{+}$ and $L=L_{-}$ are both stable.
To illustrate the results of the analysis of this case, we consider in the region $r>r^0_+$ the model of an
accretion disk made of stable particles moving on circular orbits around the central naked singularity.
We find an accretion disk composed of an interior disk contained within the radii $[\bar{r}_{lsco},r_{lsco}^+]$
in which stable particles with angular momentum $L=L_-$ co-rotate with the central singularity. A second
disk is located at $r>r_{lsco}^+$ and contains co-rotating particles with angular momentum $L=L_-$ and
counter-rotating particles with $L=-L_+$. We see that the  structure of this accretion disk is similar
to that found in Sec. \ref{yappi} for black holes. The only difference is that in the case of a naked
singularity the interior disk situated within the radii $[\bar{r}_{lsco},r_{lsco}^+]$ has a minimum size of
$r_{lsco}^+ - \bar{r}_{lsco} >8M$, whereas in the case of a black hole the size of the inner disk is always
less than $8M$ and disappears as $a\rightarrow 0$.

\subsection{The case $M<a<(3 \sqrt{3}/4)M$}
\label{subsec:min}

\begin{figure}
\centering
\begin{tabular}{cc}
\includegraphics[scale=.7]{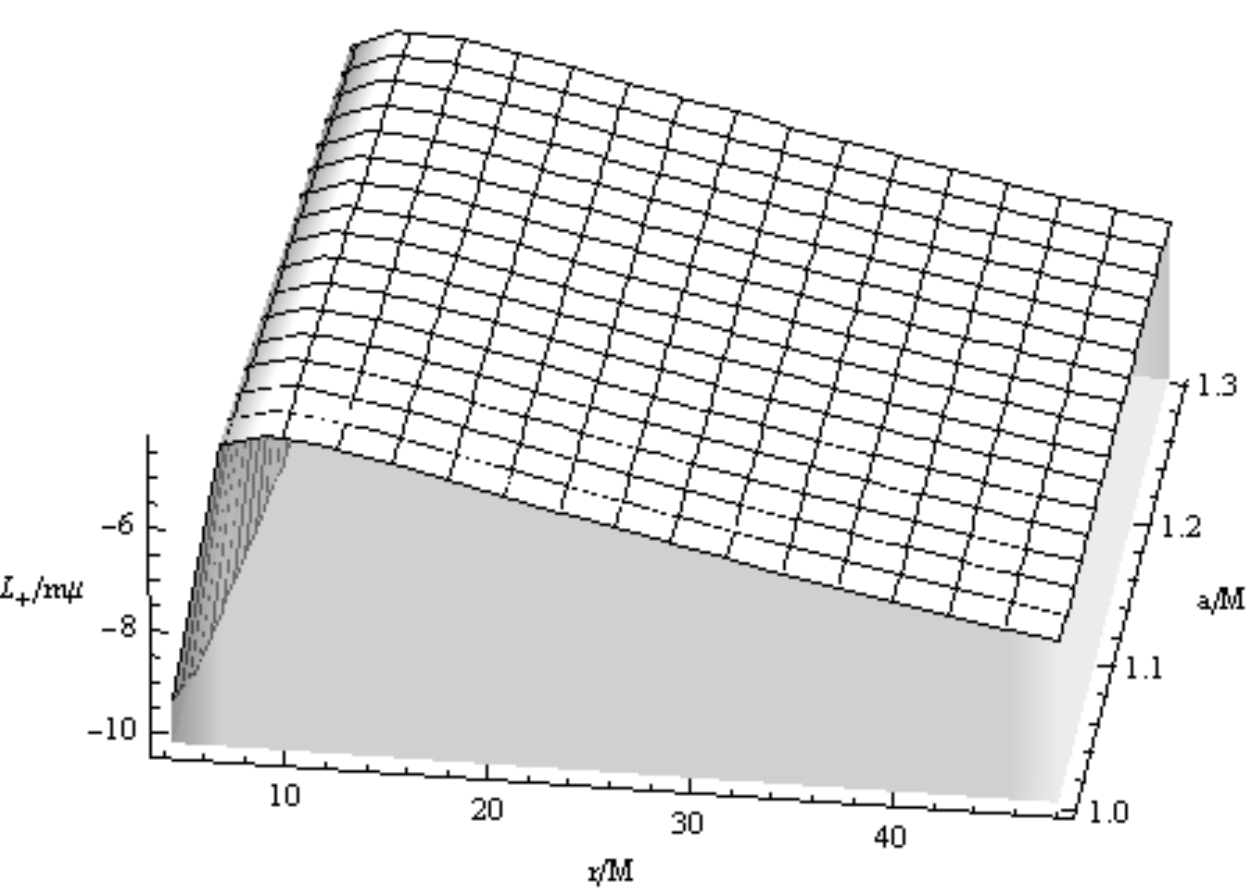}
\includegraphics[scale=.7]{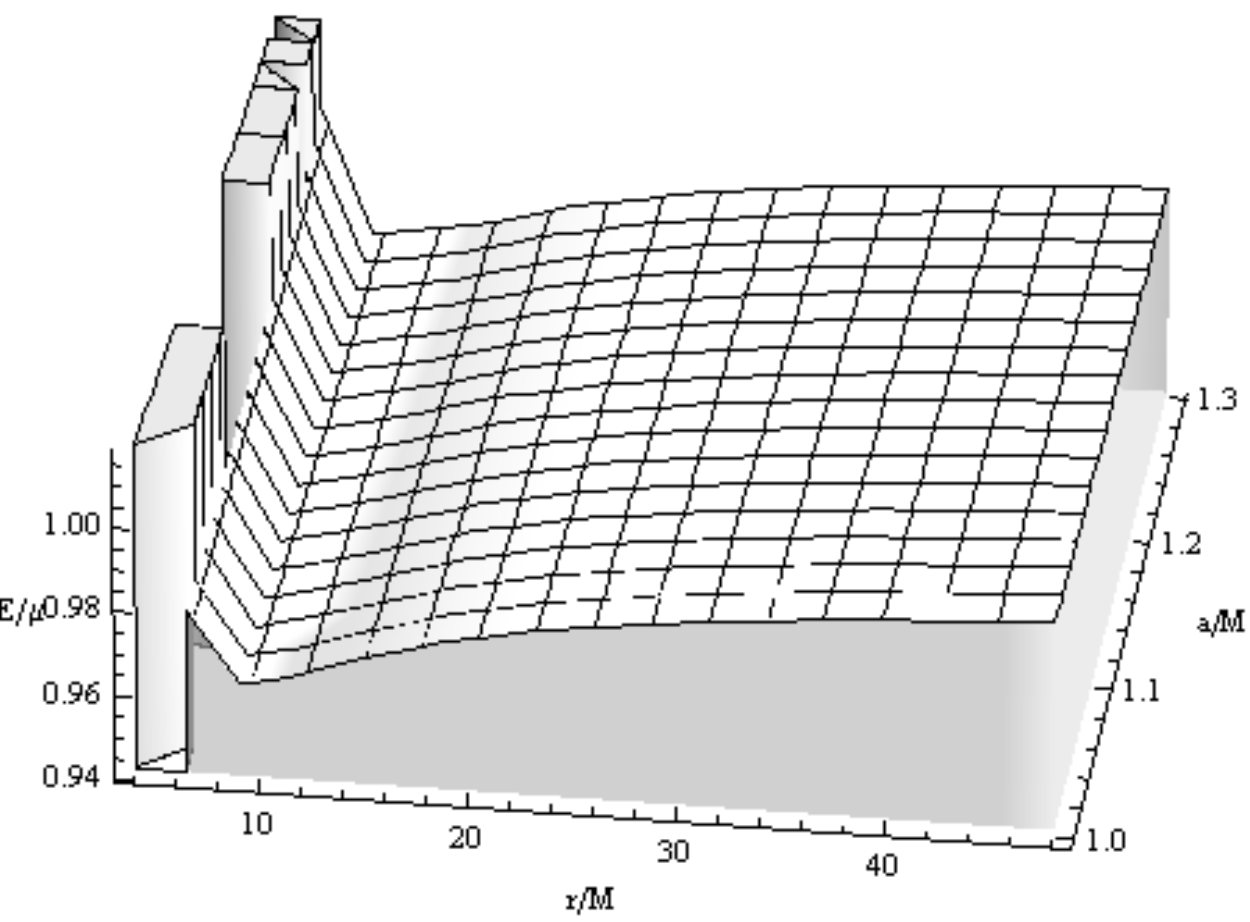}
\end{tabular}
\caption[font={footnotesize,it}]{\footnotesize{Circular motion around a naked singularity with $M<a<(3 \sqrt{3}/4)M$.
The angular momentum $L=-L_{+}$ (left plot) and the energy $E_+^{(-)} \equiv E(-L_+)$ (right plot) for circular orbits are plotted
as functions of $a$ in the range $1<a/M< 3\sqrt{3}/4$ and $r$ in the range $r>r_a$.
The particle energy is always positive with a region of minima corresponding to the  minima of $-L_{+}$.
}}
\label{Sec2mLpnew}
\end{figure}

For this range of values of the intrinsic angular momentum of the naked singularity we find
that there are circular orbits with angular momentum $L=-L_{+}$ and
energy $E(-L_{+})$ only  in the region $r>r_{a}$. In Fig.\il\ref{Sec2mLpnew} we present the
parameters for the circular orbits.

\begin{figure}
\centering
\begin{tabular}{cc}
\includegraphics[scale=.7]{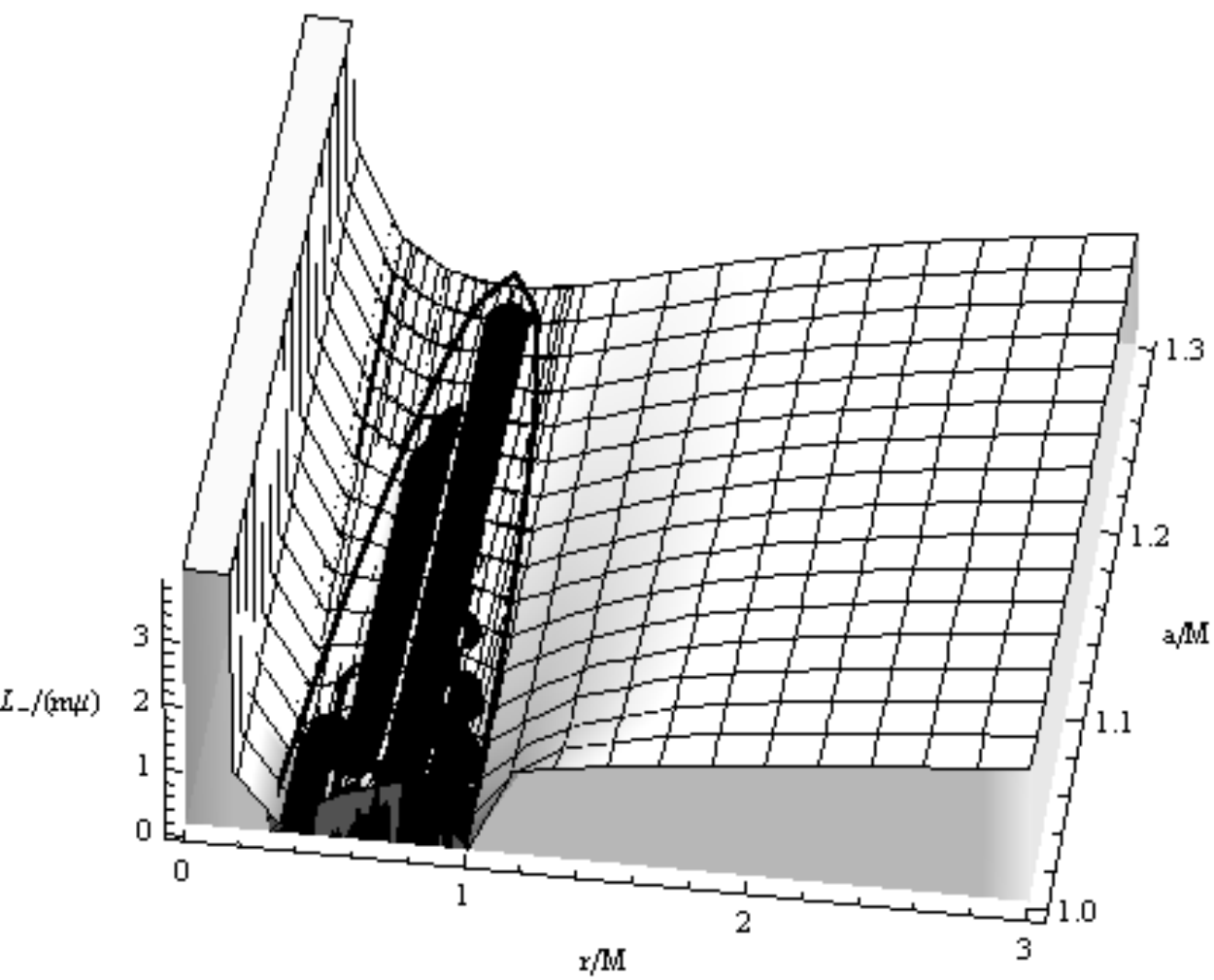}
\includegraphics[scale=.7]{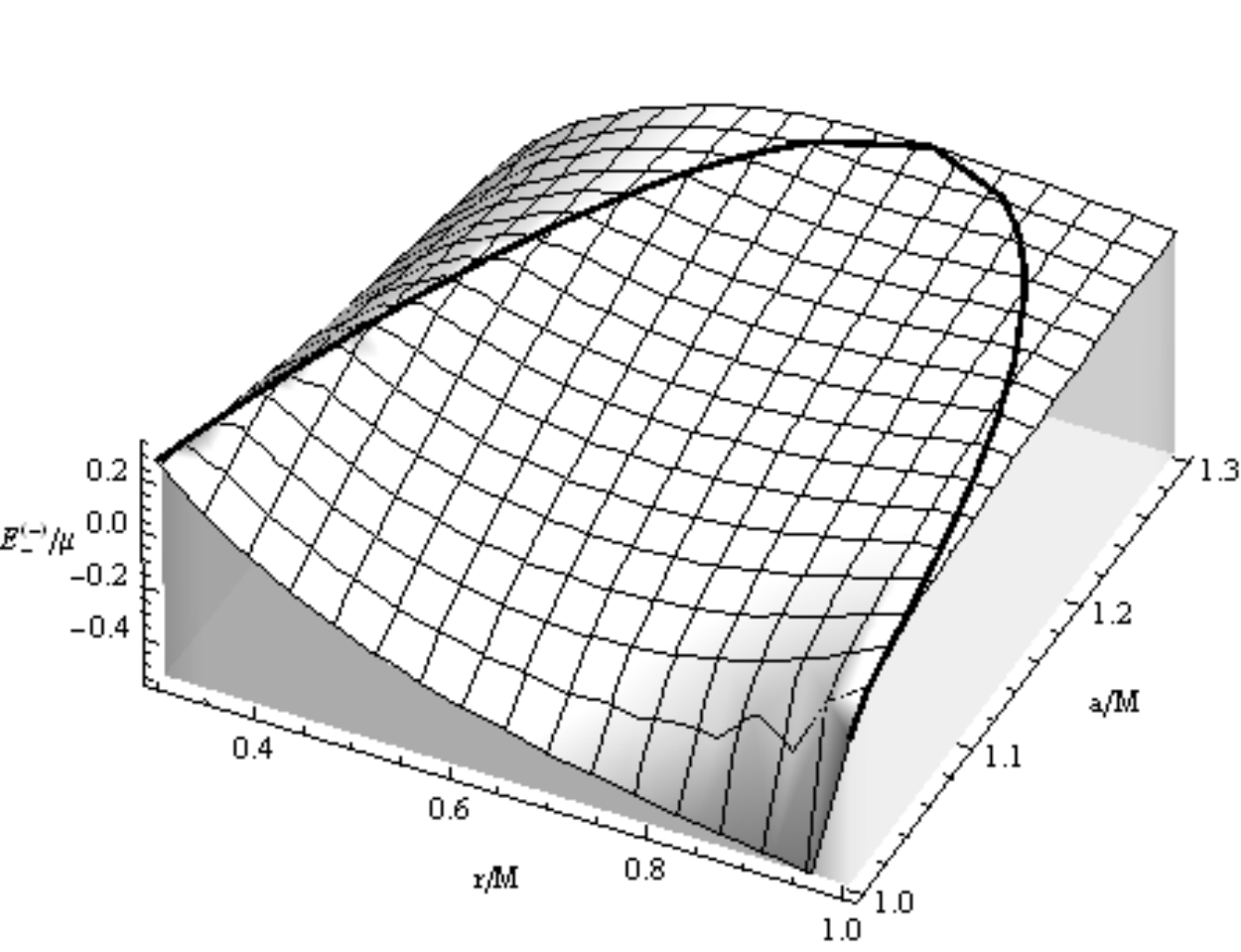}
\end{tabular}
\caption[font={footnotesize,it}]{\footnotesize{Circular motion around a naked singularity with $M<a<(3 \sqrt{3}/4)M$.
The angular momentum $L=L_{-}$ (left plot) and the energy $E_-\equiv E(L_-)$ (right plot) of circular orbits are plotted as functions of of $a$ in the range $1<a/M< 3\sqrt{3}/4$
and $r$ in the intervals $r>\hat{r}_+$ and
$0<r<\hat{r}_-$.
The region $\hat{r}_-<r<\hat{r}_+$ is represented as a dark region.
As $r/M$ approaches the singularity, the particle energy and angular momentum diverge.
As $r/M$ approaches $\hat{r}_-$ from  the left and $\hat{r}_+$ from  the right,  the particle energy and angular momentum decrease.}}
\label{ShSec1mL}
\end{figure}

\begin{figure}
\centering
\begin{tabular}{cc}
\includegraphics[scale=.7]{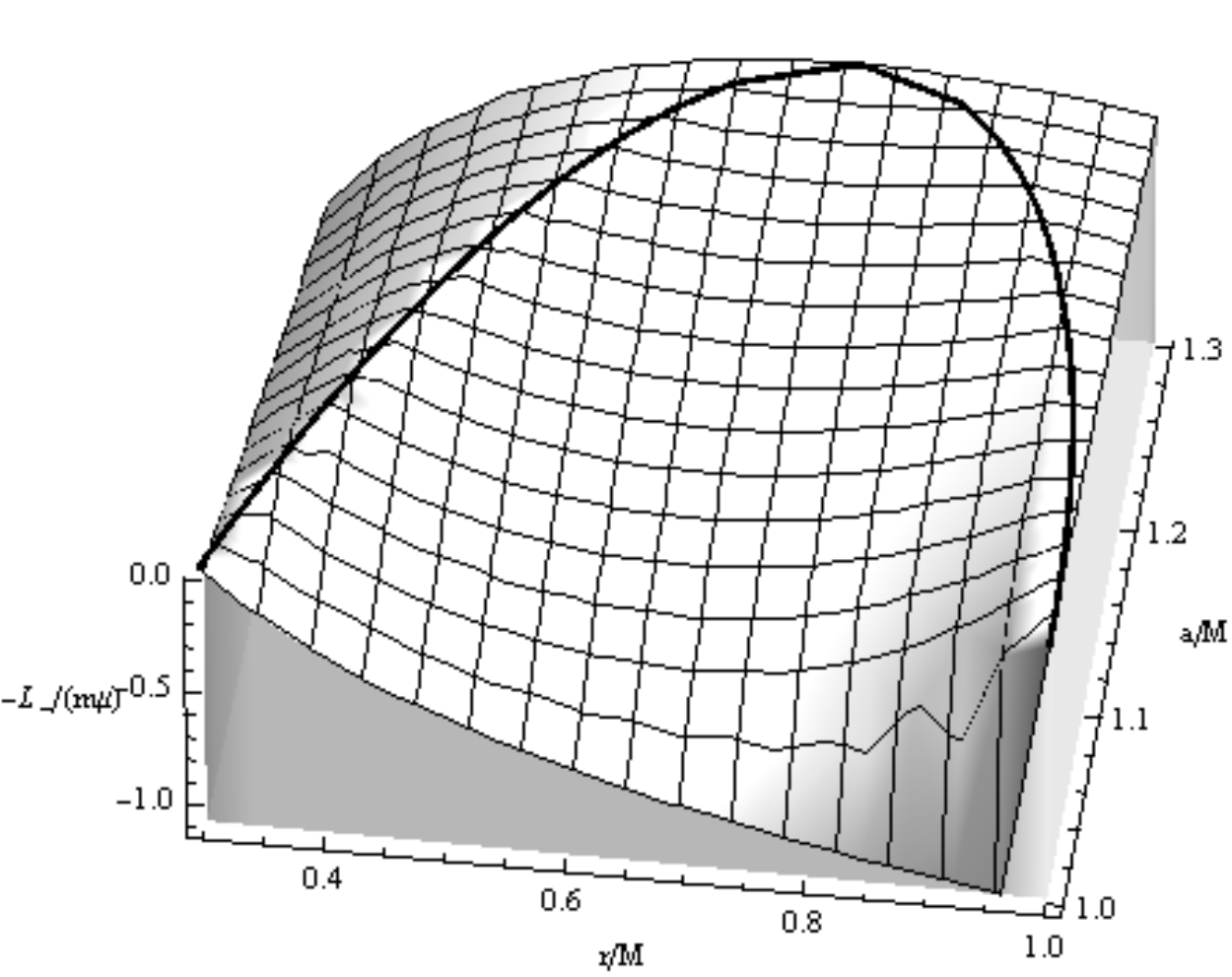}
\includegraphics[scale=.7]{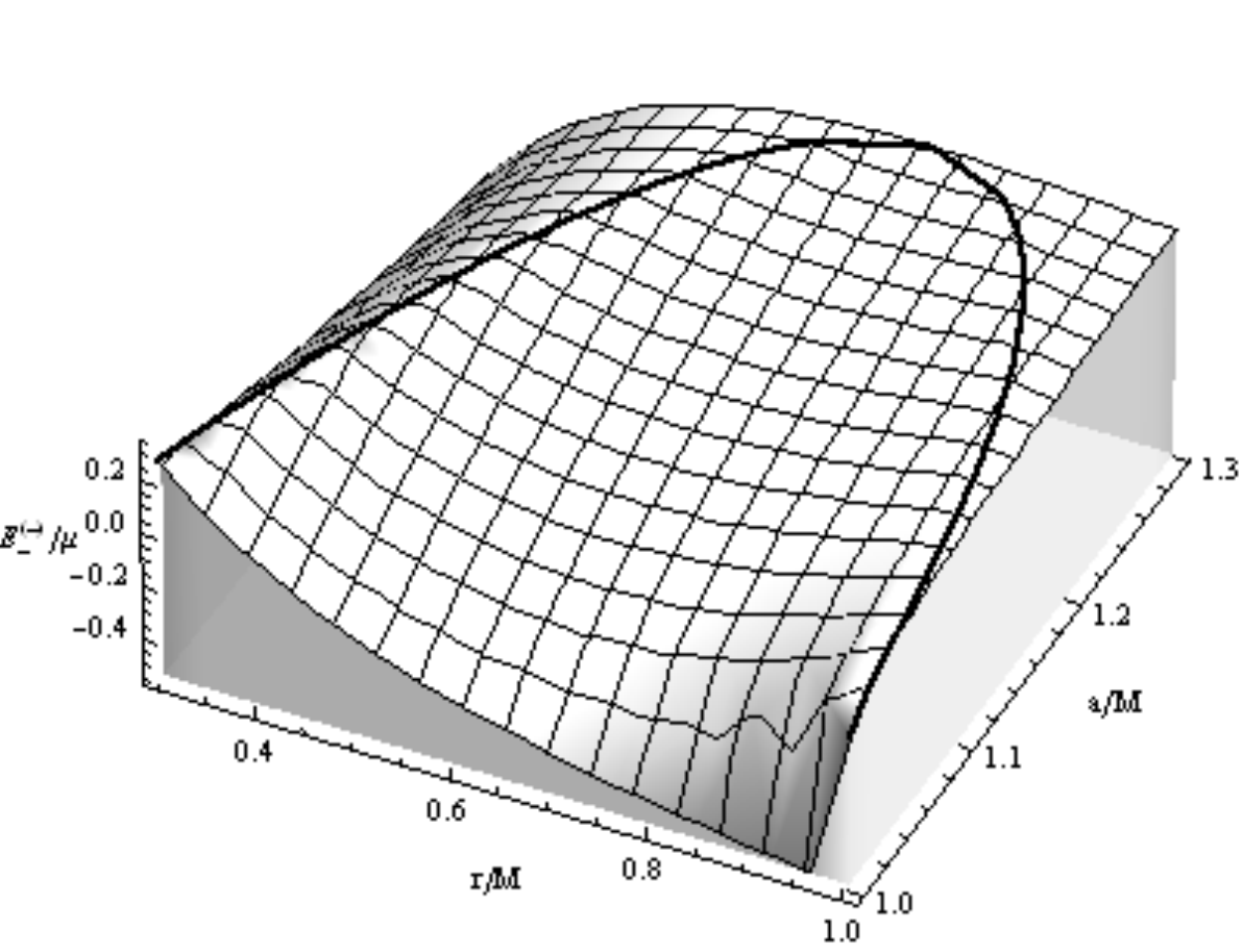}
\end{tabular}
\caption[font={footnotesize,it}]{\footnotesize{Circular motion around a naked singularity with $M<a<(3 \sqrt{3}/4)M$.
The angular momentum $L=-L_{-}$ (left plot) and the energy $E_-^-\equiv E(-L_-)$ (right plot) of circular orbits are plotted as functions of
of $a$ in the range $1<a/M< 3\sqrt{3}/4$ and of
$r$ in the interval $\hat{r}_-<r<\hat{r}_+$. The black curves represent the radii $\hat{r}_-$ and $\hat{r}_+$.
The presence of negative values for the particle energy is evident.
}}
\label{ShSec1mLm}
\end{figure}

\begin{figure}
\centering
\begin{tabular}{cc}
\includegraphics[scale=1]{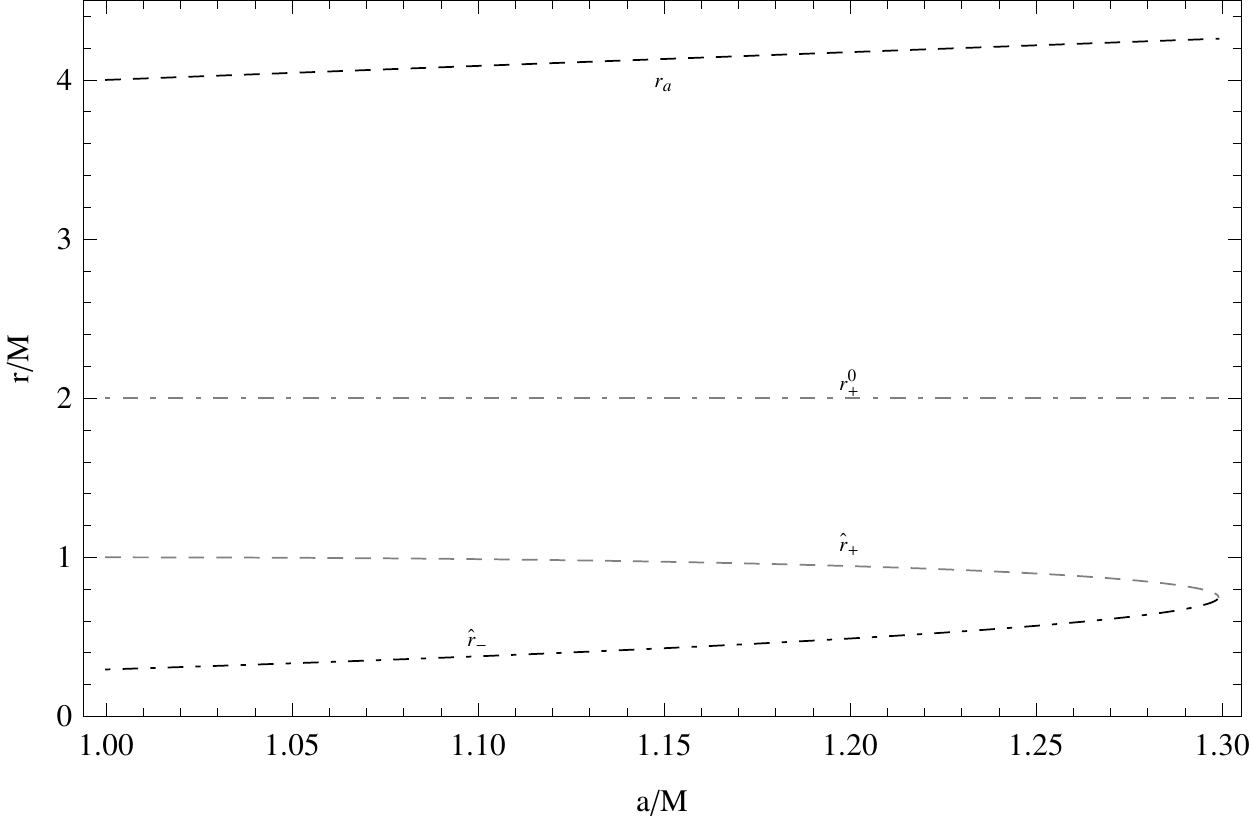}
\end{tabular}
\caption[font={footnotesize,it}]{\footnotesize{The radii $r_{a}$ and
$\hat{r}_{\pm}$ are plotted as functions of $a/M$. Circular orbits
with angular momentum $L=-L_{+}$ exist for $r>r_{a}$ , with  $L=L_{-}$ in $0<r<\hat{r}_{-}$ and
$r\geq \hat{r}_{+}$, and with $L=-L_{-}$ in $\hat{r}_-<r<\hat{r}_+$ (see text). The dotted dashed gray line represents  the outer boundary of the ergosphere $r_+^0=2M$.
}}
\label{R4R3}
\end{figure}

\begin{figure}
\centering
\begin{tabular}{c}
\includegraphics[scale=1]{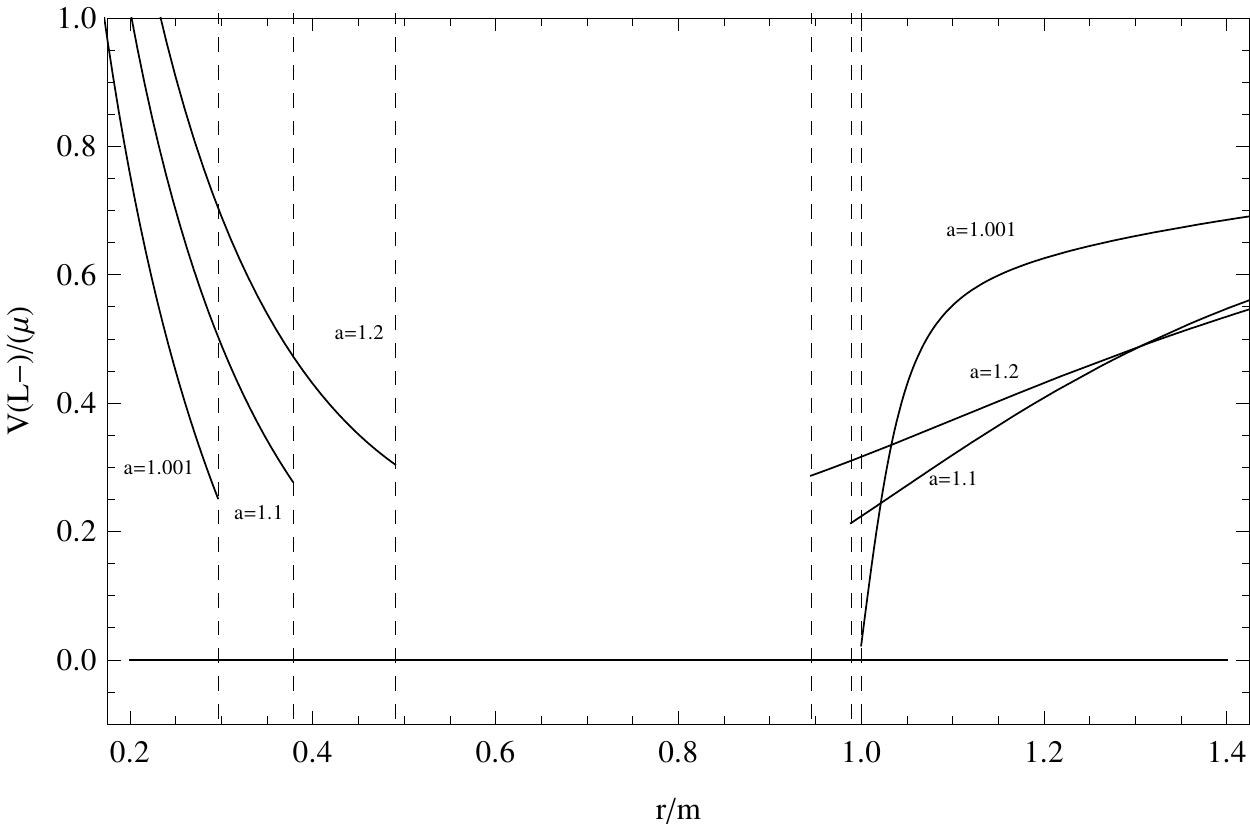}
\end{tabular}
\caption[font={footnotesize,it}]{\footnotesize{The energy $E(L_-)$ of test particles is plotted for
selected values of $a$ in the range $M<a<(3 \sqrt{3}/4)M$ and for $r>\hat{r}_+$ and
$0<r<\hat{r}_-$.}}
\label{Pmpab}
\end{figure}

\begin{figure}
\centering
\begin{tabular}{c}
\includegraphics[scale=1]{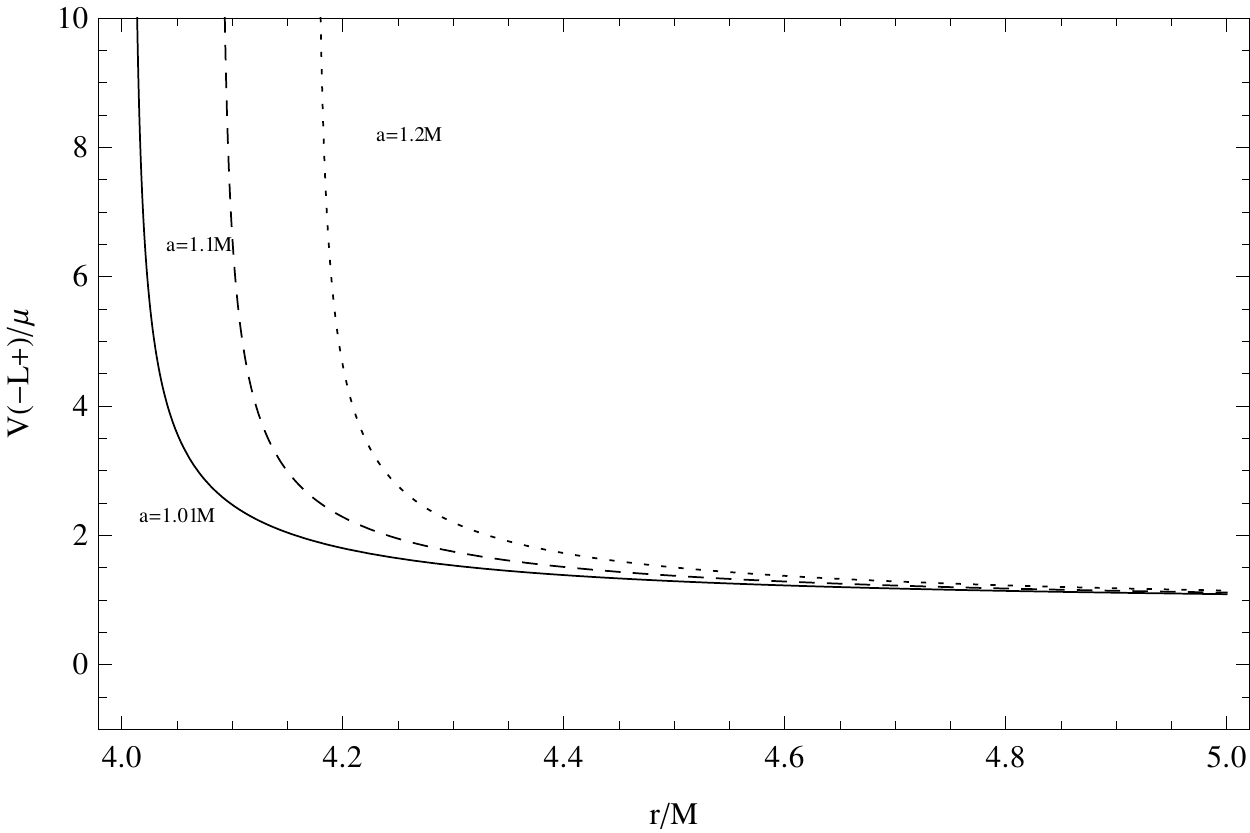}
\end{tabular}
\caption[font={footnotesize,it}]{\footnotesize{The
energy $E(-L_+)$ is plotted for selected values of $a$ in the range
$M<a<(3 \sqrt{3}/4)M$ and for $r>r_a$. The energy $E(-L_+)$ is always
strictly positive and increases as the angular momentum
$a/M$ increases.}}
\label{Pmpaa}
\end{figure}

From the expression for the effective potential and the conditions for circular motion it follows that in this case
two additional regions arise. Indeed, in the intervals
$0<r<\hat{r}_{-}$ and $r\geq\hat{r}_{+}$ there exist circular orbits with angular
momentum $L=L_{-}$ and energy $E(L_{-})$ (see Fig.\il\ref{ShSec1mL}). Moreover,
in the interval  $\hat{r}_-<r<\hat{r}_+$  we observe circular orbits with angular momentum  $L=-L_{-}$ and energy
$E(-L_{-})$ (see Fig.\il\ref{ShSec1mLm}),
where
\be\label{rpm}
\hat{r}_{\pm}\equiv\frac{1}{\sqrt{6}}\left[\Sigma \pm\sqrt{\frac{6\sqrt{6}
a^2M}{\Sigma}-\Sigma^2-6a^2}\right],
\ee
with
\be
\Sigma=\sqrt{\frac{4a^4}{\sigma^{1/3}}+\sigma^{1/3}-2a^2},
\ee
and
\be
\sigma=\left(27M^2a^4-8a^6+3M\sqrt{81M^2a^8-48a^{10}}\right)\ .
\ee
The behavior of these special radii is illustrated in Fig.\il\ref{R4R3}.
Notice that the energy of circular orbits  $E(-L_+)$ in
the interval $0<r<\hat{r}_{-}$ and in $r\geq \hat{r}_{+}$ (see Fig.\il\ref{Pmpab}),
and the energy $E(L_{-})$ in the interval $r>r_a$
are always positive (see Fig.\il\ref{Pmpaa}).
%
On the contrary, the energy $E(-L_-)$  of circular orbits within the region  $\hat{r}_-<r<\hat{r}_+$ can be
negative. In particular, we see that $E(-L_-)=0$ for $a=\bar{a}$, where
\be
\bar{a}\equiv-(r-2M) \sqrt{\frac{r}{M}}\ ,
\ee
or for the orbital radii $r=\bar{r}_{1}$ and
$r=\bar{r}_{2}$, where
\be \frac{\bar{r}_{1}}{M}\equiv\frac{8}{3}
\sin\left(\frac{1}{6} \arccos\left[1-\frac{27 a^2}{16M^2
}\right]\right)^2,
\ee
and
\be
\frac{\bar{r}_{2}}{M}\equiv\frac{4}{3}
\left(1+\sin\left[\frac{1}{3} \arcsin\left[1-\frac{27 a^2}{16M^2
}\right]\right]\right),
\ee
which are the solutions of the equation  $a=\bar{a}$.

\begin{figure}
\centering
\begin{tabular}{cc}
\includegraphics[scale=1]{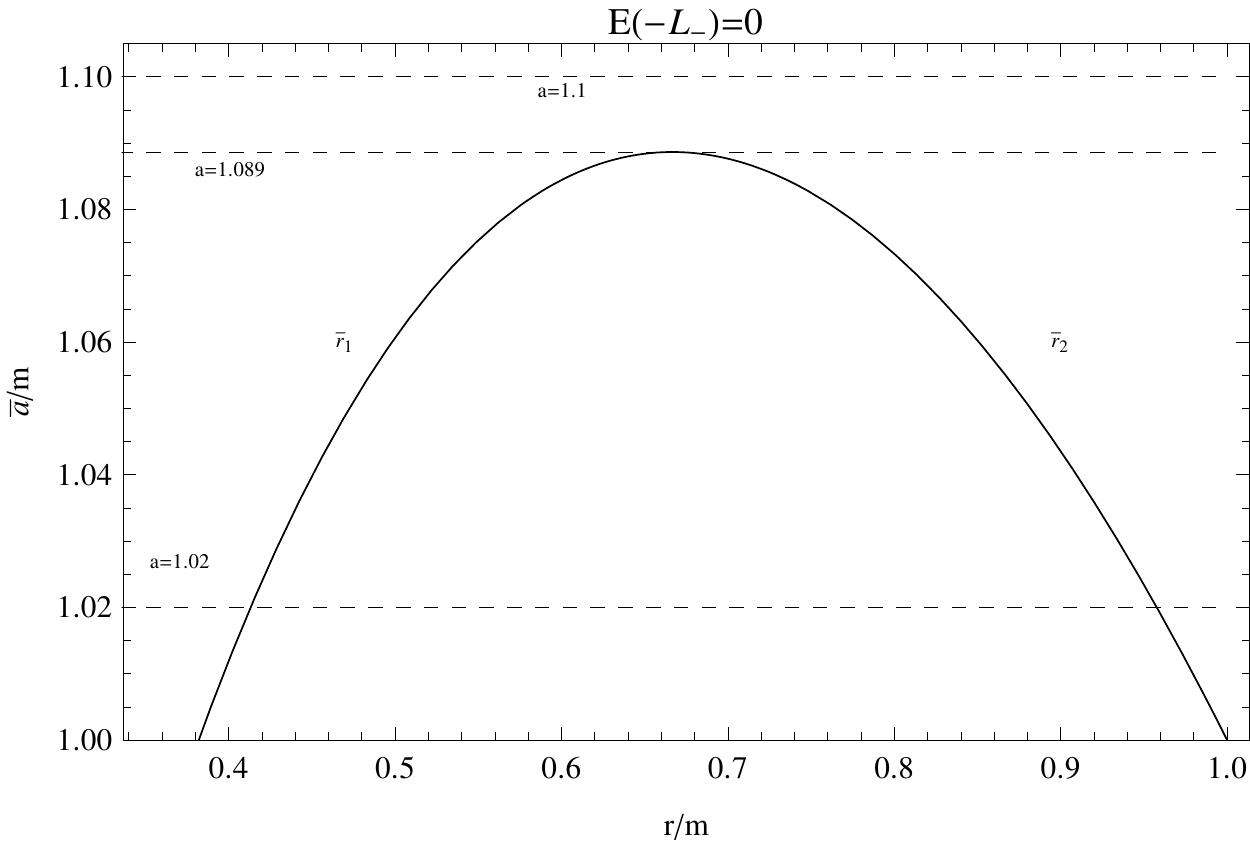}\\
\includegraphics[scale=1]{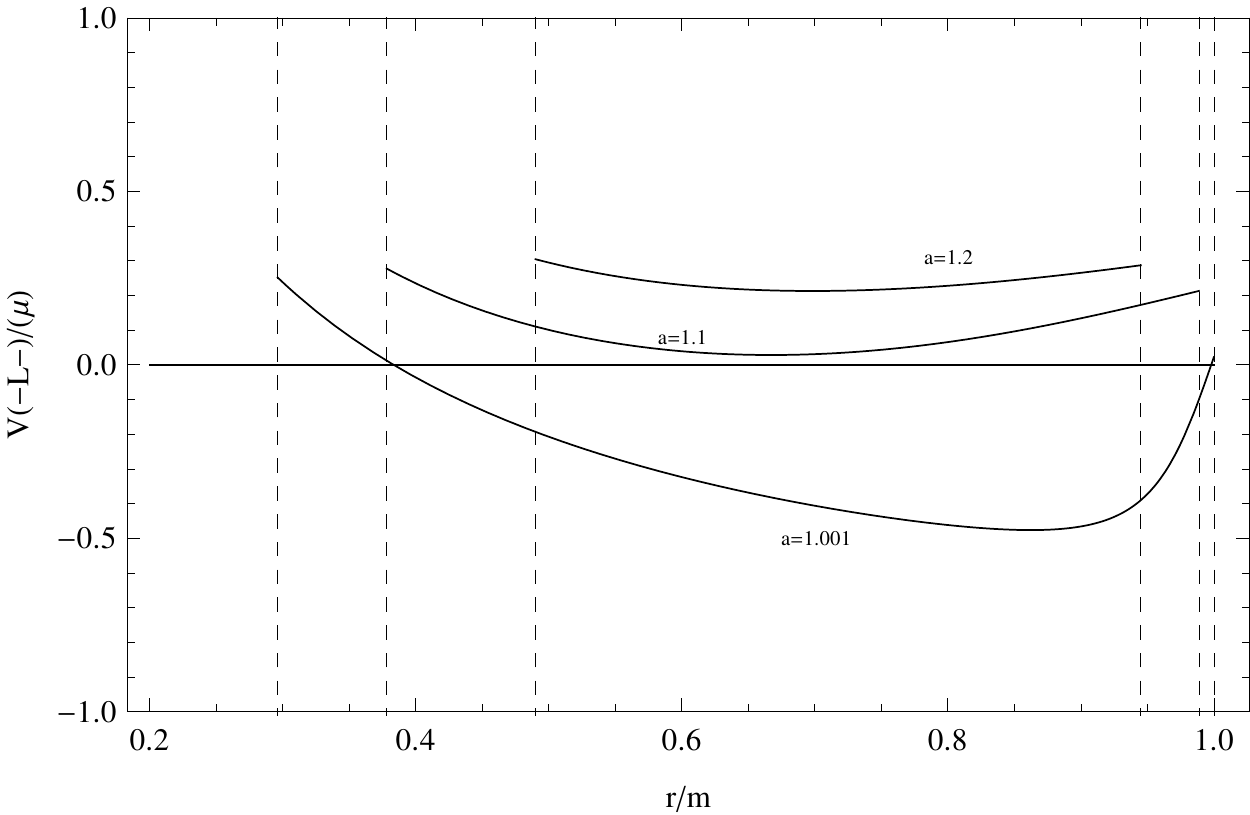}
\end{tabular}
\caption[font={footnotesize,it}]{\footnotesize{
The angular momentum $\bar{a}=-(r-2M) \sqrt{r/M}$ is
plotted as a function of $r$. The energy vanishes, $E(-L_-)=0$, for $a=\bar{a}$,
and is negative, $E(-L_-)<0$, for $1<a<\sqrt{32/27}M$ in the interval $\bar{r}_{1}<r<\bar{r}_{2}$.
For $a>\sqrt{32/27}M$ the energy  $E(-L_-)$ is always strictly positive.
For a naked singularity with momentum $a=1.02M$ the energy
$E(-L_-)=0$ at $r=0.41M$ and $r=0.96M$, and $E(-L_-)<0$ in
$0.41M<r<0.96M$.
For $a=\sqrt{32/27}M$ the energy $E(-L_-)=0$ at $r=2/3M$, whereas $E(-L_-)>0$ for $a=1.1M$.
In the upper bottom plot, the energy $E(-L_-)$ is plotted for selected values of $a/M$
in the interval $\hat{r}_-<r<\hat{r}_+$.
}}\label{Pmp}
\end{figure}

We can see that $E(-L_-)<0$ for $M<a<\sqrt{32/27}M$ in  the interval $\bar{r}_{1}<r<\bar{r}_{2}$.
Otherwise,    for $a>\sqrt{32/27}M$, the energy  $E(-L_-)$ is always strictly positive.
This behavior is illustrated in Fig.\il\ref{Pmp}.

\begin{figure}
\centering
\begin{tabular}{c}
\includegraphics[scale=1]{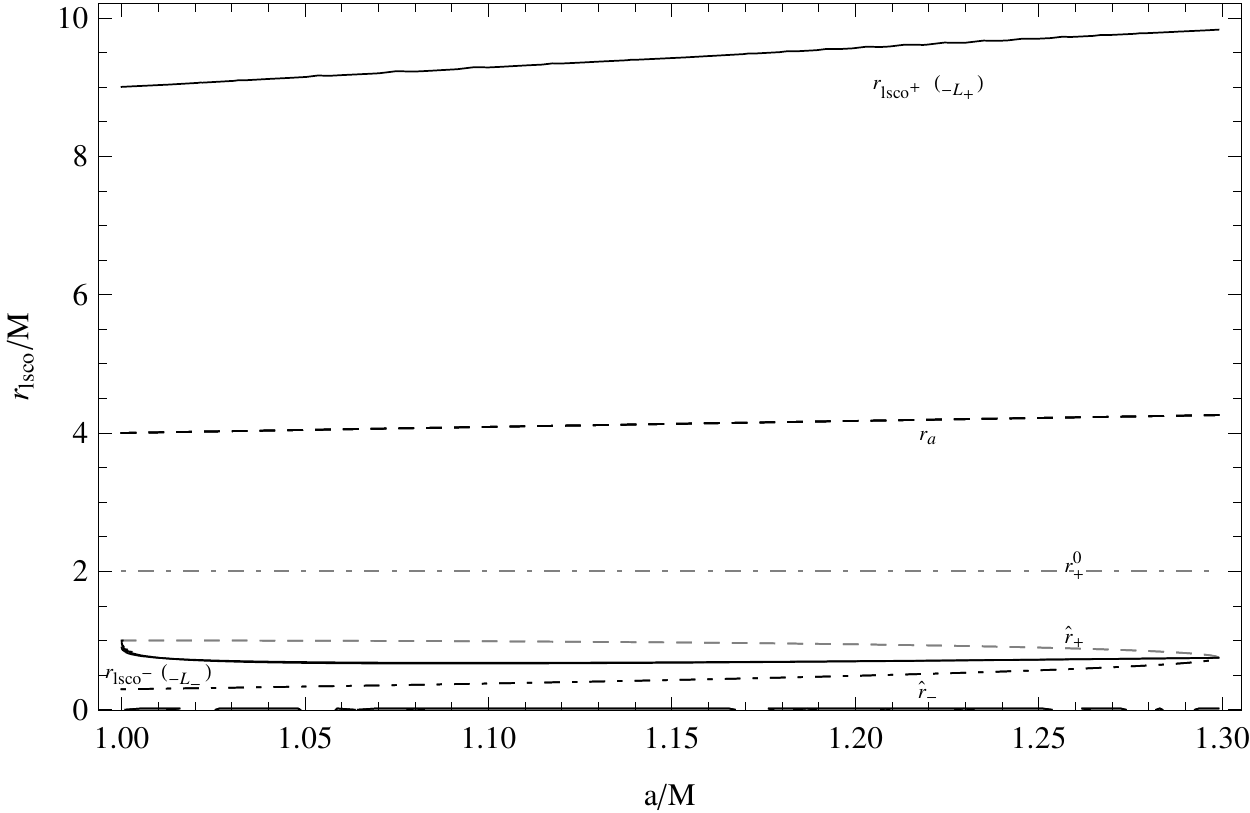}
\end{tabular}
\caption[font={footnotesize,it}]{\footnotesize{The radii  $r_{lsco}^\pm$
of the last stable circular orbits are plotted as  functions of the intrinsic
angular momentum $a$ in the interval $M<a<\frac{3 \sqrt{3}}{4}M$. The radii $r_{a}$ and $\hat r _{\pm}$ are
also plotted. The particle angular momentum $L_{\pm}$ is also denoted for
some particular radii.
 }}
\label{Stabiliint}
\end{figure}

The stability of circular orbits is determined by the turning points of the effective potential.
 For this case
we find numerically two turning points $r_{lsco}^+$ and $\tilde{r}_{lsco}^-$ with $\hat r _-<\tilde{r}_{lsco}^-   <\hat r _+$ and
$r_{lsco}^+   >r_a$ (see  Fig.\il\ref{Stabiliint}),
where
\be\label{lanza}
\tilde{r}_{lsco}^-\equiv3-Z_2-\sqrt{(3-Z_1)(3+Z_1-2Z_2)}\ .
\ee

The radii $r_{lsco}^+$ and $\tilde{r}_{lsco}^-$ correspond to the last stable circular orbits with angular $L=-L_+$ and $L=-L_-$ respectively. Then,
the distribution of circular orbits in the different regions is as follows:

\begin{itemize}
\item
In the region $0<r<\hat{r}_{-}$, the  orbits with $L=L_{-}$ are unstable.
\item
In the region $\hat{r}_-<r<\tilde{r}_{lsco}^-   $, the orbits with $L=-L_{-}$ are unstable.
\item
In the region $\tilde{r}_{lsco}^-   <r<\hat{r}_{+}$, the orbits  with  $L=-L_{-}$ are stable.
\item
In the region $\hat{r}_{+}<r<r_{a}$, the orbits with  $L=L_{-}$ are stable.
\item
In the region $r_{a}<r<r_{lsco}^+   $, the orbits with $L=-L_{+}$ are unstable and those with  $L=L_{-}$ are stable.
\item
In the region  $r>r_{lsco}^+   $, the orbits with $L=-L_+$ and $L=L_-$ are  stable.
\end{itemize}

\begin{figure}
\centering
\begin{tabular}{c}
\includegraphics[scale=1]{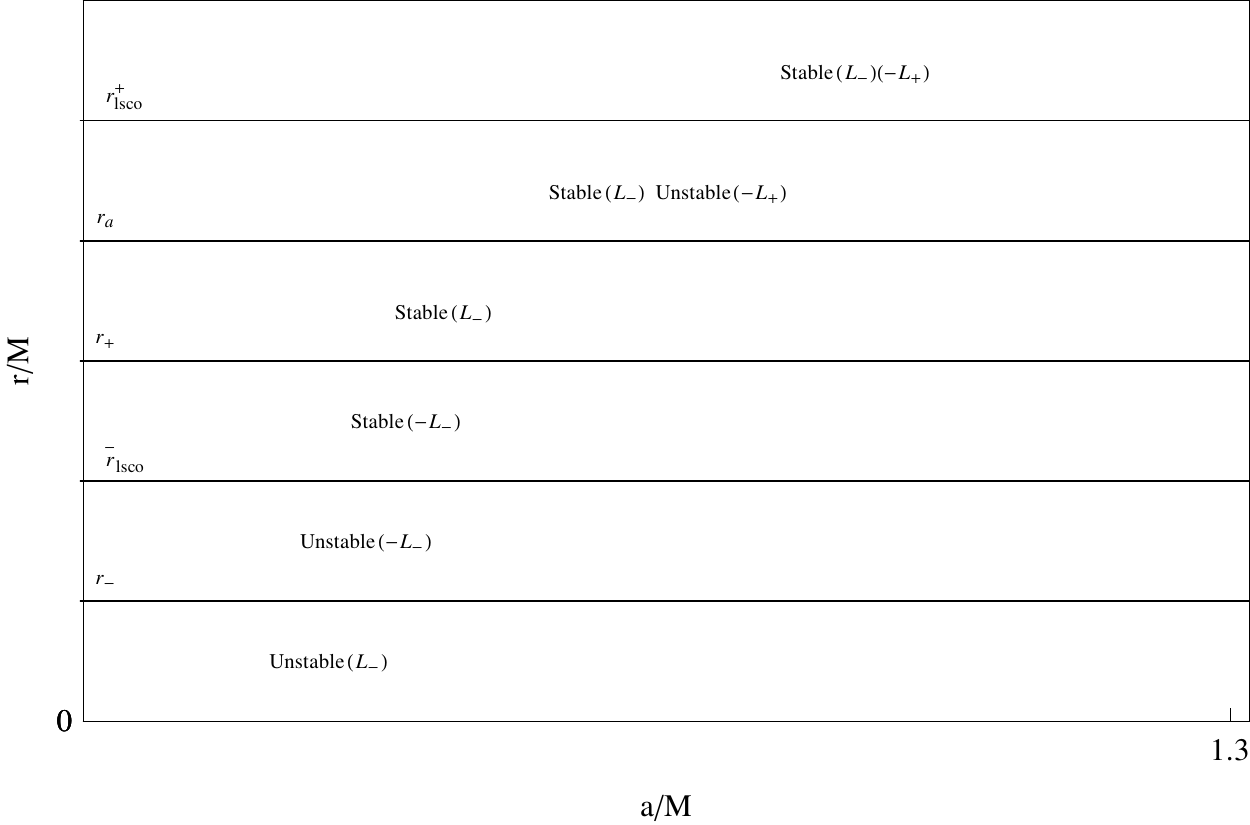}
\end{tabular}
\caption[font={footnotesize,it}]{\footnotesize{Stability of circular orbits
in a Kerr naked singularity with $M<a<\frac{3\sqrt{3}}{4}M$. The
radii $r_{lsco}^+ $  and $\tilde{r}_{lsco}^-$ of the last stable circular orbits as functions
of the ratio $a/M$.
The special radii $r_a$ and $\hat r _\pm$ are also plotted. }}
\label{Sta00}
\end{figure}

\begin{figure}
\centering
\begin{tabular}{cc}
\includegraphics[scale=.7]{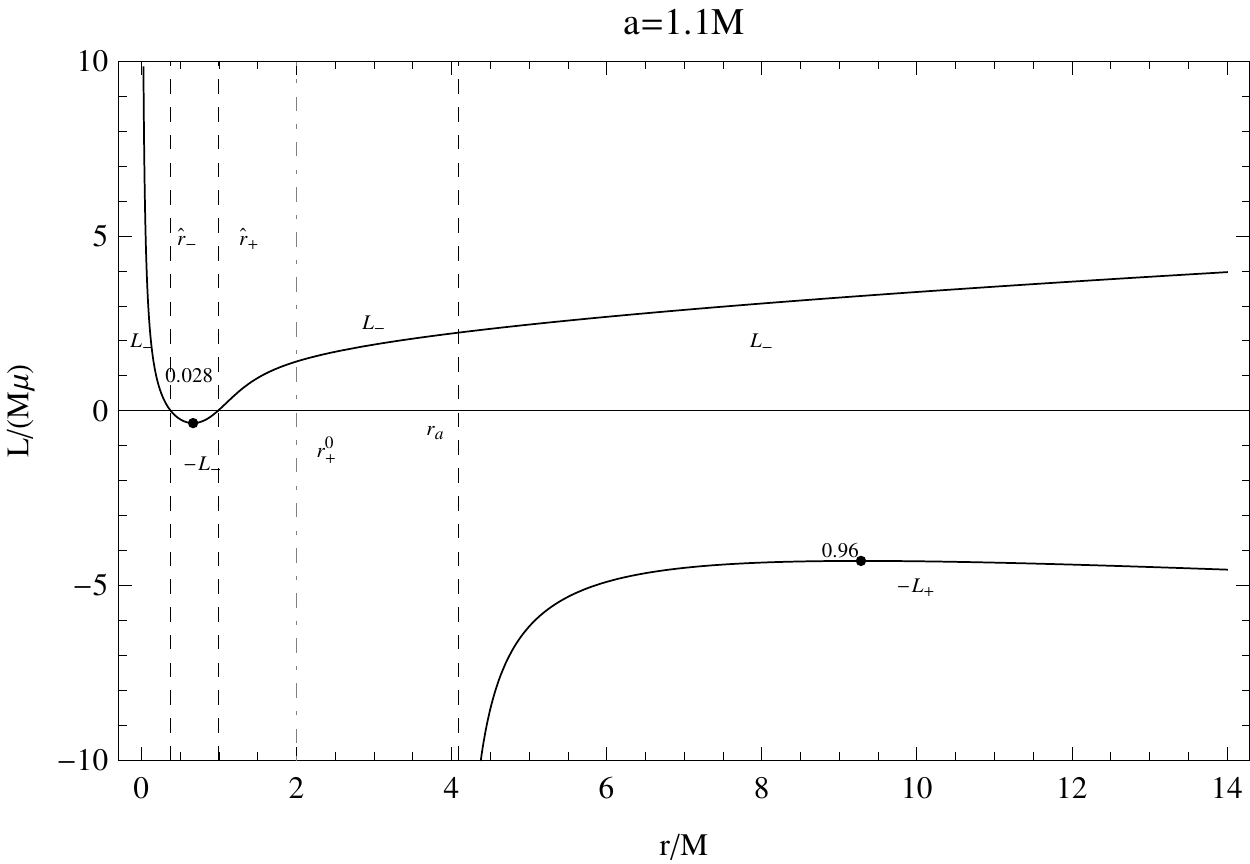}&
\includegraphics[scale=.7]{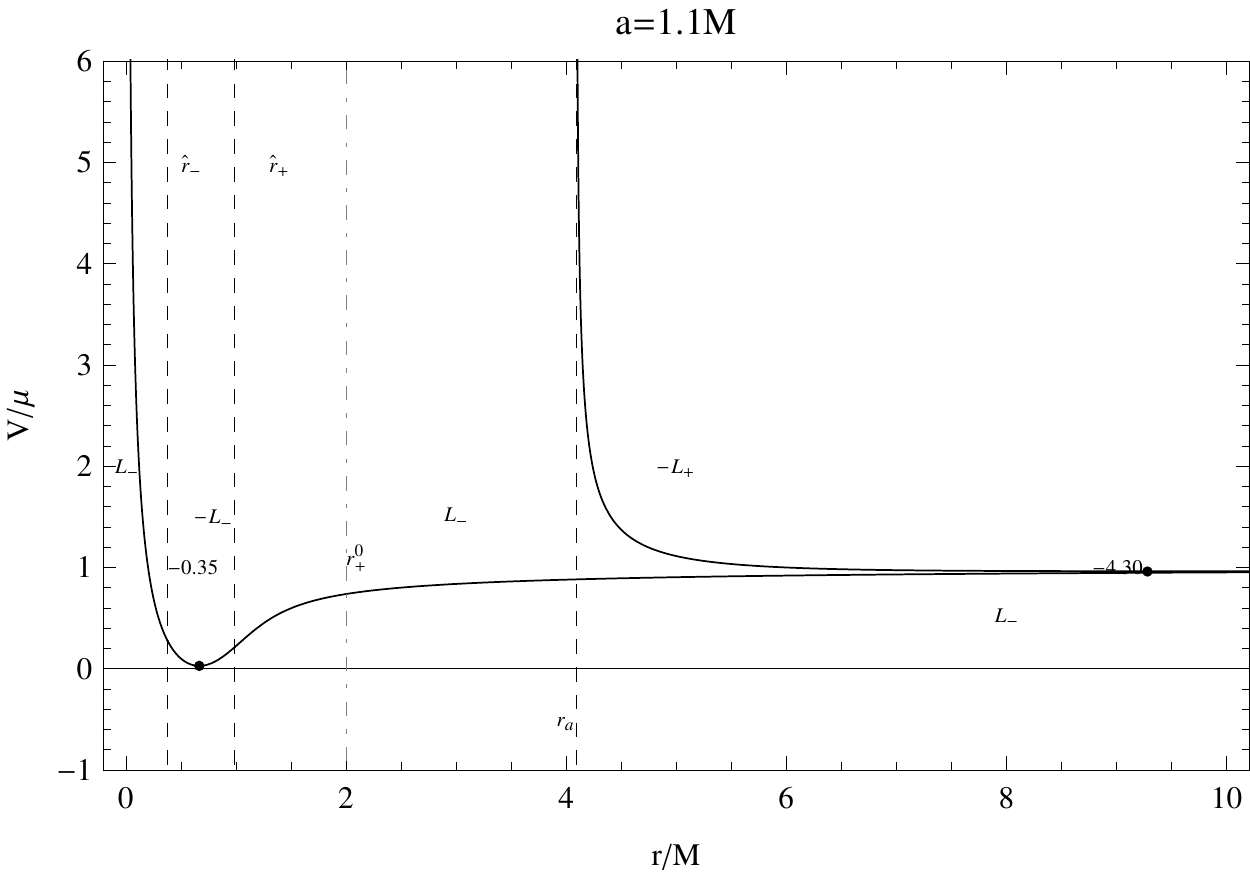}
\end{tabular}
\caption[font={footnotesize,it}]{\footnotesize{Angular momentum and energy
of circular orbits in a Kerr naked singularity with $a=1.1M$.
The dots denote the position of the last stable circular orbits, and
the numbers close to the dots indicate the value of the energy $V/\mu$ or the
angular momentum of the last stable circular orbits. In $r> r_{a}\approx4.088M$,
the particles have angular momentum $L=-L_+$;
in $0<r<\hat{r}_{-}\approx0.378M$ and $r\geq \hat{r}_+\approx0.989M$, there exist
particles with $L=L_-$;
in $\hat{r}_-<r<\hat{r}_+$, there exist particles with   $L=-L_-$.
}}
\label{Plotcira11}
\end{figure}

The summary of this case is sketched in Fig.\il\ref{Sta00}.
As a concrete example,  we investigate in detail circular motion around a naked singularity
with $a=1.1M$. The radii that determine the distribution of  test particles in this gravitational field are:
$\hat r _- \approx0.378M$,
$\tilde{r}_{lsco}^-  \approx0.989M$,
$\hat r _+ = \approx0.989M$,
$r_a \approx4.088M $, and
$r_{lsco}^+ = \approx9.280M$.
In Fig.\il\ref{Plotcira11}, we illustrate the behavior of the angular momentum and the energy
of circular orbits for this special case.

\begin{figure}
\centering
\begin{tabular}{cc}
\includegraphics[scale=.7]{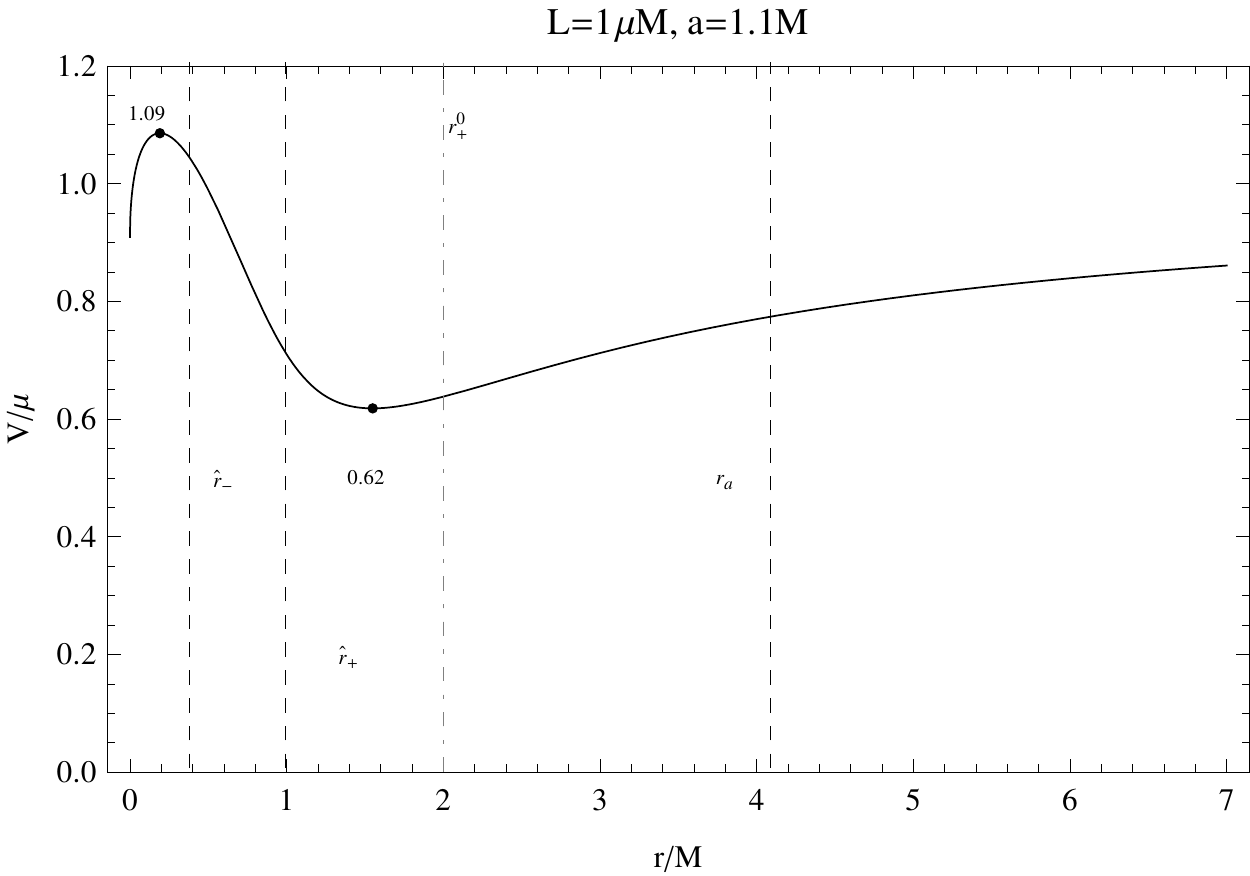}
\includegraphics[scale=.7]{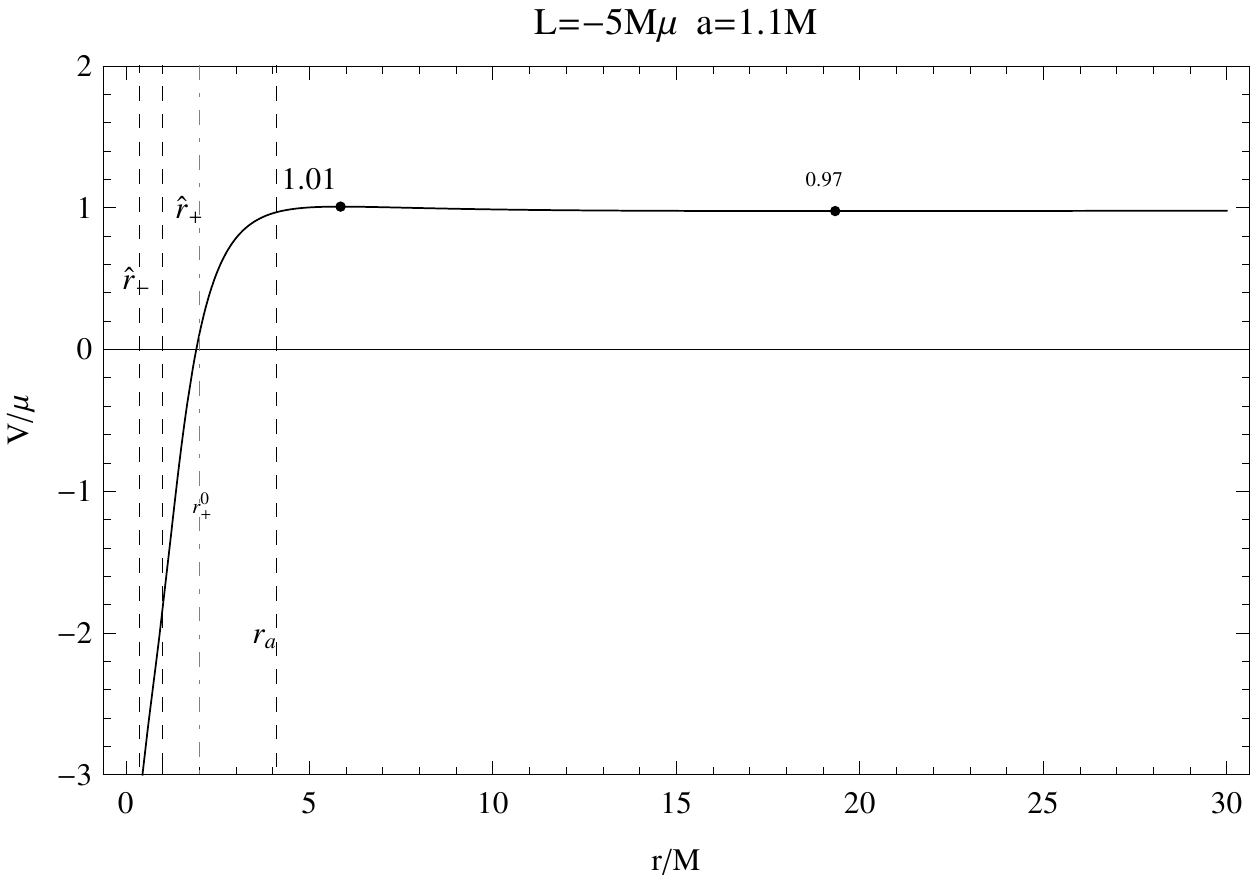}\\
\includegraphics[scale=1]{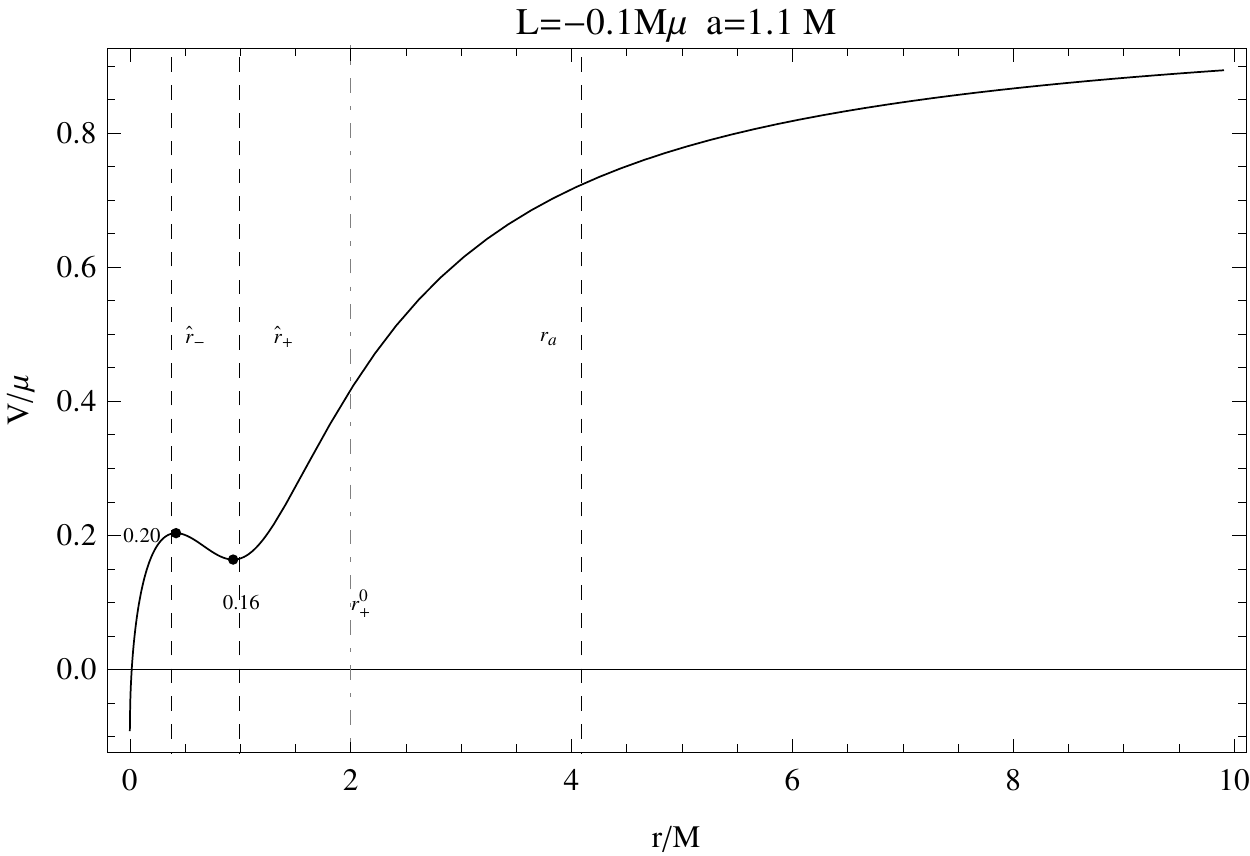}
\end{tabular}
\caption[font={footnotesize,it}]{\footnotesize{The effective
potential of a naked singularity with $a=1.1 M$ for fixed values of the particle angular
momentum $L/(M\mu)$. The radii $r_a$ and $\hat{r}_{\pm}$ are also plotted.
The dots denote the critical points of the potential. Numbers
close to the dots indicate the energy $V/\mu$ of the maxima and minima
of the effective potential.  The dotted dashed gray line represents  the outer boundary of the ergosphere $r_+^0=2M$.} }
\label{PlotVa11}
\end{figure}

In
Fig.\il\ref{PlotVa11},  we show the behavior of the effective potential for some selected  values of the orbital angular momentum.
The turning points of the effective potential are located at
$r_{lsco}^+   \approx9.280M$, where $L_{lsco}^+   \approx-4.298\mu M$ and
$V_{lsco}^+   \approx0.963\mu$, and at
$\tilde{r}_{lsco}^-  \approx 0.667M$, where  $L_{lsco}^-   \approx-0.354\mu M$ and $V_{lsco}^-   \approx0.028\mu$.

The essential results of our analysis can be described by using the
 model of an accretion disk around the central naked singularity.
Considering the properties and positions of the different radii and the positions
of the last stable circular orbits, we conclude that the stable accretion disk is composed
of three different disks. The internal disk is situated between the radii $\tilde{r}_{lsco}^-$ and $\hat r _ +$ and is
made of stable particles of counter-rotating particles with angular momentum $L=-L_-$. Particles situated
on the boundary radius $\hat r _+$ turn out to be characterized by a zero value of the angular momentum (cf. Sec. \ref{sec:L0}).
A second  disk made of stable corotating particles
with angular momentum $L=L_-$ is situated in the
region $\hat r _ + < r < r_{lsco} ^+$. Finally, the exterior stable disk is situated in the region $r>r_{lsco}^+$ and contains
corotating particles with $L=L_-$ and counter-rotating particles with $L=-L_+$.

\clearpage

\subsection{Orbits with zero angular momentum}
\label{sec:L0}
\begin{figure}
\centering
\begin{tabular}{cc}
\includegraphics[scale=.27]{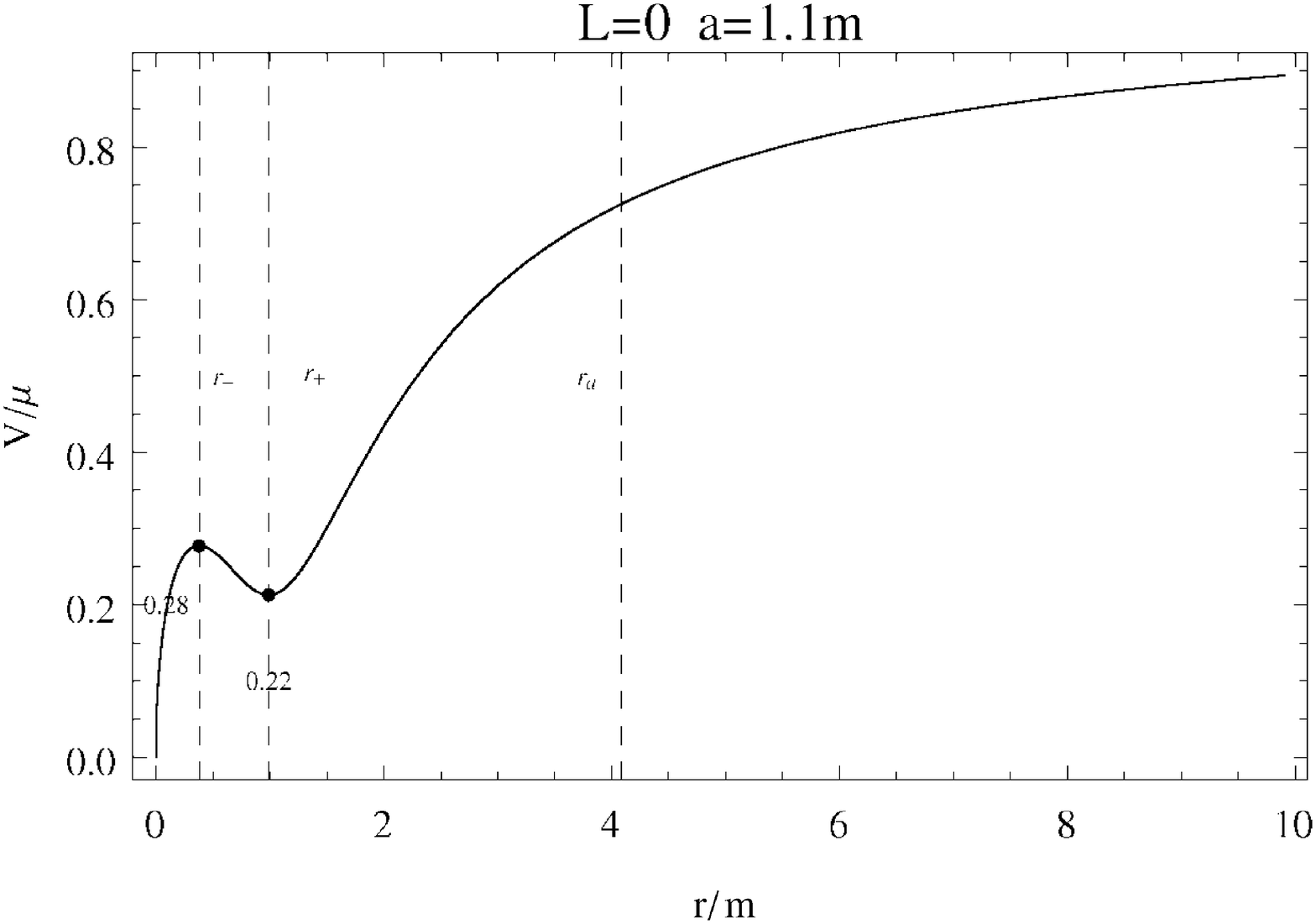}
\includegraphics[scale=.27]{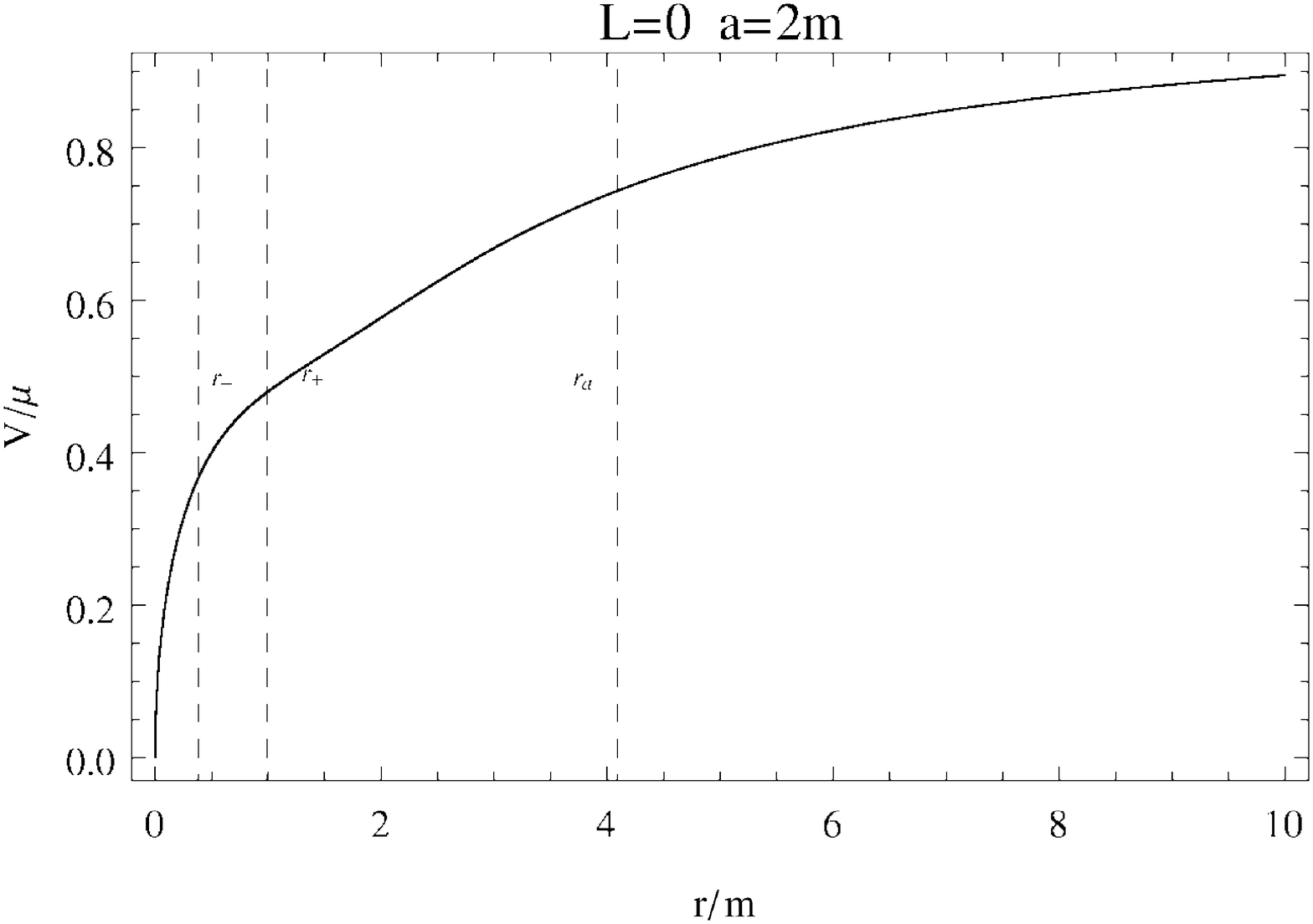}
\end{tabular}
\caption[font={footnotesize,it}]{\footnotesize{The effective
potential of a Kerr naked singularity with angular momentum
parameter $a=1.1M$ and and $a=2M$ is plotted for
the particle orbital angular momentum $L/(M\mu)=0$ as a function of
the radius $r/M$. The radii $r_a$ and $\hat{r}_{\pm}$ are also plotted for both
cases (see text). The dots represent the critical points of the potential, and
the numbers close to the dots indicate the energy $V/\mu$ of the maxima
and minima of the effective potential. In the case $a=2M$ no extreme points are observed in the
potential.
} }
\label{BL0A11}
\end{figure}

An interesting phenomenon  that occurs only in the gravitational field of naked singularity is the existence
of ``circular orbits" with zero angular momentum, as defined by the conditions
\be\label{dacons}
V=E/\mu,\quad V'(r)=0,\quad L=0\ .
\ee
This fact can be interpreted as a consequence of the repulsive gravity effects that characterize the dynamics in the field of
the naked singularity. For the repulsive gravity effects in the Kerr spacetime see, for example, \cite{F2,Preti:2008zz}.
From the expression for the angular momentum
derived in Sec. \ref{sec:orbkerr} one can show that the solution (\ref{dacons}) is allowed only for
naked singularities with intrinsic angular momentum within the interval $1<a/M\leq 3\sqrt{3}/4$. Outside this interval, i.e. for $a/M>3\sqrt{3}/4$,
no orbits exist with zero angular momentum. The behavior of the corresponding effective potential is illustrated in
Fig. \ref{BL0A11}.

\begin{figure}
\centering
\begin{tabular}{c}
\includegraphics[scale=.51]{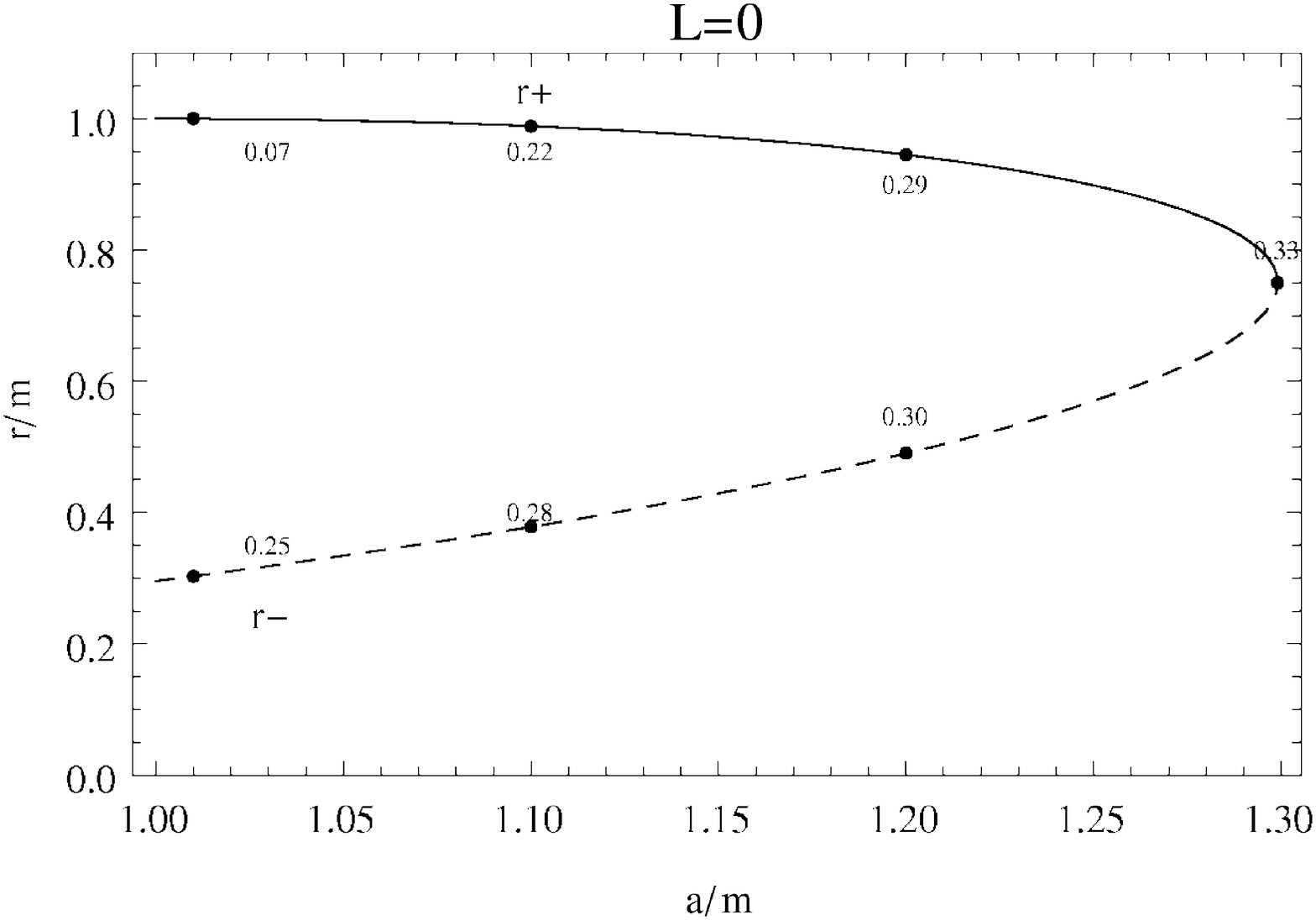}
\end{tabular}
\caption[font={footnotesize,it}]{\footnotesize{Location of particles with $L=0$ in a Kerr
naked singularity with $1<a/M\leq 3\sqrt{3}/4$. The picture plots the locus of the critical
points of the effective potential $V/\mu$ with (particle) angular
momentum $L/(M\mu)=0$. The radius of these ``circular'' orbits is plotted
as a function of the source angular momentum $a/M$. Numbers close to
the dots indicate the value of the energy $V/\mu$.
} }
\label{PlotL0AMR}
\end{figure}

A further analysis shows that the particles with $L=0$ are situated on the radii $\hat r _\pm$, and that
the radius $\hat r _-$ corresponds to unstable particles while the radius $\hat r _ +$ is withing the region
of stability. This situation is illustrated in Fig.\il\ref{PlotL0AMR}.

\begin{figure}
\centering
\begin{tabular}{cc}
\includegraphics[scale=1]{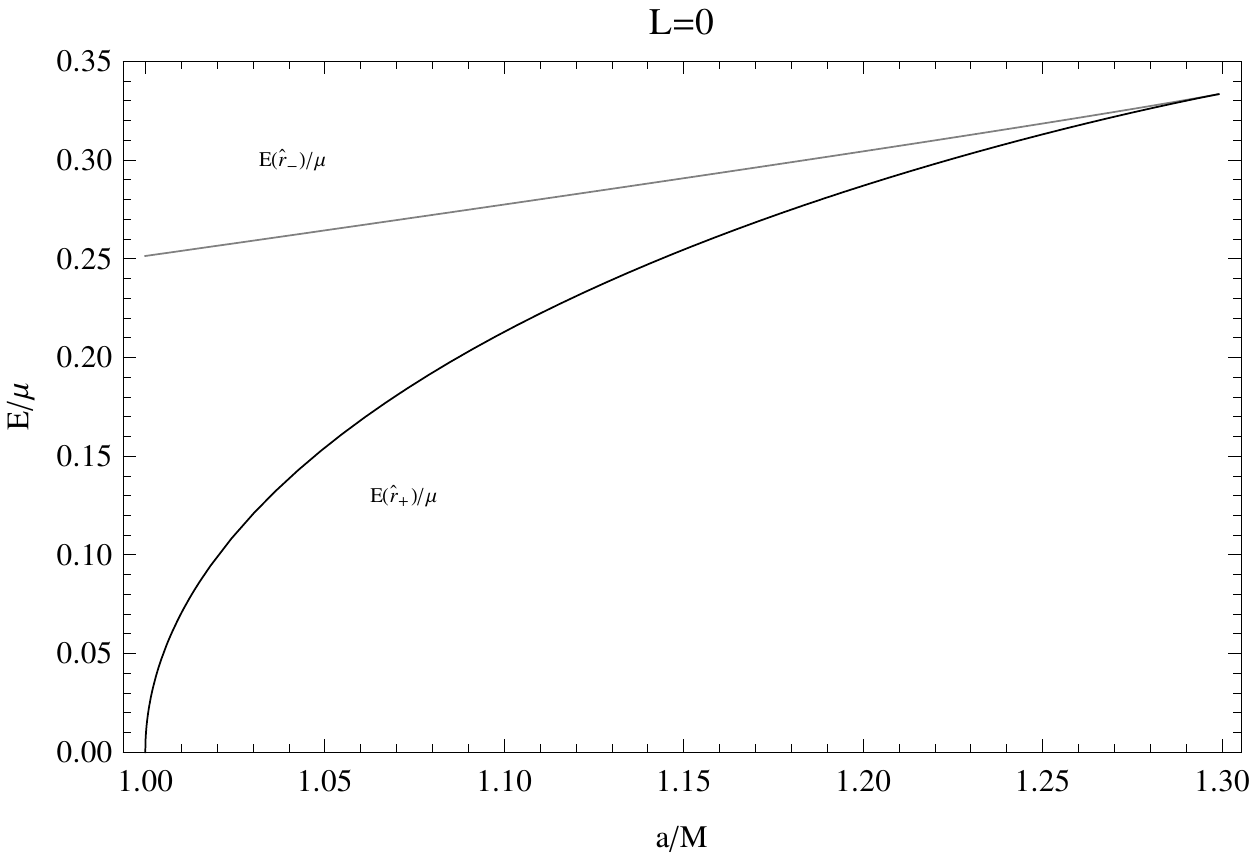}
\end{tabular}
\caption[font={footnotesize,it}]{\footnotesize{The energy particles with $L=0$ in a naked singularity with $1<a/M\leq3\sqrt{3}/4$.
The orbits are located on the radii
$r=\hat{r}_{+}$ (stable) and  $r=\hat{r}_{-}$ (unstable).
The energies $E(\hat{r}_{+})$ (black curve) and $E(\hat{r}_{-})$ (gray curve) are plotted as functions of the intrinsic angular momentum $a/M$.
It is possible to see that $E(\hat{r}_{+})<E(\hat{r}_{-})$ for $1<a/M<3\sqrt{3}/4$, and $E(\hat{r}_{+})=E(\hat{r}_{-})$ for $a/M=3\sqrt{3}/4$.
} }
\label{E4E3}
\end{figure}

The analysis of the energy of test particles with $L=0$ is presented in Fig. \ref{E4E3}. For the stable particles that are situated on the radius
$\hat r _ +$ we can note that the energy is always positive and finite. The maximum value of the energy is reached at the ratio $a/M=3\sqrt{3}/4$
and the minimum value with $E(\hat r _+)\rightarrow 0$ corresponds to the limit of the extreme black hole $a/M\rightarrow 1$.

\clearpage
\subsection{Summary of the naked singularity and black hole cases}
\label{sec:sum}

\begin{figure}
\centering
\begin{tabular}{c}
\includegraphics[scale=1]{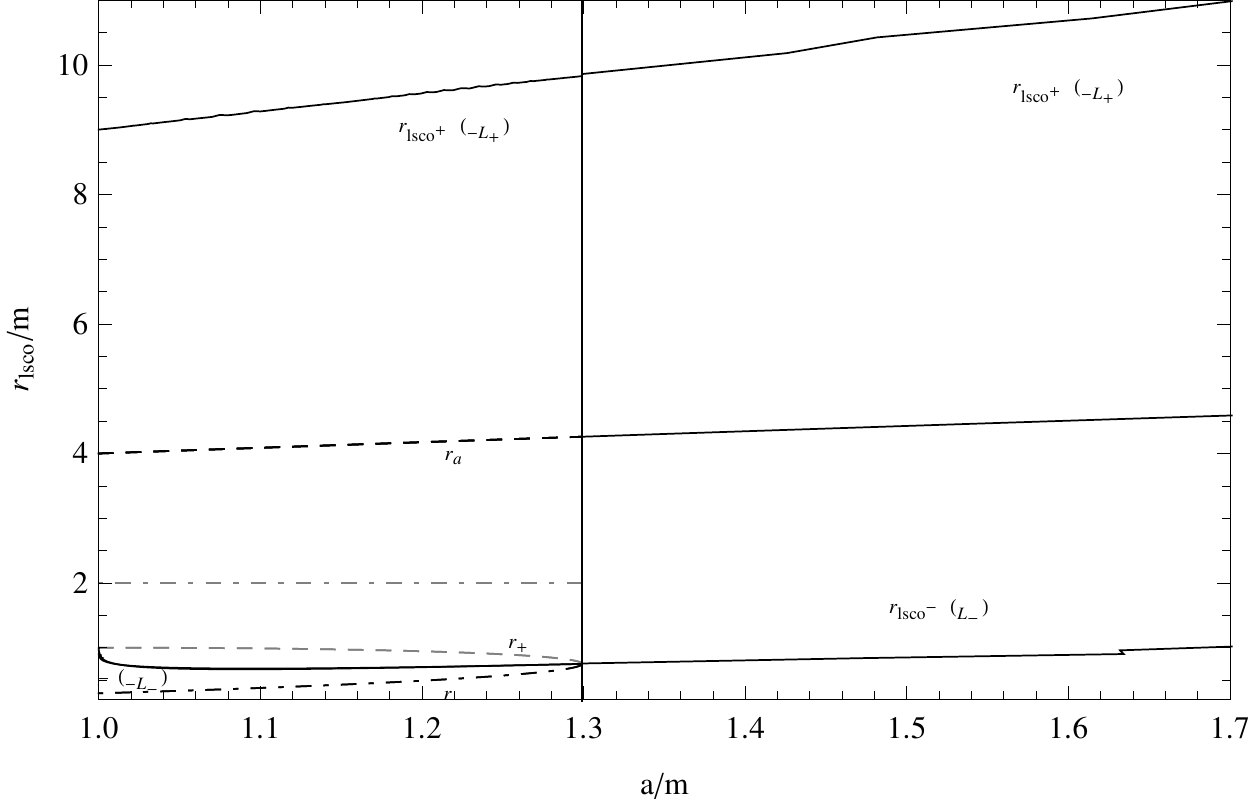}
\end{tabular}
\caption[font={footnotesize,it}]{\footnotesize{The radii $r_{lsco}$
of the turning points of the effective potential and the special radii
$r_{a}$ and $\hat{r}_{\pm}$ are plotted as functions of the ratio $a/M$ within
the interval $[1,1.7]$.
The dotted dashed gray line represents  the outer boundary of the ergosphere $r_+^0=2M$.}}
\label{Stabiliintutte}
\end{figure}

In the investigation of the circular motion of test particles around a Kerr naked singularity we found that it is necessary
to analyze separately the two regions $a\geq\frac{3 \sqrt{3}}{4}M$ and $M<a<\frac{3 \sqrt{3}}{4}M$.
The distribution of orbits depends on the position of the special radii
$\hat{r}_{\pm}$, given by Eq.\il(\ref{rpm}),   $r_a$, given by Eq.\il(\ref{rag}), and the position
of the last stable circular orbits $r_{lsco}^+$, as given in Eq.(\ref{dicadica}), $\tilde{r}_{lsco}^-$ in Eq.\il(\ref{lanza}) and $\bar{r} _{lsco}$, as given in Eq.(\ref{rbar}).
Notice that although the radius $\bar{r} _{lsco}$ is the geometric continuation of the radius $\tilde{r}_{lsco}^-$ for  the interval $a/M>3\sqrt{3}/4$,
their values are determined by different analytical expressions as follows from Eqs.(\ref{dicadica}) and (\ref{rbar}).
The arrangement of these radii in the interval $1<a/M<1.7$ is depicted in Fig.\il\ref{Stabiliintutte}.

The tables\il\ref{tab:Nemoonir}  and  \ref{tab:alpardime} summarize the distribution and stability properties of
test particles in circular motion in the field of a rotating naked singularity for the two different regions
of values of the intrinsic angular momentum.

\begin{table}
\begin{center}
\resizebox{.71\textwidth}{!}{%
\begin{tabular}{llll}
&Case: $M<a<(3\sqrt{3}/4)M$& &\\
\hline\hline
&Region&Angular momentum& Stability
\\
\hline
& $]0,\hat{r}_-[ $ &$L_-$ &$\tilde{r}_{lsco} $\\
& $]\hat{r}_-,\hat{r}_+[$ &$-L_-$ &$\tilde{r}_{lsco} $\\
& $]\hat{r}_+,\infty[$ &$L_-$ &$\tilde{r}_{lsco} $\\
& $]r_a,\infty[$ &$-L_+$ &$r_{lsco}^+ $\\
\cline{1-4}
\cline{1-4}
& $]0,\hat{r}_-[$  &$L_-$&Unstable  \\
& $]\hat{r}_-,\tilde{r}_{lsco} [$ &$-L_-$ &Unstable  \\
& $]\tilde{r}_{lsco} ,\hat{r}_+[$ &$-L_-$   &Stable\\
& $]\hat{r}_{+},r_a[$  &$L_-$ &Stable \\
& $]r_{a},r_{lsco}^+ [$ &$L_-$ $(-L_+)$  &Stable (Unstable)\\
& $]r_{lsco}^+ ,\infty[$  &($L_-$, $-L_+$) &Stable \\
\hline\hline
\end{tabular}
}
\caption[font={footnotesize,it}]{\footnotesize{Distribution and stability properties of circular orbits for a test particle in a
Kerr naked singularity with $M<a<(3\sqrt{3}/4)M$. For  each region we present the value of the orbital angular momentum of the particle as
determined by Eq.\il(\ref{LPM}). }}
\label{tab:Nemoonir}
\end{center}
\end{table}
\begin{table}
\begin{center}
\resizebox{.81\textwidth}{!}{%
\begin{tabular}{llll}
&Case: $a\geq(3\sqrt{3}/4)M$& &\\
\hline\hline
&Region&Angular momentum& Stability
\\
\hline
& $]0,\infty[ $ &$L_-$ &$\bar{r}_{lsco} $\\
& $]r_a,\infty[$ &$-L_+$ &$r_{lsco}^+ $\\
\hline
&$(3\sqrt{3}/4)M<a<9M$ ($\bar{r}_{lsco} <r_a<r_{lsco}^+ $) & &
\\ \hline\hline
& $]0,\bar{r}_{lsco} [$  &$L_-$&Unstable  \\
& $]\bar{r}_{lsco},r_a[$ &$L_-$ &Stable  \\
& $]r_a,r_{lsco}^+ [$ &$L_-$ ($-L_+$)  &Stable (Unstable) \\
& $]r_{lsco}^+ ,\infty[$  &($L_-$, $-L_+$) &Stable \\
\hline
&$a\geq9M$ ($r_a<\bar{r}_{lsco} <r_{lsco}^+ $) & &
\\ \hline\hline
& $]0,r_a[$  &$L_-$&Unstable  \\
& $]r_a,\bar{r}_{lsco} [$ &$(L_-,-L_+)$ &Unstable  \\
& $]\bar{r}_{lsco} ,r_{lsco}^+ [$ &$L_-$ ($-L_+$)  &Stable (Unstable) \\
& $]r_{lsco}^+ ,\infty[$  &($L_-$, $-L_+$) &Stable \\
\hline\hline
\end{tabular}
}
\caption[font={footnotesize,it}]{\footnotesize{
Distribution and stability properties of circular orbits for a test particle in a
Kerr naked singularity with $a\geq(3\sqrt{3}/4)M$. For  each region we present the value of the orbital angular momentum of the particle as
determined by Eq.\il(\ref{LPM}).}}
\label{tab:alpardime}
\end{center}
\end{table}

\begin{figure}
\centering
\begin{tabular}{cc}
\includegraphics[scale=.7]{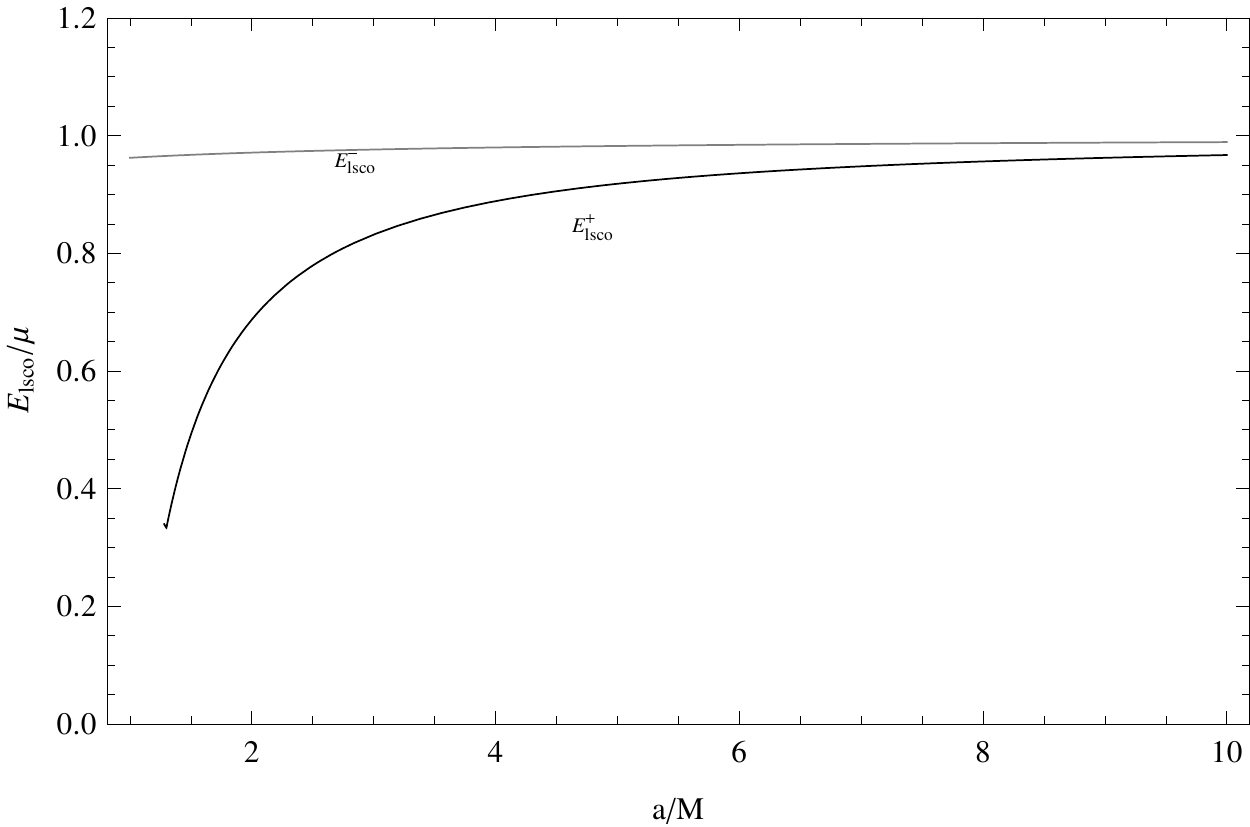}
\includegraphics[scale=.7]{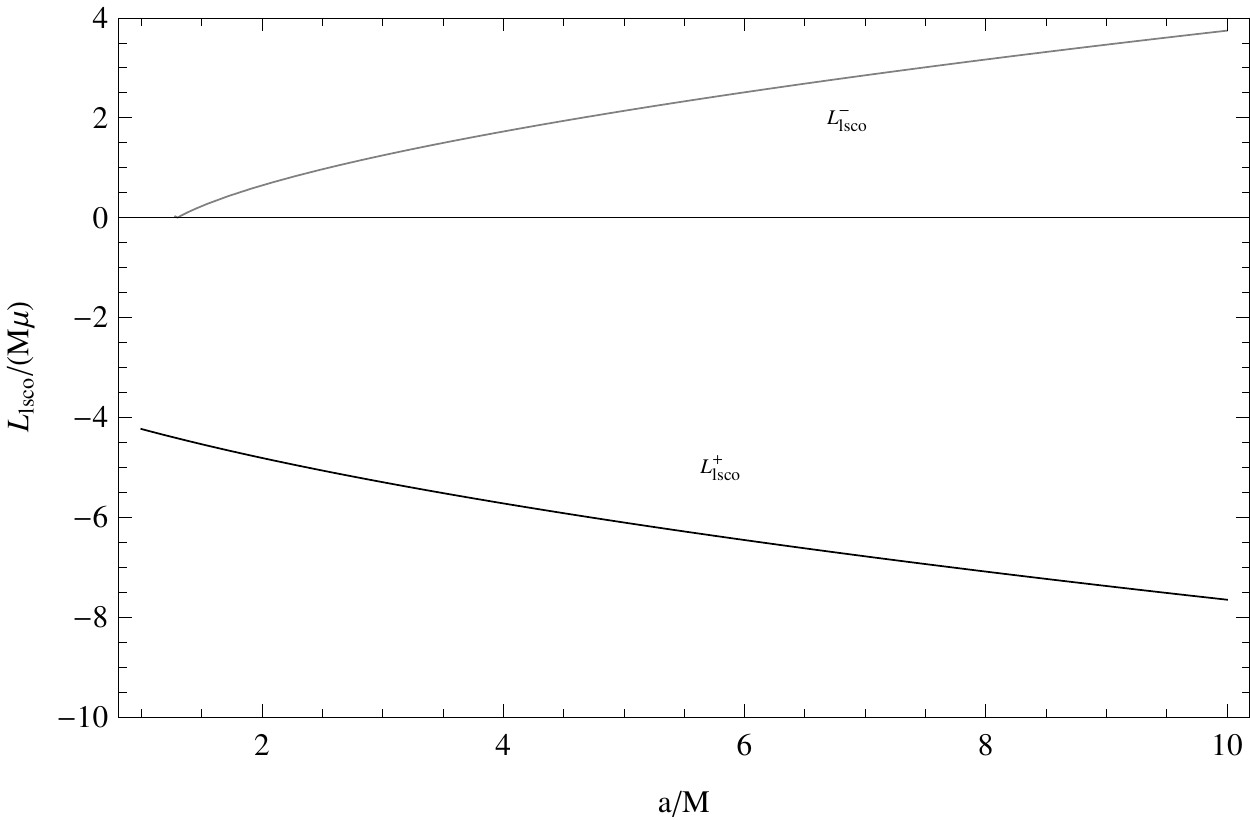}
\end{tabular}
\caption{\footnotesize{Behavior of the $E_{lsco}/\mu$ and the angular momentum for the last stable circular orbits as
functions of the intrinsic angular momentum of the naked singularity.
}}
\label{SESMPMEn}
\end{figure}

For the sake of completeness, we show in Fig.\il\ref{SESMPMEn} the behavior of  the energies $E_{lsco}^+  =E(r_{lsco}^+)$ and
$E_{lsco}^- =E({r}_{lsco}^-)$
and angular momenta
$L_{lsco}^+  =L(r_{lsco}^+)$ and $L_{lsco}^- =L({r}_{lsco}^-)$,
for the last stable circular orbits in terms of the ratio $a/M$ of the naked singularity. Notice that, as expected from a physical
viewpoint, for a fixed value of the
ratio $a/M$ the energy of the exterior last stable circular orbit $E(r_{lsco}^+)$ is always smaller than the corresponding energy
of the interior particle $E(r_{lsco}^-)$.

\begin{figure}
\centering
\begin{tabular}{c}
\includegraphics[scale=1]{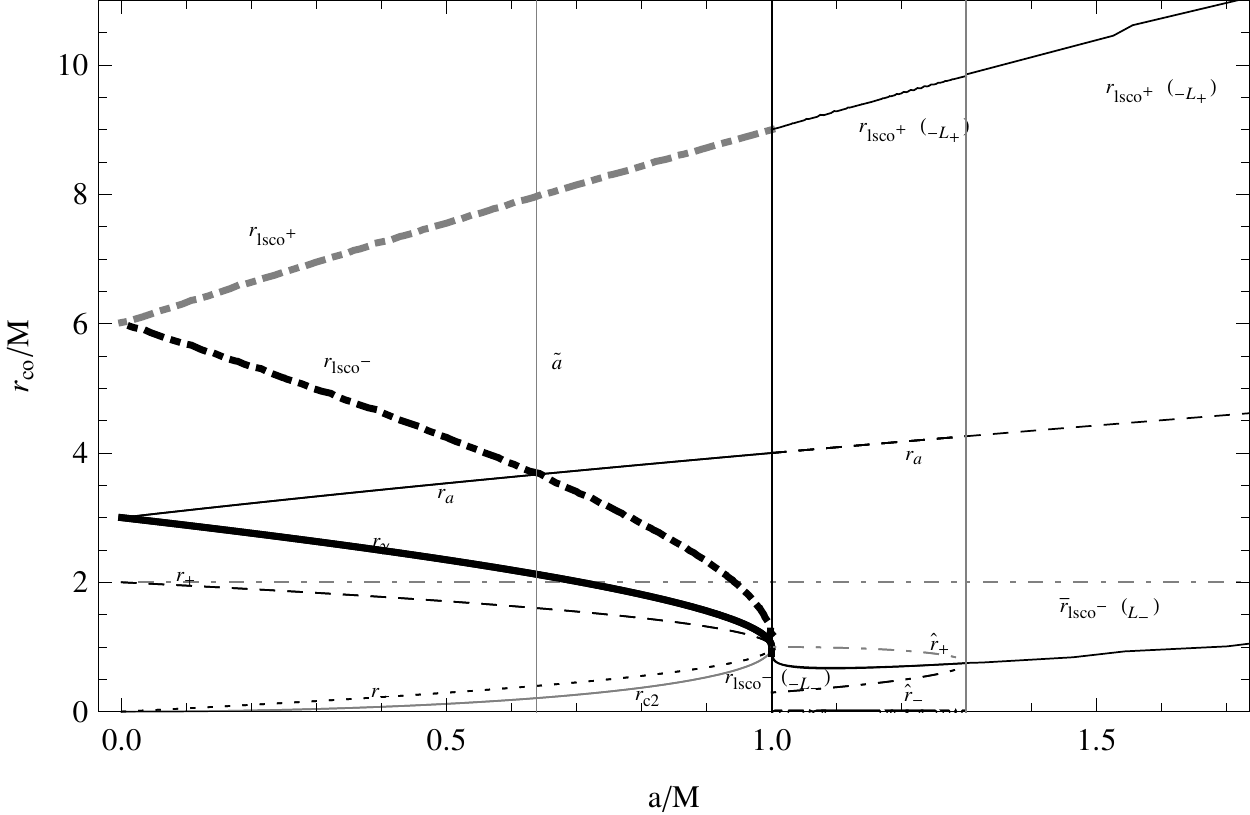}\\
\includegraphics[scale=1]{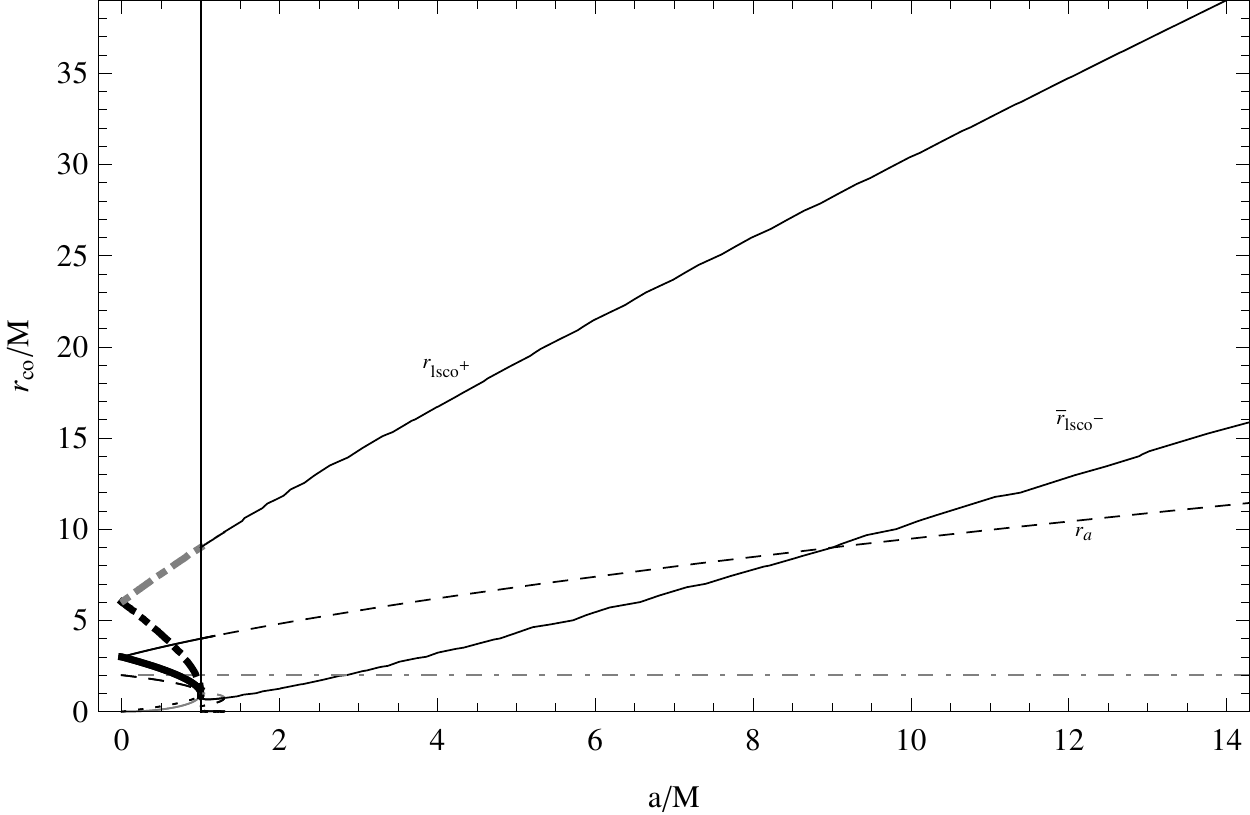}
\end{tabular}
\caption[font={footnotesize,it}]{\footnotesize{Arrangement of the radii determining the properties of circular
orbits around a rotating central mass. The upper plot is for black holes and naked singularities with
intrinsic angular momentum $a/M\in[0,1.7]$. The bottom plot is for rotating naked singularities with $a/M\in[1.7,14]$.
 The dotted dashed gray line represents  the outer boundary of the ergosphere $r_+^0=2M$.}}
\label{aggio2}
\end{figure}

Our analysis of Kerr black holes and naked singularities shows that the properties of
circular orbits depend strongly on their radial distance with respect to the central source.
The critical radii that are found in the analysis of the conditions for circular motion
determine the angular momentum and the energy of the test particles. The arrangement
of those special radii and the positions of the last stable circular orbits is depicted
in Fig. \ref{aggio2} for the relevant ranges of the ratio $a/M$.

The radii $r_a$, $r_\gamma$, $\hat r _+$, and $\hat r _-$ determine the angular momentum and
direction of rotation of test particles at a given distance from the central source.
In addition, the radii $r_{lsco}^\pm$ determine the position of the last stable circular orbits
with a given angular momentum of the test particle.

If we imagine an infinitesimal thin disk made of test particles in circular orbits around the central compact object,
the above results show that the geometric structure of the disk is sufficient to distinguish between black holes
and naked singularities. For such hypothetical disk to be a meaningful approximation of a physically realizable
disk, it is necessary that the individual particle orbits be stable with respect to infinitesimal perturbations.
In the case of radial perturbations, stability is guaranteed as a consequence of the fact that the disk is made of
stable particles in circular motion, as described above. As for perturbations out of the equatorial plane,
the analysis of stability has been performed by using the geodesic equations \cite{1970}, the phase space method \cite{palit09},
and the Rayleigh criterion \cite{letelier1,letelier2}. Although the last method has been applied only to static
central sources, the generalization to include rotating sources seems to be straightforward. All those methods show
that equatorial circular orbits around a Kerr black hole
are stable under out-of-equatorial-plane perturbations as long as the angular momentum
per unit mass of the test particles increases monotonically as the distance to the axis of symmetry increases.
A complementary analysis must be performed in the case of naked singularities; however, a brief inspection of
the analytical results obtained by using the phase space method seems to indicate that the stability does
not depend drastically on the mass-to-angular-momentum ratio of the central body. In general, one can expect
that the stability with respect to radial and out-of-equatorial-plane perturbations depends on the ratio
of source rotation to particle angular momentum.

\section{Conclusions}
\label{malditesta ... :)}

In this work, we investigated the circular motion of test particles around a rotating central mass whose gravitational
field is described by the Kerr spacetime. We limit ourselves to the study of circular orbits situated on the equatorial
plane $\theta = \pi/2$. First, we derive the conditions for the existence of circular orbits by using the
fact the geodesic motion in this case can be reduced to the motion of test particles in an effective potential. In this
procedure, two constants of motion arise, $E$ and $L$, which are interpreted as the energy and the angular momentum of the test
particles, respectively. We concentrate on the analysis of the conditions for the existence of circular orbits and their consequences
for the values of the energy and angular momentum of the test particles. Our analysis covers completely the range
of values of the intrinsic angular momentum of the central mass, including black holes and naked singularities.
We find all the regions of the equatorial plane where circular motion is allowed and analyze the behavior of the energy and
the angular momentum of the test particles in those regions. Moreover, the stability properties of all the
allowed circular orbits was investigated in detail.

For our analysis we consider separately the case of black holes with ratio $a/M\leq 1$ and naked singularities $a/M> 1$,
where $M$ is the mass and $a$ is the specific angular momentum $J/M$ of the central body. Moreover, in the case of
naked singularities it turns out that the physical properties of the circular motion depend on the value of the ratio $a/M$
so that it is necessary to explore two different ranges: $1<a/M< 3\sqrt{3}/4$ and $a/M> 3\sqrt{3}/4$.
The essential part of our results can be formulated in a plausible manner by using the model of an accretion disk
made of stable test particles which are rotating around the central mass.

In the case of a black hole $(a/M\leq 1)$, we find that the accretion disk is composed of an interior disk situated
within the radii $[r_{lsco}^-,r_{lsco}^+]$ and an exterior disk in the region $r>r_{lsco}^+$, where $r_{lsco}^\pm$
represent the radius of the last stable circular orbit with angular momentum $L=\mp L_\pm$; moreover, the value of
$L_\pm$ depends on the radius $r$ of the circular orbit and on the ratio $a/M$ of the central body [cf. Eq.(\ref{LPM})]. A similar
accretion disk is found around naked singularities with $a/M> 3\sqrt{3}/4$.
The only difference is that in the case of a naked
singularity the interior disk, situated within the radii $[\bar{r}_{lsco}^-,r_{lsco}^+]$, has a minimum size of
$r_{lsco}^+ - \bar{r}_{lsco}^- >8M$, whereas in the case of a black hole the size of the inner disk is always
less than $8M$ and disappears as $a\rightarrow 0$.

For naked singularities in the range $1<a/M\leq 3\sqrt{3}/4$ we find
that the stable accretion disk is composed
of three different disks. The internal disk is situated between the radii $\tilde{r}_{lsco}^-$ and $\hat r _ +<r_{lsco}^+$ and is
made of stable counter-rotating particles with angular momentum $L=-L_-$. The radius $\hat r_+$ corresponds
to circular orbits with zero angular momentum $(V=E/\mu,\, V'(r)=0,\, L=0)$.
A second  disk made of stable corotating particles
with angular momentum $L=L_-$ is situated in the
region $\hat r _ + < r < r_{lsco} ^+$. Finally, the exterior stable disk is situated in the region $r>r_{lsco}^+$ and contains
corotating particles with $L=L_-$ and counter-rotating particles with $L=-L_+$. We conclude that the main difference
between a rotating black hole and a rotating naked singularity consists in the different geometric structure of their accretion
disks.

The study of the dynamics of test particles around compact rotating objects is surely interesting from the point of view of the
astrophysical phenomenology. However, an immediate application of this study will be in the physics of the accretion  disks as observed
around astrophysical rotating  objects (see \cite{Tod:1976ud, Wei:2010vca,Grib:2010zs} and also
\cite{Bini:2009cg,F2,Harada:2010yv,Harada:2011xz},  for the problem concerning  the extended theories of gravity see for example  \cite{Capozziello:2009jg}).
The matter constituents, plasma elements, are the material
of the electromagnetic jets as seen in the $X$--rays and $\gamma$--ray emissions.
In this respect, a detailed and proper description of the test particle dynamics  is the first step towards
the construction of a realistic model for accretion disks around Kerr sources (see  \cite{Ledvinka:1997im,F2,Bambi:2010fe,Zhuravlev:2011pi}, and also \cite{Staicova:2010qs} and \cite{Tamburini:2011tk}) .

In this work, we also explored the physics of naked singularities (see also  \cite{Pugliese:2010he,Pugliese:2010ps,dfl,Mio1Que}).
As no naked singularity has been yet observed and furthermore the existence of these objects is still
a subject under intensive theoretical debate, the analysis of the dynamical properties of these objects is clearly important either
for a formalization of a complete theoretical picture of the  physical features of these solutions, or for observational issues \cite{Penrose,Hawking:1973uf}, \cite{Patil:2011ya,Joshi:2011zm,Joshi:2011cx,Patil:2011yb} see also \cite{DiCriscienzo:2010gh,Virbhadra:2007kw}.
We expect to generalize this work to include the physical contribution of a charged source,
therefore, exploring the Kerr--Newman metric which properly describes the spacetime of   a
rotating, electrically charged, compact object in general relativity \cite{Interessantissimo}.

\section*{Acknowledgments}
Daniela Pugliese and Hernando Quevedo would like to thank the ICRANet for support. We would like to thank Andrea Geralico for helpful comments and discussions. One of us (DP) gratefully acknowledges financial support from the A. Della Riccia Foundation.
This work was supported in part by DGAPA-UNAM, grant No. IN106110.




\begin{thebibliography}{99}
%
\bibitem{Chandra} S. Chandrasekhar, \emph{The Mathematical Theory of Black
Holes}, Clarendon Press, Oxford and Oxford University Press, New
York, 1983.
%

%
\bibitem{Compr} D. J. Raine, E. Thomas, \emph{Black holes: an introduction},
Imperial College Press, 2010.

\bibitem{GRLI}
M. P . Hobson, G . P . Efstathiou
and A . N . Lasenby,
\emph{General Relativity An Introduction for Physicists}
Cambridge University Press, 2006.

\bibitem{Bardeen:1970zz}
  J.~M.~Bardeen,
  Nature {\bf 226} (1970) 64.

\bibitem{Bardeen:1972fi}
  J.~M.~Bardeen, W.~H.~Press and S.~A.~Teukolsky,
  Astrophys.\ J.\  {\bf 178} (1972) 347.

\bibitem{Bredberg:2011hp}
  I.~Bredberg, C.~Keeler, V.~Lysov and A.~Strominger,
  arXiv:1103.2355 [hep-th].


\bibitem{ZNK1980}
Z. Stuchlik,
Bull. Astron. Inst. Czechosl. \textbf{31} (1980), 129-144.





\bibitem{MTW} C. Misner, K. Thorne, J. Wheeler, \emph{Gravitation}, Freemann and
Company, 1973.
%

\bibitem{RuRR} R.
Ruffini,
\emph{On the Energetics of Black Holes, Le Astres Occlus}
(Les Houches 1972).



\bibitem{Pugliese:2010he}
  D.~Pugliese, H.~Quevedo and R.~Ruffini,
  arXiv:1003.2687 [gr-qc].

  %




\bibitem{Pugliese:2010ps}
  D.~Pugliese, H.~Quevedo and R.~Ruffini,
  Phys.\ Rev.\  D \  Vol.83, No.2.


\bibitem{Mio1Que}
  D.~Pugliese, H.~Quevedo and R.~Ruffini,
  Phys.\ Rev.\  D {\bf 83} (2011) 104052.

\bibitem{chandra42} S. Chandrasekhar, {\it Principles of Stellar Dynamics} (Dover Publications, New York, 1942).


\bibitem{Interessantissimo}V. Balek, J. Bicak, Z. Stuchlik
%
Bull. Astron. Inst.
Czechosl. 40 (1989),133-165 Publishing House of the Czechoslovak
Academy of Sciences.


\bibitem{Bob}
D. Bini, R. T. Jantzen
Class. Quantum Grav. 17 (2000) 1637–1647.


\bibitem{Komorowski:2009cg}
  P.~G.~Komorowski, S.~R.~Valluri and M.~Houde,
  Class.\ Quant.\ Grav.\  {\bf 26} (2009) 085001.

\bibitem{Komorowski:2010we}
  P.~G.~Komorowski, S.~R.~Valluri and M.~Houde,
  Class.\ Quant.\ Grav.\  {\bf 27} (2010) 225023.

\bibitem{Komorowski:2011cd}
  P.~G.~Komorowski, S.~R.~Valluri and M.~Houde,
  arXiv:1101.0996 [gr-qc].

\bibitem{Hackmann:2009nh}
  E.~Hackmann, V.~Kagramanova, J.~Kunz and C.~Lammerzahl,
  Europhys.\ Lett.\  {\bf 88}, 30008 (2009).


\bibitem{Favata:2010ic}
  M.~Favata,
  Phys.\ Rev.\  D {\bf 83} (2011) 024028.

\bibitem{Bini:2004sy}
  D.~Bini, C.~Cherubini, G.~Cruciani and R.~T.~Jantzen,
  Int.\ J.\ Mod.\ Phys.\  D {\bf 13} (2004) 1771.

\bibitem{Bini:1997eb}
  D.~Bini, P.~Carini and R.~T.~Jantzen,
  Int.\ J.\ Mod.\ Phys.\  D {\bf 6} (1997) 143.

\bibitem{UJM}
D. Bini, C. Cherubini,
A. Geralico, R. T. Jantzen,
Gen. Rel. Grav. (2008) 40:985–1012.



\bibitem{DeFelice:1972ad}
  F.~De Felice and M.~Calvani,
  Nuovo Cim.\  B {\bf 10} (1972) 447.

\bibitem{DeFelice:1979mc}
  F.~De Felice,
  Phys.\ Rev.\  D {\bf 19} (1979) 451.




\bibitem{Gariel:2007st}
  J.~Gariel, M.~A.~H.~MacCallum, G.~Marcilhacyand N.~O.~Santos,
  arXiv:gr-qc/0702123.



\bibitem{Chicone:2006rm}
  C.~Chicone and B.~Mashhoon,
  Class.\ Quant.\ Grav.\  {\bf 23} (2006) 4021.





\bibitem{Lake:2010bq}
  K.~Lake,
  Phys.\ Rev.\ Lett.\  {\bf 104} (2010) 211102
  [Erratum-ibid.\  {\bf 104} (2010) 259903].

\bibitem{Hackmann:2010zz}
  E.~Hackmann, C.~Lammerzahl, V.~Kagramanovaand J.~Kunz,
  Phys.\ Rev.\  D {\bf 81} (2010) 044020.


\bibitem{Bini:2004md}
  D.~Bini, F.~de Felice and A.~Geralico,
  Class.\ Quant.\ Grav.\  {\bf 21} (2004) 5427.

\bibitem{Bini:2004me}
  D.~Bini, F.~de Felice and A.~Geralico,
  Class.\ Quant.\ Grav.\  {\bf 21} (2004) 5441.

\bibitem{Bini:2005xt}
  D.~Bini, A.~Geralico and R.~T.~Jantzen,
  Class.\ Quant.\ Grav.\  {\bf 22} (2005) 4729.

\bibitem{Bini:2006pc}
  D.~Bini, A.~Geralico, R.~T.~Jantzen and F.~de Felice,
  Class.\ Quant.\ Grav.\  {\bf 23} (2006) 3287.

\bibitem{Dokuchaev:2011wm}
  V.~I.~Dokuchaev,
  arXiv:1103.6140 [gr-qc].

\bibitem{Maeda} S. Suzuki and K. Maeda
Phys. Rev. D 58, 023005 (1998).
\bibitem{Gonzalez:2011fb}
  G.~A.~Gonzalez and F.~Lopez-Suspes,
  arXiv:1104.0346 [gr-qc].


\bibitem{Pani:2010jz}
  P.~Pani, E.~Barausse, E.~Berti and V.~Cardoso,
  Phys.\ Rev.\  D {\bf 82} (2010) 044009.


\bibitem{Grumiller:2009gf}
  D.~Grumiller and A.~M.~Piso,
  arXiv:0909.2041 [astro-ph.SR].
\bibitem{F2}
F. de Felice and Bradley, M., 1988,
Class. Quantum Grav. 5, 1577.


\bibitem{Preti:2008zz}
  G.~Preti and F.~d.~Felice,
  Am.\ J.\ Phys.\  {\bf 76} (2008) 671.


\bibitem{1970}
  J.~M.~Bardeen,  Astrophys.\ J.\  {\bf 161}, 103 (1970).

\bibitem{palit09} A. Palit, A. Pachenko, N. G. Migranov, A. Bhadra, and K. K. Nandi, Int. J. Theor. Phys.
{\bf 48} (2009) 1271.

\bibitem{letelier1} P. S. Letelier, Phys. Rev. D {\bf 68} (2003) 104002.

\bibitem{letelier2} J. Ramos-Caro, J. F. Pedraza, and P. S. Letelier, Mon. Not. R. Astron. Soc. {\bf 413} (2011) 3105.




\bibitem{Tod:1976ud}
  K.~P.~Tod, F.~de Felice and M.~Calvani,
  Nuovo Cim.\  B {\bf 34} (1976) 365.








\bibitem{Wei:2010vca}
  S.~W.~Wei, Y.~X.~Liu, H.~Guo and C.~E.~Fu,
  arXiv:1006.1056 [hep-th].

\bibitem{Grib:2010zs}
  A.~A.~Grib and Y.~V.~Pavlov,
  arXiv:1007.3222 [gr-qc].




\bibitem{Bini:2009cg}
  D.~Bini, A.~Geralico, O.~Luongo and H.~Quevedo,
  Class.\ Quant.\ Grav.\  {\bf 26} (2009) 225006.



\bibitem{Harada:2010yv}
  T.~Harada and M.~Kimura,
  arXiv:1010.0962 [gr-qc].

\bibitem{Harada:2011xz}
  T.~Harada and M.~Kimura,
  arXiv:1102.3316 [gr-qc].

\bibitem{Capozziello:2009jg}
  S.~Capozziello, M.~De laurentis and A.~Stabile,
  Class.\ Quant.\ Grav.\  {\bf 27}, 165008 (2010).

\bibitem{Ledvinka:1997im}
  T.~Ledvinka, M.~Zofka and J.~Bicak,
  arXiv:gr-qc/9801053.


\bibitem{Bambi:2010fe}
  C.~Bambi and N.~Yoshida,
  Phys.\ Rev.\  D {\bf 82} (2010) 124037.

\bibitem{Zhuravlev:2011pi}
  V.~Zhuravlev and P.~Ivanov,
  arXiv:1103.5739.

\bibitem{Staicova:2010qs}
  D.~R.~Staicova and P.~P.~Fiziev,
  Astrophys.\ Space Sci.\  {\bf 332} (2011)385.



\bibitem{Tamburini:2011tk}
  F.~Tamburini, B.~Thide, G.~Molina-Terriza and G.~Anzolin,
  Nature Phys.\  {\bf 7} (2011) 195.






%
\bibitem{dfl} F. de Felice
arXiv:0710.0983v1 [gr-qc] 4 Oct 2007.


%

\bibitem{Penrose}
R. Penrose: Nuovo Cimento B 1 (1969) 252.
%
\bibitem{Hawking:1973uf}
  S.~W.~Hawking and G.~F.~R.~Ellis,
{\it  Cambridge University Press, Cambridge, 1973}.
%

\bibitem{Patil:2011ya}
  M.~Patil and P.~S.~Joshi,
  arXiv:1103.1082 [gr-qc].

\bibitem{Joshi:2011zm}
  P.~S.~Joshi, D.~Malafarina and R.~Narayan,
  arXiv:1106.5438 [gr-qc].

\bibitem{Joshi:2011cx}
  P.~S.~Joshi and D.~Malafarina,
  arXiv:1105.4336 [gr-qc].

\bibitem{Patil:2011yb}
  M.~Patil and P.~S.~Joshi,
  arXiv:1103.1083 [gr-qc].






\bibitem{DiCriscienzo:2010gh}
  R.~Di Criscienzo, L.~Vanzo and S.~Zerbini,
  JHEP {\bf 1005} (2010) 092.

\bibitem{Virbhadra:2007kw}
  K.~S.~Virbhadra and C.~R.~Keeton,
  Phys.\ Rev.\  D {\bf 77} (2008) 124014.











\end{thebibliography}
\end{document}